\documentclass[prd,aps,showpacs,tightenlines,notitlepage,nofootinbib,superscriptaddress,groupedaddress]{revtex4-1}
\usepackage[colorlinks=true]{hyperref}
\usepackage{mathrsfs}
\usepackage{amsmath}
\usepackage{amsfonts}
\usepackage{array}
\usepackage{verbatim}
\usepackage{epsfig}
\usepackage{graphicx} 
\usepackage[usenames, dvipsnames]{color}
\usepackage{marginnote}
\usepackage{slashed}
\usepackage{bm}

\newcounter{mynote}

\begin{document}
\title{The role of the chiral anomaly in polarized deeply inelastic scattering II:\\
Topological screening  and transitions from emergent axion-like dynamics}
\author{Andrey Tarasov$^{1,2}$}
\author{Raju Venugopalan$^3$}

\affiliation{$^1$Department of Physics, The Ohio State University, Columbus, OH 43210, USA\\
$^2$Joint BNL-SBU Center for Frontiers in Nuclear Science (CFNS) at Stony Brook University, Stony Brook, NY 11794, USA\\
$^3$Physics Department, Brookhaven National Laboratory,
Bldg. 510A, Upton, NY 11973, U.S.A.}

\begin{abstract}
In \cite{Tarasov:2020cwl}, we demonstrated that the structure function $g_1(x_B,Q^2)$ measured in polarized deeply inelastic scattering (DIS) is dominated by the triangle anomaly in both the 
Bjorken limit of large $Q^2$ and the Regge limit of small $x_B$. In the worldline formulation of quantum field theory, the triangle anomaly arises from the imaginary part of the worldline effective action. We show explicitly how a Wess-Zumino-Witten term coupling the topological charge density to a primordial isosinglet ${\bar \eta}$ arises in this framework. We demonstrate the fundamental role played by this contribution both in topological mass generation of the $\eta^\prime$ and in the cancellation of the off-forward pole arising from the triangle anomaly in the proton's helicity $\Sigma(Q^2)$. We recover the striking result by Shore and Veneziano that $\Sigma\propto \sqrt{\chi'(0)}$, where $\chi'$ is the slope of the QCD topological susceptibility in the forward limit. We  construct an axion-like effective action for $g_1$ at small $x_B$ that describes the interplay between gluon saturation and the topology of the QCD vacuum. In particular, we outline the role of ``over-the-barrier" sphaleron-like transitions in spin diffusion at small $x_B$. Such topological transitions can be measured in polarized DIS at a future Electron-Ion Collider.
\end{abstract}

\maketitle
\section{Introduction}
In our previous paper \cite{Tarasov:2020cwl} (henceforth Paper I),  we discussed the role of the chiral anomaly in the inclusive polarized deeply inelastic scattering (DIS) process
\begin{eqnarray}
e(k) + N(P, S)\to e(k') + X\,,
\end{eqnarray}
where $k$ denotes the four-momentum of the lepton ($e$) which scatters off a polarized target hadron ($N$) with four-momentum $P = (P^+, M^2/2P^+, 0_\perp)$ and four-spin $S$ (with $S^2 = -1$) via the exchange of a virtual photon $\gamma^\ast$ with four-momentum $q = k - k'$. 
We showed, within a powerful worldline formalism, that the anomaly provides the dominant contribution to the spin-dependent structure function $g_1(x_B,Q^2)$ in the Bjorken and  Regge limits of QCD.  The  former, for center-of-mass energies $\sqrt{s}\rightarrow \infty$, corresponds to the DIS kinematics $Q^2 = -q^2 \rightarrow \infty$ and the Bjorken variable $x_B = Q^2/(2P\cdot q)$ kept fixed; the latter refers to the limit $x_B\rightarrow 0$ and fixed $Q^2$. 

We further demonstrated in Paper I that the leading  perturbative contributions to $g_1$ have a power law divergence in the Mandelstam variable $t$ in the forward scattering limit $t\rightarrow 0$ in both Bjorken and Regge asymptotics. We noted that the nonperturbative dynamics which regulates this divergence is also what resolves the $U_A(1)$ problem in QCD. 
Indeed the fundamental role of this nonperturbative dynamics was  previously argued\footnote{Note that $\frac{1}{9} \Sigma(Q^2) \approx  \int_0^1 dx_B \,g_1(x_B,Q^2)$, with the $\approx$ sign denoting that a term needs to be added on the l.h.s corresponding to a linear combination of the isotriplet and isooctet axial vector charges of the proton. These contributions are  weakly dependent on $Q^2$ and we will ignore them henceforth to focus on isosinglet contributions to $g_1(x_B,Q^2)$.} to be true~\cite{Jaffe:1989jz,Veneziano:1989ei,Shore:1990zu} for the first moment of $g_1(x_B,Q^2)$, the isosinglet quark helicity $\Sigma(Q^2)$.

In particular,  Shore and Veneziano~\cite{Shore:1990zu,Shore:1991dv} showed that $\Sigma(Q^2)\propto \sqrt{d\chi/dt}|_{t\rightarrow 0}$, where $\chi$ is the topological susceptibility of the QCD vacuum. The scale controlling this quantity is the $\eta^\prime$ mass, which is finite even in the chiral limit~\cite{Witten:1979vv,Veneziano:1979ec}. The derivation in \cite{Shore:1990zu,Shore:1991dv} extensively employed functional chiral Ward identities that follow from the Wess-Zumino action for QCD coupled to external sources~\cite{Zumino,Wess:1971yu}. 

Further, invoking QCD sum rule arguments to compute $\chi(t)$, Narison, Shore and Veneziano~\cite{Narison:1994hv,Narison:1998aq} showed that their results for $\Sigma(Q^2)$ are in good agreement with HERMES~\cite{Airapetian:2007mh} and COMPASS~\cite{Alekseev:2010ub} data. A comprehensive review of this ``topological screening" picture of why the quark helicity is anomalously small (when compared to simple quark model expectations based on the OZI rule~\cite{Ellis:1973kp}) can be found in \cite{Shore:2007yn}. 

Another nonperturbative approach to computing the proton's helicity follows 't Hooft's seminal work~\cite{tHooft:1976snw,tHooft:1986ooh} relating classical instanton configurations in the QCD vacuum to $U_A(1)$ breaking and the origin of the mass of the $\eta^\prime$. This description of the anomaly in the language on instantons, while by no means unique, is consistent with Veneziano's approach~\cite{Veneziano:1979ec}. For the instanton picture of the quark helicity in polarized DIS, we refer the reader to 
\cite{Forte:1989qq,Forte:1990xb,Dorokhov:1993ym,Qian:2015wyq,Schafer:1996wv}.

In this paper, we will develop an alternative formulation of the problem in a generalization of the worldline framework we discussed in Paper I to include the coupling of the Dirac fields to scalar, pseudoscalar and axial vector fields representing low energy degrees of freedom in the QCD effective action. As we noted in Paper I, the coupling of the isosinglet axial vector current to low energy dynamics of gauge fields (represented by the topological charge density) arises from the imaginary part of the  worldline effective action. The generalization of the worldline action to include all possible low energy degrees of freedom will therefore include  additional such imaginary terms that must be taken into account to fully describe the dynamics of the anomaly.

In a certain sense, as will become apparent, this worldline approach to the role of the chiral anomaly in proton spin threads a line between the two aforementioned approaches, the Shore-Veneziano approach employing 
chiral Ward identities and that of instanton based approaches. This third way will prove especially beneficial when we turn our attention to $g_1(x_B,Q^2)$ at small $x_B$. 

Our first objective is to understand in detail the cancellation of the anomaly pole in $\Sigma(Q^2)$; we will show how one recovers the Shore-Veneziano results.
Specifically, in the Bjorken limit, the anomaly requires we replace the  isosinglet current $J_5^\mu$ (whose expectation value in the polarized proton ground state is $\Sigma(Q^2)$)  with 
\begin{equation}
\label{eq:anomaly}
 J_5^\mu\rightarrow \frac{l^\mu}{l^2} \Omega\,,   
\end{equation}
where $\Omega=\frac{\alpha_s}{4\pi} {\rm Tr}\left(F {\tilde F}\right)$ is the topological charge density expressed in terms of the field strength tensor and its dual $F_{\mu\nu}$ defined as $\tilde{F}_{\mu\nu} = \frac{1}{2}\,\epsilon_{\mu\nu\rho\sigma}F^{\rho\sigma}$. Here $l= P^\prime-P$ is the four vector corresponding to the four-momentum transfer from the proton to the DIS probe, with $l^2=t$, the Mandelstam variable. The r.h.s of Eq.\,(\ref{eq:anomaly}) corresponds to a massless exchange from the topological charge density $\Omega$ which couples nonperturbatively to the proton. However this massless exchange can also be mediated (in the chiral limit) by a massless flavor singlet pseudoscalar
field ${\bar \eta}$, with a similar nonperturbative coupling to the proton. One may in the chiral limit, and at large $N_c$, interpret ${\bar \eta}$ as the ``primordial" ninth Goldstone boson. The absence of this field in the hadron spectrum is of course the $U_A(1)$ problem. It is resolved by the nontrivial susceptibility of the QCD vacuum which generates the large mass of the $\eta^\prime$~\cite{Witten:1979vv,Veneziano:1979ec}.

We will derive explicitly in our approach the Wess-Zumino-Witten (WZW)~\cite{Wess:1971yu,Witten:1983tw} term in the imaginary part of the effective action that couples the isosinglet pseudoscalar field ${\bar \eta}$ to the topological charge density $\Omega$. This term plays a fundamental role in the cancellation of the anomaly pole because an identical pole exists in the ${\bar \eta}$ exchange with the proton. We will show in detail how this cancellation arises in our approach both at leading order in the $\Omega$ and ${\bar \eta}$ exchange, and to all orders. We will further show how this interplay results in an anomalous Goldberger-Treiman relation~\cite{Veneziano:1989ei} and in topological mass generation of the $\eta^\prime$. Thus as noted, the role played by the chiral anomaly in the proton's spin is deeply tied to the resolution of the $U_A(1)$ problem. 

While these conclusions, if not the approach, are familiar from the work of Veneziano and collaborators, our framework can be extended to the  computation of $g_1(x_B,Q^2)$ in the Bjorken and Regge asymptotics of QCD. This is because, as demonstrated in Paper I, the triangle anomaly in Eq.\,(\ref{eq:anomaly}) is the dominant contribution to $g_1(x_B,Q^2)$ in both limits. In the Bjorken limit, as we will discuss briefly,  the formalism of Shore and Veneziano goes through for $g_1(x_B,Q^2)$, precisely as for its first moment $\Sigma(Q^2)$. The corresponding matrix element can be computed on the lattice similarly to prior lattice computations~\cite{Liang:2018pis,Alexandrou:2019brg,Mejia-Diaz:2017hhp,Lin:2018obj}. For discussions of how to extract $\Sigma(Q^2)$ directly from the slope (with $t$) of the topological susceptibility, see \cite{Giusti:2001xh}; a recent review of lattice extractions of the topological susceptibility can be found in \cite{Bali:2021qem}. 

The situation is quite different at small $x_B$ because of the phenomenon of gluon saturation and the emergence of a corresponding saturation scale $Q_S$~\cite{Gribov:1984tu,Mueller:1985wy}. In Regge asymptotics, this scale is larger than the scales governing intrinsically nonperturbative dynamics in QCD. In the unpolarized proton, the gauge configurations representing the saturated state are static classical configurations~\cite{McLerran:1993ka,McLerran:1993ni,McLerran:1994vd} and their dynamics is described by the Color Glass Condensate (CGC) Effective Field Theory (EFT)~\cite{Gelis:2010nm,Kovchegov:2012mbw}. In the polarized proton, such configurations can be dynamical on the time scales of spin diffusion. Thus the classical configurations responsible for spin diffusion in this high energy asymptotics are not the energy degenerate instanton solutions describing tunneling between different $\theta$-vacua (each corresponding to distinct integer valued Chern-Simons number) but ``over the barrier" topological transitions that are enhanced by the large dynamical saturation scale. A well-known example of such transitions are the sphaleron solutions~\cite{Klinkhamer:1984di} conjectured to play a major role in electroweak baryogenesis~\cite{Kuzmin:1985mm}. Similar sphaleron-like topological transitions have also been discussed in the QCD context both in-and out-of equilibrium~\cite{McLerran:1990de,Moore:2010jd,Shuryak:2002qz,Mace:2016svc}. 

With these considerations in mind, we will write down an ``axion-like" effective action for $g_1(x_B,Q^2)$ at small $x_B$ that captures both the physics of gluon saturation and spin diffusion, which are respectively controlled by $Q_S$ and the Yang-Mills topological susceptibility $\chi_{\rm YM}$. Depending on their relative magnitude (specifically of $Q_S$ and $m_{\eta^\prime}$, the $\eta^\prime$ mass), the gauge field configurations are either, as noted,  ``conventionally" sphaleron-like (for $m_{\eta^\prime} > Q_S$) or novel topological shock wave configurations (for $Q_S > m_{\eta^\prime}$).
We will discuss the consequences of this interplay and other qualitative features of the dynamics captured by the effective action.  A quantitative study 
of QCD evolution in this framework, and phenomenological consequences thereof for polarized DIS measurements at the Electron-Ion Collider (EIC)~\cite{Accardi:2012qut,Aschenauer:2017jsk}, will be discussed in follow-up work~\cite{PaperIII}. 

The paper is organized as follows. In the next section, we will briefly recapitulate the worldline derivation of $g_1(x_B,Q^2)$ in Paper I which demonstrated the dominance of the triangle graph of the anomaly in both Bjorken and Regge asymptotics. We will emphasize that the triangle graph, and indeed all dynamical effects of the anomaly, can be computed directly from the imaginary part of the worldline effective action. In Section III, we will discuss the extension of the imaginary part of the worldline effective action  to include, in addition to gauge fields, the coupling of the fermions to scalar, pseudoscalar, and axial vector fields, which capture the dynamics of low energy modes in the QCD effective action. We will then show in this formalism how the WZW  term coupling ${\bar \eta}$ to $\Omega$  arises. The profound consequences of this result for the cancellation of the pole of the anomaly is discussed in detail in Section IV. In Section V, we will write down the effective action for $g_1(x_B,Q^2)$ and sketch its key features. 
We end in Section VI with a summary, outlook on future work, and a discussion of some of the larger implications of our work. 

In Appendix A, we provide details of the derivation of the WZW term from the imaginary part of the worldline effective action. In Appendix B, we will outline the derivation of the CGC effective action in the worldline formalism for the case where the hadron is not polarized. In Appendix C, we will extend this discussion to the polarized proton case. In particular, we will present an argument for the failure of high energy expansions in perturbative QCD for operators in the polarized proton that are sensitive to the anomaly. 

\section{Anomaly dominance of $g_1(x_B,Q^2)$}
\label{sec:g1-worldline}

We will first briefly recapitulate\footnote{We refer interested readers to Paper I  for more details~\cite{Tarasov:2020cwl}.} here the worldline derivation in Paper I where we demonstrated that the triangle graph of the anomaly dominates $g_1(x_B,Q^2)$ in both the Bjorken and Regge limits of DIS. We will also discuss the result in  Paper I showing that the triangle graph can be recovered directly from the imaginary part of the worldline effective action. This  will serve to motivate our focus in the rest of this paper on the role of the imaginary part of the effective action. As is well-known~\cite{Schubert:2001he}, its contributions can be fundamentally understood as arising from the noninvariance of the measure of the QCD path integral under a global chiral rotation~\cite{Fujikawa:1979ay}. 

The $g_1(x_B,Q^2)$ structure function can be extracted most generally from the antisymmetric piece of the hadron tensor~\cite{Anselmino:1994gn}, which can be expressed as 
\begin{eqnarray}
\tilde{W}_{\mu\nu}(q, P, S)= \frac{2M_N}{P\cdot q}\epsilon_{\mu\nu\alpha\beta}\, q^\alpha\Big\{ S^\beta g_1(x_B, Q^2) + \Big[S^\beta - \frac{(S\cdot q)P^\beta}{P\cdot q}\Big]g_2(x_B, Q^2)\Big\}\,,
\label{WA}
\end{eqnarray}
where $M_N$ denotes the proton mass and the totally antisymmetric Levi-Civita tensor $\epsilon_{\mu\nu\alpha\beta}$  is defined with $\epsilon_{0123} = -1$. For a  longitudinally polarized target, $S^\mu(\lambda) \simeq \frac{2{\tilde \lambda}_P}{M_N}P^\mu$, with ${\tilde \lambda}_P = \pm \frac{1}{2}$ representing the proton's helicity; in this case, the $g_2$ structure function does not contribute. 

The full hadron tensor itself can be expressed as the imaginary part of the expectation value of the polarization tensor: 
\begin{eqnarray}
W^{\mu\nu}(q, P, S) = \frac{1}{\pi e^2}{\rm Im}\  \int d^4x~ e^{iqx}\langle P,S| \frac{\delta^2 \Gamma[a, A]}{\delta a_\mu(\frac{x}{2})\delta a_\nu(-\frac{x}{2})} |P,S\rangle\,,
\label{TprodEffex}
\end{eqnarray}
 where $\Gamma[a,A]$ is the QED+QCD worldline effective action,  $a_\mu(x)$ denotes the QED electromagnetic field and $A$ is the four-vector denoting the QCD gauge field. Its antisymmetric piece, which appears on the l.h.s of Eq.\,(\ref{WA}), can be written as 
 \begin{eqnarray}
i\tilde{W}^{\mu\nu}(q, P, S) = \frac{ 1 }{ 2 \pi e^2}\,{\rm Im}\ \int d^4x \,e^{-iqx} 
 \int \frac{d^4k_1}{(2\pi)^4} \int \frac{d^4k_3}{(2\pi)^4} e^{-ik_1 \frac{x}{2}} e^{ik_3 \frac{x}{2} }\langle P,S| {\tilde{\Gamma}}^{\mu\nu}_A[k_1, k_3] |P,S\rangle\,.
\label{WModFu}
\end{eqnarray}
Here ${\tilde{\Gamma}}^{\mu\nu}_A[k_1, k_3] \equiv {\tilde{\Gamma}}^{\mu\nu}[k_1, k_3] - (\mu\leftrightarrow\nu)$, with 
\begin{eqnarray}
{\tilde{\Gamma}}^{\mu\nu}[k_1, k_3] \equiv \int d^4z_1 d^4z_3 \frac{\delta^2 \Gamma[a, A]}{\delta a_\mu(z_1)\delta a_\nu(z_3)}|_{a=0} \,e^{ik_1 z_1} e^{ik_3 z_3}\,,
\label{secder}
\end{eqnarray}
where $k_1$ and $k_3$ denote the incoming photon four-momenta. Because the r.h.s of Eq.\,(\ref{WModFu}) corresponds to the same in-out ground state of the proton,  $k_1=-k_3=-q$ in the forward limit. However to extract the infrared pole of the anomaly, as discussed at length in Paper I, one needs to keep the incoming photon momenta distinct in computing the off-forward matrix element $\langle P'|\dots|P\rangle$ in Eq.\,(\ref{TprodEffex}), with $P'-P\equiv l$ and $t=l^2$, and then subsequently take $t\to0$ in the final expression. 

To compute $i\tilde{W}^{\mu\nu}(q, P, S)$, and hence $g_1(x_B,Q^2)$, we employed a powerful worldline formalism in Paper I; to one loop accuracy, the  QED+QCD effective action in this formalism\footnote{As we will discuss shortly, and in greater detail in Section~\ref{sec:WZW}, we will replace $\Gamma\rightarrow {\cal W}$, a more general expression which includes additional couplings with scalar, pseudoscalar and axial vector fields; we will split the latter into real and imaginary pieces.} can be expressed  as~\cite{Schubert:2001he}
\begin{eqnarray}
&&\Gamma [a, A] = -\frac{1}{2} \int^T_0 \frac{dT}{T} {\rm Tr_c} \int \mathcal{D}x \int \mathcal{D} \psi \exp\Big\{-\int^T_0 d\tau \Big(\frac{1}{4} \dot{x}^2 + \frac{1}{2}\psi_\mu\dot{\psi}^\mu + ig\dot{x}^\mu (A_\mu + a_\mu) - ig \psi^\mu \psi^\nu F_{\mu\nu}(A+a)\Big)\Big\}\,,
\nonumber\\
\label{MLag}
\end{eqnarray}
where $x^\mu(\tau)$  and $\psi^\mu(\tau)$ are respectively 0+1-dimensional scalar coordinate and Grassmann variables coupled to the background electromagnetic ($a_\mu$) and gluon ($A_\mu$) fields. Note that the scalar functional integral has periodic (P) boundary conditions while the Grassmannian functional integral has anti-periodic (AP) boundary conditions. 

In this formalism,  ${\tilde{\Gamma}}^{\mu\nu}_A[k_1, k_3]$ on 
the r.h.s of Eq.\,(\ref{WModFu}) can be written, to one loop accuracy, as 
\begin{eqnarray}
&&{\tilde \Gamma}^{\mu\nu}_A[k_1, k_3] = \int \frac{d^4k_2}{(2\pi)^4} \int \frac{d^4k_4}{(2\pi)^4}~
\Gamma^{\mu\nu\alpha\beta}_A[k_1, k_3, k_2, k_4]~ {\rm Tr_c}({\tilde A}_\alpha(k_2) {\tilde A}_\beta(k_4))\,,
\label{Hadgamma}
\end{eqnarray}
where the $ {\tilde A}$ denote the Fourier transforms of the background gauge fields, the trace is over their color degrees of freedom and 
the box diagram in the r.h.s takes the form,
\begin{eqnarray}
&&\Gamma^{\mu\nu\alpha\beta}_A[k_1, k_3, k_2, k_4] = -\frac{g^2e^2 e_f^2 }{2}\int^\infty_0\frac{dT}{T} ~ \int \mathcal{D}x\int \mathcal{D}\psi~\exp\Big\{-\int^T_0 d\tau \Big(\frac{1}{4} \dot{x}^2 + \frac{1}{2}\psi\cdot \dot{\psi} \Big)\Big\}
\nonumber\\
&&\times \prod^4_{k=1}\int^T_0 d\tau_k~ \Big[\sum^9_{n=1}\mathcal{C}^{\mu\nu\alpha\beta}_{n;(\tau_1,\tau_2,\tau_3,\tau_4)}[k_1, k_3, k_2, k_4] - (\mu\leftrightarrow\nu)\Big]e^{i\sum^4_{i=1}k_i x_i} \,.
\label{GammawithC}
\end{eqnarray}
The coordinate ($x_i \equiv x(\tau_i)$) and Grassmann variables ($\psi_i\equiv \psi(\tau_i)$) in the coefficients $\mathcal{C}^{\mu\nu\alpha\beta}_{n;(\tau_1,\tau_2,\tau_3,\tau_4)}[k_1, k_3, k_2, k_4]$ on the r.h.s depend on the proper time coordinates of the interaction of the worldlines with the external electromagnetic and gauge fields. Fully general expressions for these coefficients were provided in Paper I.

The worldline representation of the box diagram provides a useful intuition by mapping the ordering of the momentum labels of the four vertices to that of the corresponding proper times. This allows one to understand the usual triangle limit of the box diagram in Bjorken asymptotics as ``pinching" $\tau_1\rightarrow \tau_3$  as $k_1\rightarrow k_3$ (with corrections of order $1/Q$). More unexpectedly, it allows one to interpret  Regge asymptotics as $\tau_2\rightarrow \tau_4$ in the shockwave limit $k_2\rightarrow k_4$ of the gauge fields with corrections of order $O(\frac{x_B}{x})$, as $x_B\rightarrow 0$ for  gluon momenta carrying a finite but small fraction $x$ of the hadron's large ``+" momentum. This gives rise to an  ``inverted triangle" which too is sensitive to the anomaly. 

Indeed, we showed in Paper  I that the computation of Eq.\,(\ref{Hadgamma}) in either limit gives identically, 
\begin{eqnarray}
\label{eq:g1-Bj-Regge}
 S^\mu g_1(x_B, Q^2) = \sum_f e_f^2 \frac{\alpha_s}{i\pi M_N}
 \int^1_{x_B} \frac{dx}{x} 
~ \Big( 1  - \frac{x_B}{x} \Big) \int \frac{d\xi}{2\pi} e^{-i\xi x }\lim_{l_\mu \to 0} \frac{ l^\mu }{l^2  }\langle P^\prime,S| {\rm Tr_c} F_{\alpha\beta}(\xi n) \tilde{F}^{\alpha\beta}(0) |P,S\rangle\,.
\end{eqnarray}
Here $\alpha_s$ is the QCD coupling,  the sum is over $n_f$  quark  flavors\footnote{We will assume these to be massless and $n_f=3$ for our discussion.} with electric charge $e_f$.
 The structure of the r.h.s is dominated by the triangle graph in either limit; therefore the operator that governs the r.h.s is the topological charge density $\Omega=\frac{\alpha_s}{4\pi} {\rm Tr}\left(F {\tilde F}\right)$, with $\tilde{F}_{\mu\nu} = \frac{1}{2}\,\epsilon_{\mu\nu\rho\sigma}F^{\rho\sigma}$. While the structure of the operator is identical in both limits, we will see that one obtains qualitatively different results in the two limits. The other noteworthy feature of Eq.\,(\ref{eq:g1-Bj-Regge}) is the pole $l^\mu/l^2$, which is a consequence of the anomaly equation for $J_\mu^5$; the cancellation of this pole will be the topic of Section~\ref{sec:Shore-Veneziano}. 

We also showed in Paper I that the anomaly can be extracted directly and with relative ease~\cite{Schubert:2001he,Mueller:2017arw} from the imaginary part of the effective action. Contributions to the imaginary part ${\cal W}_I$ (as we will discuss shortly) can be extracted by adding auxiliary terms to the worldline effective action that contain odd powers of $\gamma_5$. To extract the triangle graph, it is sufficient to include the interaction term with the axial vector field $\slashed{B}\gamma_5 $~\cite{McKeon:1998et,Mondragon:1995ab,DHoker:1995aat,Schubert:2001he,Mueller:2017arw,Mueller:2017lzw}:
\begin{eqnarray}
&&{\cal W}_I[A, B] = -\frac{1}{2} {\rm Tr_c} \int^\infty_0 \frac{dT}{T} \, \int \mathcal{D}x \int_{AP} \mathcal{D} \psi
\nonumber\\
&&\times\exp\Big\{-\int^T_0 d\tau \Big(\frac{1}{4} \dot{x}^2 + \frac{1}{2}\psi_\mu\dot{\psi}^\mu + ig\dot{x}^\mu A_\mu - ig \psi^\mu \psi^\nu F_{\mu\nu} - 2i\psi_5 \dot{x}^\mu \psi_\mu \psi_\nu B^\nu + i \psi_5 \partial_\mu B^\mu + (D-2) B^2\Big)\Big\}\,,
\label{A5action}
\end{eqnarray}
where $D$ is the number of spacetime dimensions and $\psi_5$ is the Grassmann counterpart of  the $\gamma_5$ matrix in the worldline framework. The structure of the $B$-dependent terms, as we will briefly review in Section~\ref{sec:WZW}, comes from exponentiating the phase of the Dirac determinant in the QCD effective action~\cite{McKeon:1998et,Mondragon:1995ab,DHoker:1995aat}.

Since $J_5^\mu$ couples to $B^\mu$, its expectation value is obtained by taking  the functional derivative of ${\cal W}_I[A, B]$ with respect to $B$ and then setting the latter equal to zero: 
\begin{eqnarray}
\langle P^\prime,S| J^\kappa_5 |P,S\rangle = \int d^4y\, \frac{\partial {\cal W}_I[A, B]}{\partial B_{\kappa}(y)}\Big|_{B_\kappa=0} e^{ily} \equiv \Gamma^\kappa_5[l]\,,
\end{eqnarray}
which gives,
\begin{eqnarray}
\Gamma^\kappa_5[l] &=& \frac{i}{2} {\rm Tr_c} \int^\infty_0 \frac{dT}{T} \, \int \mathcal{D}x \int_{AP} \mathcal{D} \psi
~\int^T_0 d\tau_l~ \psi_5 \Big(i  l^\kappa + 2 \psi^\kappa_l \dot{x}_l \cdot\psi_l \Big) e^{ilx_l}
\nonumber\\
&\times& \exp\Big\{-\int^T_0 d\tau \Big(\frac{1}{4} \dot{x}^2 + \frac{1}{2}\psi_\mu\dot{\psi}^\mu + ig\dot{x}^\mu A_\mu - ig \psi^\mu \psi^\nu F_{\mu\nu} \Big)\Big\}\,,
\label{trianglebefexp}
\end{eqnarray}
where $\tau_l$ is the proper time coordinate of the $B$-field insertion into the worldline, and $l$ is the incoming momentum.  We use the shorthand notation $x_l\equiv x(\tau_l)$, $\psi_l\equiv\psi(\tau_l)$.

\begin{figure}[htb]
 \begin{center}
 \includegraphics[width=120mm]{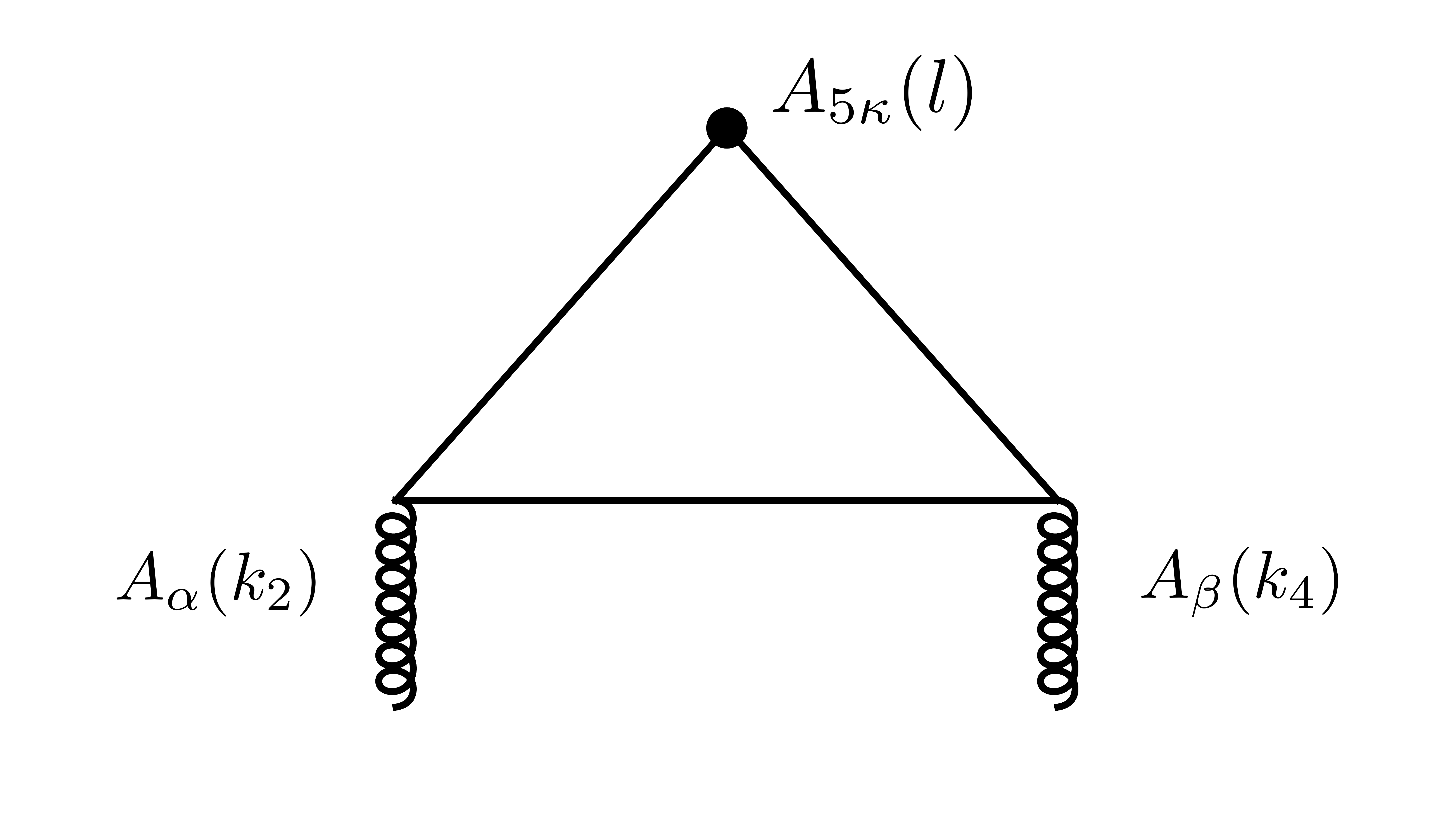}
 \end{center}
 \caption{\label{fig:tringlegraph}The  triangle graph representing the vector-vector-axial vector (VVA) coupling of the chiral anomaly.}
 \end{figure}
 
 Expanding the phase in Eq.\,(\ref{trianglebefexp}) to  second order in the coupling constant,
\begin{eqnarray}
\label{triangleaftexp}
&&\Gamma^\kappa_5[l] = -\frac{ig^2}{2} {\rm Tr_c} \int^\infty_0 \frac{dT}{T} \, \int \mathcal{D}x \int_{AP} \mathcal{D} \psi
~\int^T_0 d\tau_l~ \psi_5 \Big(i  l^\kappa + 2 \psi^\kappa_l \dot{x}_l \cdot\psi_l \Big) e^{ilx_l}\int^T_0 d\tau_2 \int^T_0 d\tau_4 
\nonumber \\
&&\times \Big( \dot{x}^\alpha_2 A_\alpha(x_2) -  2 \psi^\lambda_2 \psi^\alpha_2 \partial_\lambda A_\alpha(x_2) \Big) \Big( \dot{x}^\beta_4 A_\beta(x_4) -  2\psi^\eta_4 \psi^\beta_4 \partial_\eta A_\beta(x_4) \Big)
\exp\Big\{-\int^T_0 d\tau \Big(\frac{1}{4} \dot{x}^2 + \frac{1}{2}\psi_\mu\dot{\psi}^\mu \Big)\Big\}\,,
\nonumber\\
\end{eqnarray}
we can rewrite this equation as 
\begin{eqnarray}
&&\Gamma^\kappa_5[l] = \int \frac{d^4k_2}{(2\pi)^4} \int \frac{d^4k_4}{(2\pi)^4} ~\Gamma^{\kappa\alpha\beta}_5[l,k_2,k_4]~{\rm Tr_c} A_\alpha(k_2)  A_\beta(k_4) \,,
\label{trianglFour}
\end{eqnarray}
where the VVA vertex function shown in Fig. \ref{fig:tringlegraph} can be expressed as
\begin{eqnarray}
&&\Gamma^{\kappa\alpha\beta}_5[l,k_2,k_4] \equiv -\frac{ig^2}{2}  \int^\infty_0 \frac{dT}{T} \, \int \mathcal{D}x \int_{AP} \mathcal{D} \psi
~\int^T_0 d\tau_l~ \psi_5 \Big(i  l^\kappa + 2 \psi^\kappa_l \dot{x}_l \cdot\psi_l \Big) e^{ilx_l}
\label{trianglFourArg}\\
&&\times\int^T_0 d\tau_2 \int^T_0 d\tau_4 \Big( \dot{x}^\alpha_2  +  2 i \psi^\alpha_2 \psi^\lambda_2 k_{2\lambda} \Big)  e^{ik_2x_2} \Big( \dot{x}^\beta_4 + 2 i \psi^\beta_4 \psi^\eta_4 k_{4\eta} \Big) e^{ik_4 x_4}
\exp\Big\{-\int^T_0 d\tau \Big(\frac{1}{4} \dot{x}^2 + \frac{1}{2}\psi_\mu\dot{\psi}^\mu \Big)\Big\}\,.
\nonumber
\end{eqnarray}

This three-point function $\Gamma^{\kappa\alpha\beta}_5[l,k_2,k_4]$  has a $\psi_5$ in the argument of the Grassmannian functional integral which changes the boundary condition  from being antiperiodic (AP) to being periodic (P).  As a result, 
the Grassmann variables in the functional integral acquire a zero mode, which can 
be separated out from the nonzero modes in the action, and in the measure, as
\begin{eqnarray}
\label{eq:zero}
\psi^\mu(\tau) = \psi^\mu_0 + \xi^\mu(\tau)\,;\ \ \ \int_P \mathcal{D}\psi = \int d^4\psi_0 \int_P \mathcal{D}\xi\,;\ \ \ \int^T_0 d\tau \,\xi(\tau) = 0\,.
\end{eqnarray}
Separating out the zero mode thus, we obtain 
\begin{eqnarray}
&&\Gamma^{\kappa\alpha\beta}_5[l,k_2,k_4] \equiv -\frac{ig^2}{2}  \int^\infty_0 \frac{dT}{T} \, \int \mathcal{D}x \int d^4\psi_0 \int_P \mathcal{D}\xi
~\int^T_0 d\tau_l~ \Big(i  l^\kappa + 2 \psi^\kappa_l \dot{x}_l \cdot\psi_l \Big) e^{ilx_l}
\nonumber\\
&&\times\int^T_0 d\tau_2 \int^T_0 d\tau_4 \Big( \dot{x}^\alpha_2  +  2 i \psi^\alpha_2 \psi^\lambda_2 k_{2\lambda} \Big)  e^{ik_2x_2} \Big( \dot{x}^\beta_4 + 2 i \psi^\beta_4 \psi^\eta_4 k_{4\eta} \Big) e^{ik_4x_4}
\exp\Big\{-\int^T_0 d\tau \Big(\frac{1}{4} \dot{x}^2 + \frac{1}{2}\psi_\mu\dot{\psi}^\mu \Big)\Big\}\Big|_{\psi = \psi_0+\xi}\,.
\nonumber\\
\end{eqnarray}

The evaluation of the functional integrals over $x$ and $\xi$, as well as the integral over zero mode $\psi_0$, is straightforward and discussed at length in 
\cite{Tarasov:2019rfp,Tarasov:2020cwl}. We obtain, 
\begin{eqnarray}
\label{eq:VVA1}
&&\Gamma^{\kappa\alpha\beta}_5[l, k_2, k_4] =  \frac{ 1 }{2\pi^2}\frac{k^\kappa_{2} + k^\kappa_{4}}{ (k_2 + k_4)^2 } ~ \epsilon^{\alpha\sigma\beta\lambda} k_{2\sigma} k_{4\lambda}  (2\pi)^4 \delta^4(l + k_2 + k_4)\,,
\end{eqnarray}
which agrees with the result given in Ref.~\cite{Schubert:2001he}.

Substituting the VVA vertex function back into Eq.\,(\ref{trianglFour}), we  obtain, 
\begin{eqnarray}
&&\Gamma^\kappa_5[l] = \frac{ 1 }{4\pi^2} \frac{l^\kappa}{ l^2 } \int \frac{d^4k_2}{(2\pi)^4} \int \frac{d^4k_4}{(2\pi)^4} ~ {\rm Tr_c} F_{\alpha\beta}(k_2)  \tilde{F}^{\alpha\beta}(k_4)~(2\pi)^4 \delta^4(l + k_2 + k_4)\,,
\end{eqnarray}
the operator structure of which, up to kinematic factors,  is identical to the  expressions that lead the principal result of Paper I (given here in Eq.\,(\ref{eq:g1-Bj-Regge})). 

A common interpretation of the first moment $\Sigma$ is that of a local operator because the  corresponding integral over $x_B$ can be written as the local operator $J_\mu^5$, which is often interpreted as being qualitatively distinct from the operator in Eq.\,(\ref{eq:g1-Bj-Regge}). However, as also emphasized previously in \cite{Jaffe:1989jz}, the presence of the infrared pole ensures that $J_\mu^5$ receives an intrinsically nonlocal contribution; indeed, as we shall see, $J_\mu^5$ can be expressed in terms of the QCD  topological susceptibility, which is manifestly nonlocal. Further corroboration  follows from our result that the anomaly dominates both in Bjorken  and Regge asymptotics. While the operator product expansion may be employed in the former limit, it cannot be presumed to hold in the latter. One must therefore interpret the r.h.s
of Eq.\,(\ref{eq:g1-Bj-Regge}) as a smearing of the topological charge density $\Omega$. The treatment of $\Omega$ as an intrinsic low energy degree of freedom, on par with the Goldstone modes of chiral symmetry breaking, is discussed extensively in \cite{Veneziano:1989ei,Shore:1990zu,Shore:1991dv,Shore:2007yn} and will be be addressed in the following sections.

\section{ WZW ${\bar \eta}$ term from the imaginary part of the worldline effective action}
\label{sec:WZW}

From Eq.\,(\ref{eq:g1-Bj-Regge}), we see that the triangle anomaly generates a 
contribution, proportional to the topological charge density, that diverges in the 
forward limit. This is of course untenable and there must be other nonperturbative contributions that cancel this power law divergence. As observed previously in the literature~\cite{Jaffe:1989jz,Veneziano:1989ei,Shore:1990zu,Shore:1991dv}, this cancellation can be understood as arising (in the chiral limit and at large $N_c$) from the exchange of a  ``primordial" ninth Goldstone boson ${\bar \eta}$ arising from the spontaneous symmetry breaking of the flavor group  $U_L(3)\times U_R(3)$ to the vector group $U_V(3)$. 

There is of course no ${\bar \eta}$ Goldstone pole just as there is no anomaly pole in the QCD spectrum; the appearance of both, and their cancellation,  are features of a particular limit of the theory that do not survive when one fully accounts for its rich nonperturbative dynamics.  The important point to note however is that the same physics that ensures the former by generating a massive $\eta^\prime$ meson (the famous ``$U_A(1)$ problem") is also what ensures the latter\footnote{The $U_A(1)$ problem is resolved by the nontrivial susceptibility of the QCD vacuum; as noted, an attractive mechanism that generates this susceptibility is provided by instanton mediated interactions~\cite{tHooft:1976snw,tHooft:1986ooh}.}. In other words, the dynamical interplay between the physics of the anomaly, and that of the isosinglet pseudoscalar $U_A(1)$ sector of QCD resolves both problems simultaneously: the lifting of the ${\bar \eta}$ pole by topological mass generation of the $\eta^\prime$ and the cancellation of the anomaly pole. This will be shown explicitly in Section~\ref{sec:Shore-Veneziano}. 

Before we get there, we will first show how such contributions arise in the worldline formalism. In particular, we will derive the Wess-Zumino-Witten (WZW) term that couples the pseudoscalar isosinglet ${\bar \eta}$ field to the topological charge density. The presence of this term is crucial for our discussion in Section~\ref{sec:Shore-Veneziano}. 

The interplay of perturbative and nonperturbative dynamics is captured in the worldline formalism by parametrizing the low frequency modes of the Dirac operator in terms of scalar, pseudoscalar, and axial vector fields\footnote{For a nice  discussion of the underpinnings of this approach, see \cite{Leutwyler:1992yt}. Note further that we have implicitly in mind the separation of the gauge field configurations in $A_
\mu$ into high energy gluon modes and low energy nonperturbative modes which could be glueball or instanton configurations. In this light, the topological charge density $\Omega$ must be viewed as an intrinsically nonperturbative degree of freedom.}.
Restricting ourselves to the isosinglet pseudoscalar sector of interest, the QCD fermion action can be written as\footnote{We will follow here, for convenience, the conventions and notations of \cite{DHoker:1995aat,DHoker:1995uyv} since some of the key results in these papers are central to this work. For another  discussion, with similar features, we refer the reader to \cite{Mondragon:1995ab,Mondragon:1995va,McKeon:1998et}; both approaches are reviewed in  \cite{Schubert:2001he}.}
\begin{eqnarray}
S_{\rm fermion}[{\bar \Psi},\Phi,\Pi,A,B,\Psi] = \int d^4 x\,{\bar \Psi}^I\left[i\slashed{\partial}-\Phi+i\gamma^5 \Pi + \slashed{A} + \gamma^5 \slashed{B} \right]^{IJ}\Psi^J\,.
\end{eqnarray}
Here $\Phi$, $\Pi$, $A$ and $B$ denote respectively scalar, pseudoscalar, vector and axial vector fields, whose couplings to the higher frequency fermion fields ${\bar \Psi}$, $\Psi$, are absorbed into the field definitions. The superscripts $I$ and $J$ denote the internal quantum numbers of the fermion multiplet as well as those of the matrix valued ``source" fields. 

Since the worldline effective action corresponds to computing a quark loop in an arbitrary number of background fields, the corresponding perturbative expression is simply
\begin{eqnarray}
{\cal W}[\Phi,\Pi,A,B] &=& \sum_{n=1}^\infty \frac{1}{n} \int \frac{d^4 k_1}{(2\pi)^4}\cdots \frac{d^4 k_n}{(2\pi)^4}\, \delta^{(4)}(k_1+\cdots +k_n)\int \frac{d^4 q}{(2\pi)^4}\, \frac{\slashed{q}+ i m}{q^2+m^2}\nonumber \\
&\times&\left(i{\tilde \Phi}_1 + \gamma_5\, {\tilde \Pi}_1+ {\tilde {\slashed{A}}}_1 + \gamma_5 {\tilde {\slashed{B}}}_1\right)\frac{\slashed{q}-\slashed{k}_1-\cdots\slashed{k}_n+im}{\left(q-k_1-\cdots k_n\right)^2 + m^2}\cdots \left(i{\tilde \Phi}_n + \gamma_5\, {\tilde \Pi}_n+ {\tilde {\slashed{A}}}_n+ \gamma_5{\tilde {\slashed{B}}}_n \right)\,.
\end{eqnarray}
Even numbers of insertions of scalar, pseudoscalar and axial vector fields contribute to the real part of ${\cal W}$ while odd numbers contribute to the imaginary part. The map between the worldline and Feynman diagram computations of these was discussed previously in \cite{DHoker:1995uyv,Mondragon:1995ab}; for DIS specifically, it was discussed in \cite{Tarasov:2019rfp}.

The (Euclidean) fermion effective action in the presence of these sources can in general be written as 
\begin{equation}
    -{\cal W}[A,B,\Phi,\Pi] = {\rm Ln}\,{\rm Det }\,\left[{\cal D}\right]\,,
\end{equation}
with the Dirac operator, 
\begin{equation}
    {\cal D} = \slashed{p} -i\Phi(x) - \gamma_5\,\Pi\,-\slashed{A}-\gamma_5 \slashed{B}\,.
    \label{eq:Dirac-operator}
\end{equation}
This effective action can split into real and 
imaginary parts, with~\cite{DHoker:1995aat}, 
\begin{eqnarray}
    {\cal W}_R = - \frac{1}{2} {\rm Ln} \left({\cal D}^\dagger {\cal D}\right)\,\,\,;\,\,\,
    {\cal W}_I = \frac{1}{2}{\rm Arg}\, {\rm Det}\left({\cal D}^2\right)\,.
    \label{eq:Real-Imaginary-action}
\end{eqnarray}
Since the anomaly is sensitive to the imaginary part of the effective action, we will focus on the latter alone in the rest of this paper\footnote{The derivation of the real part of the effective action in  the presence of sources is discussed at length in \cite{DHoker:1995aat,DHoker:1995uyv}.}. 

An important observation in \cite{DHoker:1995aat} is that substituting Eq.\,(\ref{eq:Dirac-operator}) into Eq.\,(\ref{eq:Real-Imaginary-action}) leads to terms linear in the Grassmann variables which are physically unappealing; the solution (which does not alter the effective action), is to double the degrees of freedom in both the real and imaginary parts of the effective action as 
\begin{eqnarray}
{\cal W}_R = -\frac{1}{2}\,{\rm Ln}\,{\rm Det}\left({\cal \theta}\right)\,\,\,\,
{\rm with }\,\,\,\,
  {\cal \theta} = \begin{pmatrix} 0 & {\cal D} \\ {\cal D}^\dagger & 0
  \end{pmatrix}\,,
\end{eqnarray}
and 
\begin{eqnarray}
{\cal W}_I = -\frac{1}{2}\,{\rm Arg}\,{\rm Det}\left({\tilde {\cal \theta}}\right)\,\,\,\,
{\rm with }\,\,\,\,
 {\tilde {\cal \theta}} = \begin{pmatrix} 0 & {\cal D} \\ {\cal D} & 0
  \end{pmatrix}\,,
\end{eqnarray}
In this ``doubling' framework, one also likewise replaces the $4\times 4$ gamma matrices with the $8\times 8$ matrices,
\begin{eqnarray}
\Gamma_\mu=  \begin{pmatrix} 0 & \gamma_\mu\\ \gamma_\mu & 0
  \end{pmatrix}\,\,\,;\,\,\, \Gamma_5 = \begin{pmatrix} 0 & \gamma_5\\ \gamma_5 & 0
  \end{pmatrix}
  \,\,\,;\,\,\,
 \Gamma_6= \begin{pmatrix} 0 & i I\\ -i I & 0
  \end{pmatrix}\,,
\end{eqnarray}
where $I$ is the $4\times 4$ unit matrix and the six Hermitean $\Gamma$-matrices satisfy 
$\left\{\Gamma_A,\Gamma_B\right\}= 2\delta_{AB} \,I_{8\times 8}$. 

In this representation, d'Hoker and Gagn\'{e}, derived a remarkable expression for the argument of the Dirac determinant~\cite{DHoker:1995uyv}:
\begin{eqnarray}
{\cal W}_\mathcal{I} = - \frac{i}{32}\int^1_{-1}d\alpha \int^\infty_0 dT\, \mathcal{N} \int_{{\rm PBC}} \mathcal{D}x\, \mathcal{D}\psi~ {\rm tr} ~\chi\, \bar{\omega}(0) \exp\Big[-\int^T_0 d\tau \mathcal{L}_{(\alpha)}(\tau)\Big]\,,
\label{eq:Imag-W}
\end{eqnarray}
 which has a structure very similar to that of the real part, albeit with some key differences we shall enumerate\footnote{We have set here, and elsewhere, the value of the einbein $\mathcal{E}=2$. Further, 
\begin{equation}
    {\cal N}(T) = \int [Dp]\, \exp\left(-\int_0^T d\tau p^2(\tau)\right)\,,
\end{equation}
is a field-independent normalization factor. }. Firstly, $\mathcal{D}\psi = \mathcal{D}\psi_\mu \mathcal{D}\psi_5$ and ${\rm PBC}$ denotes periodic boundary conditions for {\it both} cooordinate and Grassmann variables. This  qualitatively differs from  ${\cal W}_R$, where the Grassmann integrals have anti-periodic boundary conditions. Other differences to the real part of the effective action are,\\ i) the integration over 
$\alpha$, which explicitly breaks global chiral invariance for $|\alpha|<1$, \\ii) the proper time measure
$dT \rightarrow dT/T$ in the real part, \\iii) and not least, the factor $\chi \,{\bar \omega}(0)$, which is a direct consequence of the anomaly. 

This last term is given by 
\begin{eqnarray}
\label{eq:Jacobian}
&&\chi\bar{\omega}(0) = 4 \times
\nonumber\\
&&\begin{pmatrix} - i\psi^5_0 ( \partial^\mu B_\mu(x_0) - 2 \Phi(x_0) \Pi(x_0)) + \frac{4 i }{\mathcal{E}} \psi^\mu_0\psi^\nu_0\psi^5_0 \dot{x}_{0\mu} B_\nu(x_0) & \frac{ 2 }{\mathcal{E}} \psi^\mu_0 \dot{x}_{0\mu} \Phi(x_0) \\ \frac{2}{\mathcal{E}} \psi^\mu_0 \dot{x}_{0\mu} \Phi(x_0) & - i \psi^5_{0} ( \partial^\mu B_\mu(x_0) - 2 \Phi(x_0) \Pi(x_0) )  + \frac{4 i }{\mathcal{E}} \psi^\mu_0\psi^\nu_0\psi^5_{0} \dot{x}_{0\mu} B_\nu(x_0)\end{pmatrix}\,.\nonumber\\
\end{eqnarray}

Strikingly, the worldline Lagrangian for the imaginary part of the effective action  is nearly identical to that for the real part except for the chiral symmetry breaking ``regulator" $\alpha$ multiplying 
$\Phi$ and $B_\mu$:
\begin{eqnarray}
\label{eq:chiral-worldline}
\mathcal{L}_{(\alpha)}(\tau) = \mathcal{L}(\tau)\Big|_{\Phi\to \alpha\Phi, B\to \alpha B}\,,
\end{eqnarray}
where the worldline Lagrangian for the real part of the effective action is 
\begin{eqnarray}
\label{eq:worldline-Lagrangian}
\mathcal{L}(\tau) = \frac{\dot{x}^2}{2\mathcal{E}} + \frac{1}{2}\psi\dot{\psi} - i \dot{x}^\mu \mathcal{A}_\mu + \frac{\mathcal{E}}{2}\mathcal{H}^2 + i\mathcal{E}\psi^\mu\psi_5 \mathcal{D}_\mu \mathcal{H} + \frac{i\mathcal{E}}{2} \psi^\mu \psi^\nu \mathcal{F}_{\mu\nu}\,.
\end{eqnarray}
We have adopted here a two component notation combining respectively scalar and pseudoscalar source fields, and likewise for the vector and axial vector fields, 
These fields are defined as
\begin{eqnarray}
\mathcal{A}_\mu \equiv \begin{pmatrix} A^L_\mu & 0 \\ 0 & A^R_\mu \end{pmatrix} = \begin{pmatrix} A_\mu + B_\mu & 0 \\ 0 & A_\mu - B_\mu \end{pmatrix}\,.
\end{eqnarray}
\begin{eqnarray}
\mathcal{H} \equiv \begin{pmatrix} 0 & iH \\ -iH^\dag & 0 \end{pmatrix} = \begin{pmatrix} 0 & i \Phi + \Pi \\ -i \Phi + \Pi & 0 \end{pmatrix}\,.
\end{eqnarray}

Note that if we turn off the scalar and pseudoscalar sources (set ${\cal H}=0$), we will recover precisely\footnote{One must first separate out the zero and nonzero modes, as in Eq.\,(\ref{eq:zero}), in the latter.} the ${\cal W}_I$ in Eq.\,(\ref{A5action}) that was employed  to compute the triangle graph and to relate it to the topological charge density $\Omega$.

As noted earlier, in addition to the triangle graph, the imaginary part of the worldline action contains all other anomalous contributions allowed by the symmetries of the theory. This can be appreciated immediately by observing the similarities between the worldline form of the effective action in the presence of sources to the Wess-Zumino action~\cite{Wess:1971yu}.  As is well-known, variations of the latter with respect to these sources generates functional Ward identities (including anomalous ones)~\cite{Shore:2007yn}. 

When we add the scalar and pseudoscalar sources to the mix, 
d'Hoker and Gagn\'{e} \cite{DHoker:1995aat} showed explicitly that ${\cal W}_I$ (when expanded out to order $O(\Pi^5)$) reproduces precisely the  Wess-Zumino-Witten term (WZW)~\cite{Wess:1971yu,Witten:1983tw}  governing $\pi^0\rightarrow 2\,\gamma$. One sees in this derivation that it is essential that the scalar $\Phi$ have a nonzero vacuum expectation value (vev). This is apparent from Eq.\,(\ref{eq:Jacobian}) which contributes the odd power of $\Pi$ in the WZW action. The presence of this vev can be interpreted as that acquired by the $\sigma$-field in the linear sigma model after the spontaneous breaking of chiral symmetry\footnote{As shown by Weinberg~\cite{Weinberg:1966fm}, in the limit of large vev masses, one  recovers the non-linear sigma model which provides the scaffolding for chiral perturbation theory. The computation by d'Hoker and Gagn\'{e} of the WZW-$\pi^0$ term is performed in this large mass limit.}. 

We can likewise show that, expanding  $W_{\mathcal{I}}$ up to order $\Pi\, A^2$,  gives the relation 
\begin{eqnarray}
\label{eq:WZW-worldline-eta0}
&&{\cal W}_{\mathcal I}[\Pi A^2] = \frac{ig^2 \,2 n_f}{16\pi^{2}} \frac{1}{\Phi} \,{\rm tr}_c \int d^4x \,\Pi(x)\, F_{\mu\nu}(x) \tilde{F}^{\mu\nu}(x) \,. 
\end{eqnarray}
The explicit derivation\footnote{Here we  have analytically continued the result in Eq.\,(\ref{resappA}) to Minkowski space, restored the gauge coupling, and taken into account the sum over quark flavors.} is given in Appendix \ref{sec:WZW-appendix} and is in agreement\footnote{To see this, note that $\Pi= \psi/3$ in \cite{Kaiser:2000gs}.} with the corresponding ${\cal L}_{\rm WZW}$ term in \cite{Kaiser:2000gs} which was derived from chiral perturbation theory for the $U_V(3)$ nonet.

We define the primordial  ${\bar \eta}$ isosinglet field as 
\begin{eqnarray}
{\bar \eta} = -\sqrt{2 n_f}\, \frac{\Pi}{\Phi}\,F_{{\bar\eta}}\,,
\end{eqnarray}
where the relation 
\begin{equation}
\label{eq:eta0-decay}
\langle 0 |J^\mu_5|{\bar \eta}\rangle = i\sqrt{2n_f}\,l^\mu\, F_{{\bar\eta}}(l^2)\,,
\end{equation}
defines the ${\bar \eta}$ decay constant  $F_{{\bar\eta}}$ in the forward limit $l^2\rightarrow 0$. 

We digress here briefly to note that in the description of the isosinglet sector of the QCD chiral Lagrangian\footnote{See \cite{Kaiser:2000gs} for the relevant discussion.},  the massless ${\bar \eta}$ field is understood as the prodigal ninth Goldstone boson of $U_V(3)$ that survives the spontaneous symmetry breaking of the global 
$U_R(3)\times U_L(3)$ symmetry that generates the chiral condensate. The ground state of the broken symmetry phase is invariant only under $U_V(3)$ and the dynamical fields of the low energy effective theory can be represented by the matrix $U(x)$ that transforms under this group. In the large $N_c$ limit, the ${\bar \eta}$ field corresponds to the phase of the determinant of $U(x)$. This description is therefore completely consistent with our worldline construction where we showed that 
Eq.\,(\ref{eq:WZW-worldline-eta0}) follows from the phase of the determinant, as seen in 
Eq.\,(\ref{eq:Real-Imaginary-action}). 

In the large $N_c$ limit, this Goldstone description in terms of the ${\bar \eta}$ is exact and the decay constants of all the nonet pseudo-Goldstone bosons are identical: $F_{{\bar \eta}} = F_\pi \approx 93$ MeV. For finite quark masses and finite $N_c$, the decay constants mix amongst each other; their values can be obtained from Dashen--Gell-Mann-Oakes-Renner relations and the rates for radiative decays of $\eta$, $\eta^\prime$ to photons. Remarkably, the diagonal components corresponding to the isosinglet decay to $\eta^\prime$ constant and the isooctet decay to $\eta$ are  extracted to be very close to $F_\pi$, about $15\%$ larger. For a detailed discussion, we refer the reader to \cite{Shore:2007yn,Gan:2020aco}.

The WZW effective action for the ${\bar \eta}$ field is then given by 
\begin{eqnarray}
\label{eq:WZW-eta0}
S_{\rm WZW}^{{\bar \eta}} = -i \frac{\sqrt{2\,n_f}}{F_{{\bar \eta}}}\int d^4x\, {\bar \eta}\,\Omega\,,
\end{eqnarray}
where the ``i" indicates its origin in the imaginary part of the effective action in 
Eq.\,(\ref{eq:Imag-W}).

This WZW term that couples the ${\bar \eta}$ to the topological charge density $\Omega$ plays a fundamental role in QCD. Firstly, the mixing of ${\bar \eta}$ and $\Omega$ ensures that the ${\bar \eta}$ is ``eaten up" by the latter, leaving a physical massive $\eta^\prime$ meson in the spectrum. Secondly, this term plays a key role the cancellation of the anomaly pole, and in generating $\Sigma(Q^2)$, the proton's helicity. We will now discuss the relation of these two fundamental issues, namely,  topological mass generation\footnote{For an elegant discussion of topological mass generation, see \cite{Dvali:2005an,Dvali:2005ws}.} and the  proton's helicity.

\section{Topological screening of the proton helicity}
\label{sec:Shore-Veneziano}

We noted previously two early bodies of work relevant to our discussion here that addressed the role of anomaly cancellation in the proton's helicity $\Sigma (Q^2)$. Jaffe and Manohar~\cite{Jaffe:1989jz} argued that the infrared pole obtained in the perturbative computation of the triangle anomaly must be cancelled by a like contribution in the pseudoscalar sector but did not discuss this mechanism in detail, specifically its relation to topological mass generation. In contrast, the approach of Shore and Veneziano~\cite{Shore:1990zu,Shore:1991dv,Shore:2007yn}, developing 
previous work\footnote{See also related work in \cite{Hatsuda:1989bi,Ji:1990fj,Liu:1991ni}.} by Veneziano~\cite{Veneziano:1989ei}, was fully nonperturbative, extensively employing chiral Ward identities derived from the Wess-Zumino~\cite{Wess:1971yu} effective action. They did not however discuss the cancellation of the anomaly pole, which is only implicit in their approach. 

We will discuss here a diagrammatic treatment that reconciles these two approaches and in particular, recovers the key results of Shore and Veneziano's ``topological screening" description of the proton's helicity. Since the triangle graph gives the dominant contribution to $g_1(x_B,Q^2)$ in both Bjorken and Regge asymptotics, we will for simplicity focus here on $\Sigma(Q^2)$; we will take up the question of the $x_B$ dependence of the triangle graph in Section~\ref{sec:Axion}.

\subsection{$\Sigma (Q^2)$ and the anomalous Goldberger-Treiman relation}
We begin by examining closely how the cancellation of the infrared pole in the matrix element of the axial vector current occurs. Our starting point is the general decomposition of the off-forward matrix element of the flavor singlet axial vector current\footnote{We thank Elliot Leader for a discussion of off-forward matrix elements and for bringing \cite{Bakker:2004ib} to our attention  where the properties of such matrix elements are discussed.}; we will eventually take the forward limit. 
Introducing the spinor $u(P, S)$ for the nucleon target of mass $M_N$ with momentum $P$ and spin $S$, we can write the matrix element of $ J^\mu_5$ in terms of the axial and pseudoscalar form factors $G_A$ and $G_P$ defining the coupling of the current to the target at finite momentum transfer as
\begin{eqnarray}
\langle P', S| J^\mu_5 |P, S\rangle = \bar{u}(P',S)\Big[\gamma^\mu \gamma_5G_A(l^2) +  l^\mu \gamma_5 G_P(l^2)\Big]u(P,S)\,.
\label{melexpcoup}
\end{eqnarray}
Here $l^\mu = {P^\prime}^\mu - P^\mu$ is the momentum transfer between the outgoing and incoming nucleon. 

\begin{figure}[htb]
\begin{center}
\includegraphics[width=140mm]{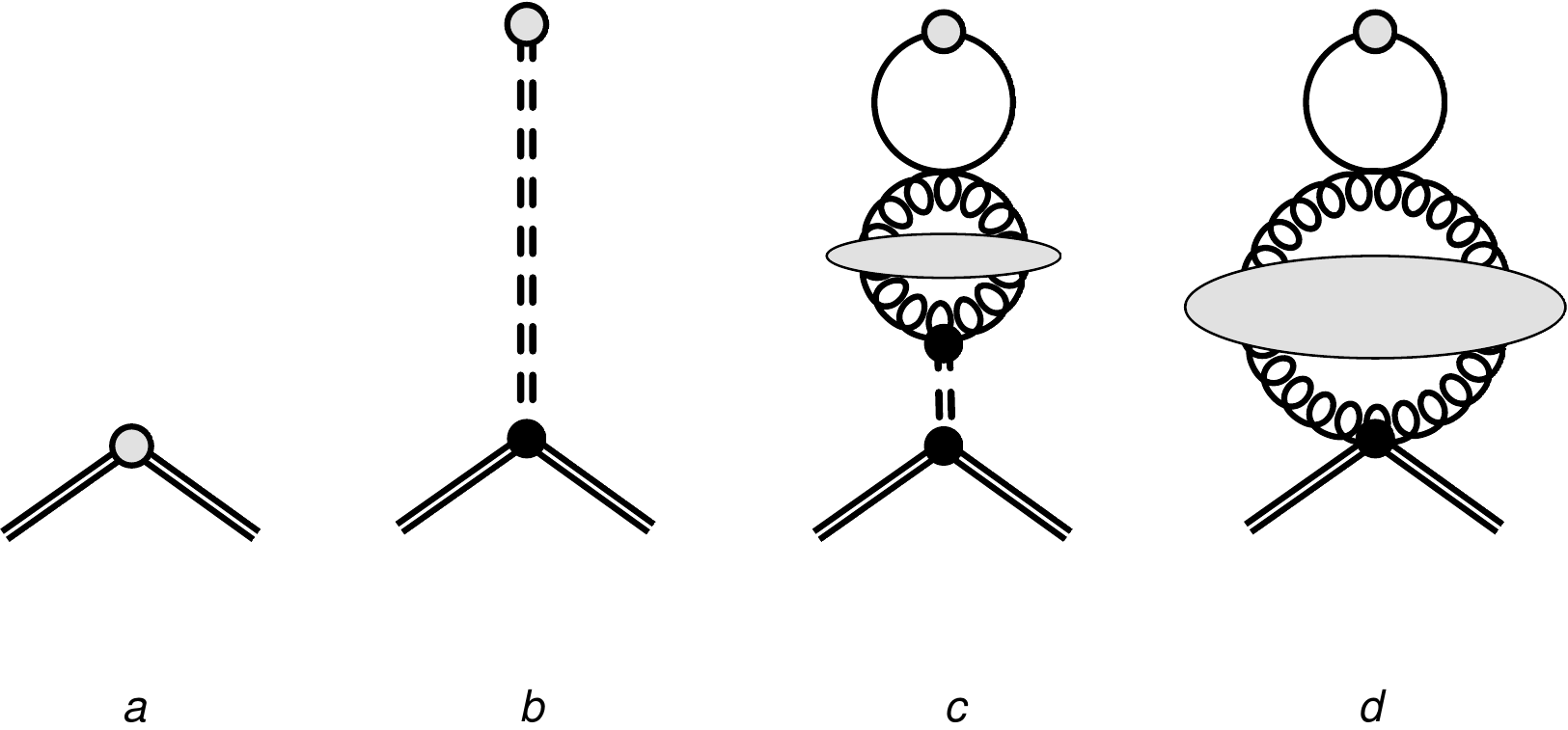}
\end{center}
\caption{Diagrams representing the coupling of the axial vector current $J^\mu_5$ to the nucleon target. See text for details. 
\label{fig:2nn}}
\end{figure}

 Fig. \ref{fig:2nn} shows the contributions to the form factors $G_A$ and $G_P$ and we will discuss each of these at length. We first note that the diagrams in Figs. \ref{fig:2nn}a and \ref{fig:2nn}b representing the coupling of the isosinglet axial vector current to the nucleon target are fully analogous to similar diagrams representing the  couplings of the non-anomalous axial currents to the nucleon. In contrast, the diagrams in Figs. \ref{fig:2nn}c and \ref{fig:2nn}d are generated by the triangle anomaly.

The diagram in Fig. \ref{fig:2nn}a represents the direct coupling of the isosinglet axial vector current to the nucleon target. It is fundamentally different from other diagrams in Fig. \ref{fig:2nn} since it is the only diagram which contributes to the axial form factor $G_A(l^2)$:
\begin{equation}
\langle P^\prime,S| J^\mu_5 |P,S\rangle|_{\rm Fig. \ref{fig:2nn}a} = G_A(l^2)\, \bar{u}(P',S)\,\gamma^\mu\gamma_5\, u(P,S)\,,
\label{dircontr}
\end{equation}
while the diagrams in Figs. \ref{fig:2nn}b-d contribute to the pseudoscalar form factor $G_P(l^2)$.
With regard to the latter diagrams, we first observe that as a consequence of  the dynamical breaking of $U_A(1)$, no massless isosinglet pseudoscalar particle can exist in the physical spectrum. The requirement that the form factor $G_P(l^2)$ cannot have a pole at 
$l^2 = 0$ can be expressed as 
\begin{eqnarray}
&&\lim_{l\to 0}\Big[ \langle P^\prime,S| J^\mu_5|P,S\rangle|_{\rm Fig. \ref{fig:2nn}b} + \langle P^\prime,S| J^\mu_5|P,S\rangle|_{\rm Figs. \ref{fig:2nn}c+\ref{fig:2nn}d}\Big] = 0\,.
\label{F3cancel}
\end{eqnarray}
This of course implies that in the forward limit the matrix element in Eq.\,(\ref{melexpcoup}) is solely determined by the contribution in Fig. \ref{fig:2nn}a:
\begin{equation}
\langle P,S| J^\mu_5 |P,S\rangle = \langle P,S| J^\mu_5 |P,S\rangle|_{\rm Fig. \ref{fig:2nn}a} = 2M_N \,G_A(0) \,S^\mu\,.
\label{fmel4a}
\end{equation}
As this expression indicates, $\Sigma(Q^2) = 2\,G_A(0)$: the proton's helicity is equal to twice its isosinglet axial vector form factor, which can be extracted from the first moment of $g_1$ in combination with results for the isotriplet ($G_A^{(3)}$) and isooctet ($G_A^{(8)}$) axial vector charges extracted respectively from nucleon and hyperon beta decay. From these extractions, the COMPASS collaboration determines that $G_A = 2\, \Sigma (Q^2)=0.32 \pm 0.03 \,(stat.) \pm 0.03\, (syst.)$ at $Q^2 = 3$ GeV$^2$~\cite{COMPASS:2010hwr} which is compatible with the HERMES collaboration's extraction~\cite{HERMES:2006jyl} at $Q^2 = 5$ GeV$^2$ of $G_A = 2\,\Sigma(Q^2) = 0.330\pm 0.011 \,(th.)\pm 0.025 \,(exp.)\pm 0.028 \, (evol.)$.

One could view Eq. (\ref{fmel4a}) as our final result. However it by itself provides little insight into  the numbers quoted, and in particular, the ``spin puzzle"~\cite{Kuhn:2008sy} of why it is much smaller from the so-called OZI expectation~\cite{Ellis:1973kp} that $G_A|_{\rm OZI} = 2\,\sqrt{3}\,G_A^{(8)}\equiv 2\,\sqrt{3}\times (0.167\pm 0.006) =0.579\pm 0.021$. To understand how it can be computed, and its connection to the $U_A(1)$ problem, we need to delve more deeply into the dynamics underlying the individual contributions in Fig. \ref{fig:2nn} and further, into relations that can be deduced\footnote{An alternate decomposition of the pseudoscalar contributions into a sum of terms with one proportional to the gluon helicity and the other proportional to a dimension six operator has been proposed in the literature~\cite{Hatta:2020ltd}. However as also discussed in some detail in \cite{Jaffe:1989jz}, we believe such a  decomposition must be interpreted with care.} amongst these.  

Indeed one of the relations, as we will discuss later, explicitly ensures  that 
Eq.\,(\ref{F3cancel}) is satisfied.  Another relation that must be satisfied is of course the anomaly equation for the divergence of the singlet axial current in the chiral limit,
\begin{eqnarray}
\langle P', S| \partial_\mu J^\mu_5 |P, S\rangle = \langle P', S|2\,n_f\,\Omega |P, S\rangle\,.
\label{aneq}
\end{eqnarray}
Recall that $\Omega \equiv \frac{\alpha_s}{4\pi} {\rm Tr} (F\tilde{F})$ is the topological charge density.
Since the l.h.s of Eq.\,(\ref{aneq}) is given by the sum of the diagrams in Fig. \ref{fig:2nn}, it imposes an important constraint on the individual dynamical contributions. 

Turning now to these, the diagram in Fig. \ref{fig:2nn}b represents the exchange, between the axial current and the nucleon, of an  ($\langle 0|u{\bar u}+d{\bar d}|0\rangle$)  $\eta_0$ projection of the ($\langle 0|u{\bar u}+d{\bar d}+s {\bar s}|0\rangle$) ${\bar \eta}$ field we discussed in the previous section. This is necessary because ${\bar \eta}$ contains an $s{\bar s}$ component that cannot couple directly to the nucleon and requires ``OZI violating" gluon exchanges to propagate~\cite{Liu:1991ni}. Nevertheless, the projection of $J_\mu^5$ on the $\eta_0$ state is nonzero and one can therefore define, in analogy to Eq.\,(\ref{eq:eta0-decay}),
\begin{eqnarray}
\label{eq:eta-decay-function}
\langle 0 |J^\mu_5|\eta_0\rangle = i\sqrt{2 {\tilde n}_f}\,l^\mu\, F_{\bar \eta}(l^2)\,,
\end{eqnarray}
where $l$ denotes the four-momentum of the intermediate $\eta_0$ field. Here ${\tilde n}_f=2$, is used to represent the two up and down flavors\footnote{We do so to avoid carrying factors of 2 and 3 (denoting $n_f=3$ in the ${\bar \eta}$ case) around in intermediate steps.}. In other words, since $J_\mu^5$ is flavor blind, the only change is the normalization with respect to the number of flavors. 

The contribution of  Fig. \ref{fig:2nn}b can thus be expressed as 
\begin{equation}
\langle P^\prime,S| J^\mu_5 |P,S\rangle|_{\rm Fig. \ref{fig:2nn}b} = g_{\eta_0 NN}\,\bar{u}(P',S)\,\gamma_5 \,u(P,S) \cdot \frac{i}{l^2}\cdot i \sqrt{2 {\tilde n}_f}\, l^\mu F_{\bar \eta}(l^2)\,,
\label{etacontr}
\end{equation}
where we have further parametrized the $\eta_0$-nucleon interaction by the $\eta_0 N N$ coupling $g_{\eta_0 NN}$ in the effective Lagrangian, 
\begin{eqnarray}
&&\Delta \mathcal{L} = i g_{\eta_0 NN}\,\eta_0\,\bar{N}\,\gamma_5\, N\,.
\end{eqnarray}

As their structure suggests, the diagrams in Figs. \ref{fig:2nn}c and \ref{fig:2nn}d are generated by the triangle anomaly and following our discussion in Sec.~\ref{sec:g1-worldline}, their contribution can be formally written in terms of the matrix element of $\Omega$ as 
\begin{equation}
\langle P^\prime,S| J^\mu_5|P,S\rangle|_{\rm Figs. \ref{fig:2nn}c+\ref{fig:2nn}d} = -i \frac{l^\mu}{l^2}\langle P', S| 2\,n_f\,\Omega |P, S\rangle\,.
\label{anomconts}
\end{equation}

Substituting the divergence of each of  Eqs.\,(\ref{dircontr}), (\ref{etacontr}) and (\ref{anomconts}) into the l.h.s of Eq.\,(\ref{aneq}) gives, 
\begin{eqnarray}
&&i G_A(l^2) \bar{u}(P',S)\slashed{l}\gamma_5 u(P,S) - i g_{\eta_0 NN}\bar{u}(P',S)\gamma_5 u(P,S) \sqrt{2 {\tilde n}_f} F_{\bar \eta}(l^2)  + \langle P', S| 2n_f \Omega |P, S\rangle = \langle P', S| 2n_f \Omega |P, S\rangle\,.
\end{eqnarray}
Then using  equation of motion for the spinor $u(P,S)$, we can rewrite this equation as
\begin{eqnarray}
&&\bar{u}(P',S) \Big[ 2 M_N \,G_A(l^2) -  g_{\eta_0 NN}  \sqrt{2 {\tilde n}_f} F_{\bar \eta}(l^2) \Big] \gamma_5\, u(P,S) = 0\,,
\end{eqnarray}
This equation relates $G_A$ to $g_{\eta_0 NN}\, F_{\bar \eta}$ which yields, in the forward limit $l\to0$, 
\begin{eqnarray}
&&G_A(0) = \frac{\sqrt{2 {\tilde n}_f}}{2M_N} \,F_{\bar \eta} \,g_{\eta_0 NN}\,.
\label{Fandetacoup}
\end{eqnarray}
This expression is the generalization of the well-known Goldberger-Treiman relation to the isosinglet  axial vector current, as first suggested by Veneziano~\cite{Veneziano:1989ei} and developed further by Shore and Veneziano~\cite{Shore:1990zu,Shore:1991dv}. 

If we substitute Eq.\,(\ref{Fandetacoup}) into Eq.\,(\ref{fmel4a}), we can  alternatively formulate our result in Eq.\,(\ref{fmel4a}) as  
\begin{equation}
\langle P,S| J^\mu_5 |P,S\rangle = \sqrt{2 {\tilde n}_f}\, F_{\bar \eta}\, g_{\eta_0 NN} S^\mu\,.
\label{axcursecform}
\end{equation}
Thus the anomalous Goldberger-Treiman relation allows us to express the matrix element of $ J^\mu_5$ equivalently in terms of the vacuum decay constant of the primordial pseudoscalar 
$\eta_0$ field and its coupling thereof to the polarized nucleon.

We obtained Eq.\,(\ref{axcursecform}) from Eq.\,(\ref{fmel4a}) due to the relation between the  diagrams in Figs. \ref{fig:2nn}a and \ref{fig:2nn}b and making further use of the anomaly relation in Eq. (\ref{aneq}). In the forward limit, we can also employ the constraint in Eq.\,(\ref{F3cancel}) relating the diagrams in Fig. \ref{fig:2nn}b-d, to rewrite Eq.\,(\ref{axcursecform}) in yet another form. Substituting Eqs.\,(\ref{etacontr}), and (\ref{anomconts}) into Eq.\,(\ref{F3cancel}) we obtain
\begin{eqnarray}
\label{eq:eta0-Omega}
&&\lim_{l\to 0}\Big[ g_{\eta_0 NN}\bar{u}(P',S)\gamma_5 u(P,S) \cdot \frac{i}{l^2}\cdot i \sqrt{2 {\tilde n}_f} l^\mu F_{\bar \eta}(l^2) -i \frac{l^\mu}{l^2}\langle P', S| 2n_f \Omega |P, S\rangle \Big] = 0\,. 
\end{eqnarray}
Thus the pole of the triangle anomaly that we computed perturbatively in Paper I can be understood as being cancelled,  as conjectured in \cite{Jaffe:1989jz}, by the t-channel exchange of a massless primordial $\eta_0$ meson, with the further proviso that there be no such pole in the physical spectrum.

We will now take into account the fact that the matrix element of $\Omega$ is given by the two diagrams in Figs. \ref{fig:2nn}c-d that couple the anomaly to the nucleon. The diagrams in Figs. \ref{fig:2nn}c-d can be written formally as\footnote{Here, and later in the text, the correlators $\langle 0|T\,\Omega \eta_0|0\rangle$ and $\langle 0|T\,\Omega \Omega|0\rangle$ are to be understood as the Fourier transforms of the 
corresponding correlators in coordinate space.}
\begin{equation}
\langle P^\prime,S| J^\mu_5|P,S\rangle|_{\rm Fig. \ref{fig:2nn}c} = -i \frac{l^\mu}{l^2}\langle P', S| 2n_f \Omega |P, S\rangle|_{\rm Fig. \ref{fig:2nn}c} = i \frac{l^\mu}{l^2} 2n_f \cdot \langle 0|T\,\Omega \eta_0|0\rangle \cdot g_{\eta_0 NN}\,\bar{u}(P',S)\,\gamma_5 \,u(P,S)\,,
\label{jmu5ometa}
\end{equation}
and
\begin{equation}
\langle P^\prime,S| J^\mu_5|P,S\rangle|_{\rm Fig. \ref{fig:2nn}d} = -i \frac{l^\mu}{l^2}\langle P', S| 2n_f \Omega |P, S\rangle|_{\rm Fig. \ref{fig:2nn}d} = i \frac{l^\mu}{l^2} 2n_f \cdot \langle 0|T\,\Omega \Omega|0\rangle \cdot g_{\Omega NN}\,\bar{u}(P',S)\,\gamma_5\,u(P,S)\,.
\label{jmu5omom}
\end{equation}
Here ``T" denotes that the vacuum correlators $\langle 0| T\,\Omega \eta_0|0\rangle$ and $\langle 0|T\,\Omega \Omega|0\rangle$ are time-ordered.

Substituting these equations into Eq. (\ref{eq:eta0-Omega}), we obtain,
\begin{eqnarray}
\label{eq:decayconst-Omega}
&&\sqrt{2 {\tilde n}_f}\, F_{\bar \eta}\,g_{\eta_0 NN} =  2n_f\, \lim_{l\to 0} \Big[i\,\langle 0|T\,\Omega \eta_0|0\rangle\, g_{\eta_0 NN} + i\,\langle 0|T\,\Omega \Omega|0\rangle\, g_{\Omega NN}\Big]\,. 
\end{eqnarray}

In Section~\ref{sec:ACTS}, we will employ the WZW term to derive general expressions for the correlators in the r.h.s. of Eq. (\ref{eq:decayconst-Omega}). We  will show in particular\footnote{It is useful to compare this result with that in \cite{Liu:1991ni}. Firstly, in that work, our $g_{\Omega NN}$ is denoted as $g_{{\bar \eta}NN}$. Secondly, in the relevant discussion of this contribution, a ``subtraction term" is introduced to impose by hand that this contribution be nonzero in order to recover the anomalous Goldberger-Treiman relation. This is because \cite{Liu:1991ni} did not explicitly consider the contribution shown in Fig.~\ref{fig:2nn}b, which as we saw, naturally gives the anomalous Goldberger-Treiman relation.} that the correlator $\langle 0|T\,\Omega \Omega|0\rangle$ (whose Fourier transform is the topological susceptibility) doesn't survive in the forward limit due to a shift of the infrared pole of the ${\bar \eta}$ to $m^2_{\eta'}$. Anticipating this result,  we obtain 
\begin{eqnarray}
\label{eq:decayconst-Omega2}
&&\sqrt{2 {\tilde n}_f}\, F_{\bar \eta} =  2n_f\, \lim_{l\to 0} i\,\langle 0|T\,\Omega \eta_0|0\rangle\,. 
\end{eqnarray}
Eq.\,(\ref{eq:decayconst-Omega2}) relates the ${\bar \eta}$ decay constant to the vacuum correlator $\langle 0|T\,\Omega \eta_0|0\rangle$ in the forward limit. Substituting Eq. (\ref{eq:decayconst-Omega2}) into Eq. (\ref{axcursecform}), we obtain
\begin{equation}
\langle P,S| J^\mu_5 |P,S\rangle = 2n_f\, \lim_{l\to 0} i\,\langle 0|T\,\Omega \eta_0|0\rangle\, g_{\eta_0 NN} \, S^\mu\,.
\label{jmumatrom}
\end{equation}
Eqs.\,(\ref{jmumatrom}), (\ref{fmel4a}) and (\ref{axcursecform}) represent different equivalent expressions for the matrix element of the axial vector current in the forward limit. Each of these expressions provides unique insight into the nonperturbative dynamics that generates the proton's helicity.

 Eq. (\ref{jmumatrom}) makes explicit the connection of the proton's helicity to the topological charge density of the QCD vacuum.  As we will
 further show, the r.h.s of  Eq. (\ref{jmumatrom}) is proportional to the slope of the QCD topological susceptibility at $l^2\rightarrow 0$. In the chiral limit, the topological susceptibility of the QCD vacuum is strictly zero; its slope is therefore small. Thus the screening of the topological charge explains  why the proton's helicity is  small providing a natural resolution to the proton spin puzzle. We turn now to a more detailed discussion of topological screening.

\subsection{Anomaly cancellation and topological screening}
\label{sec:ACTS}
We will compute here respectively $\langle 0|T \Omega \Omega|0\rangle$ and $\langle 0|T \Omega \eta_0|0\rangle$ employing the WZW term in Eq.\,(\ref{eq:WZW-eta0}). Specifically, we'll study the effect of the WZW coupling in the two-point Green functions and demonstrate how it generates a nonzero $m^2_{\eta'}$. We'll then show how the pole cancellation in Eq.\,(\ref{eq:eta0-Omega}) gives us the result noted earlier. 

\subsubsection{The WZW-${\bar \eta}$ term and the topological susceptiblity}
\begin{figure}[htb]
\begin{center}
\includegraphics[width=60mm]{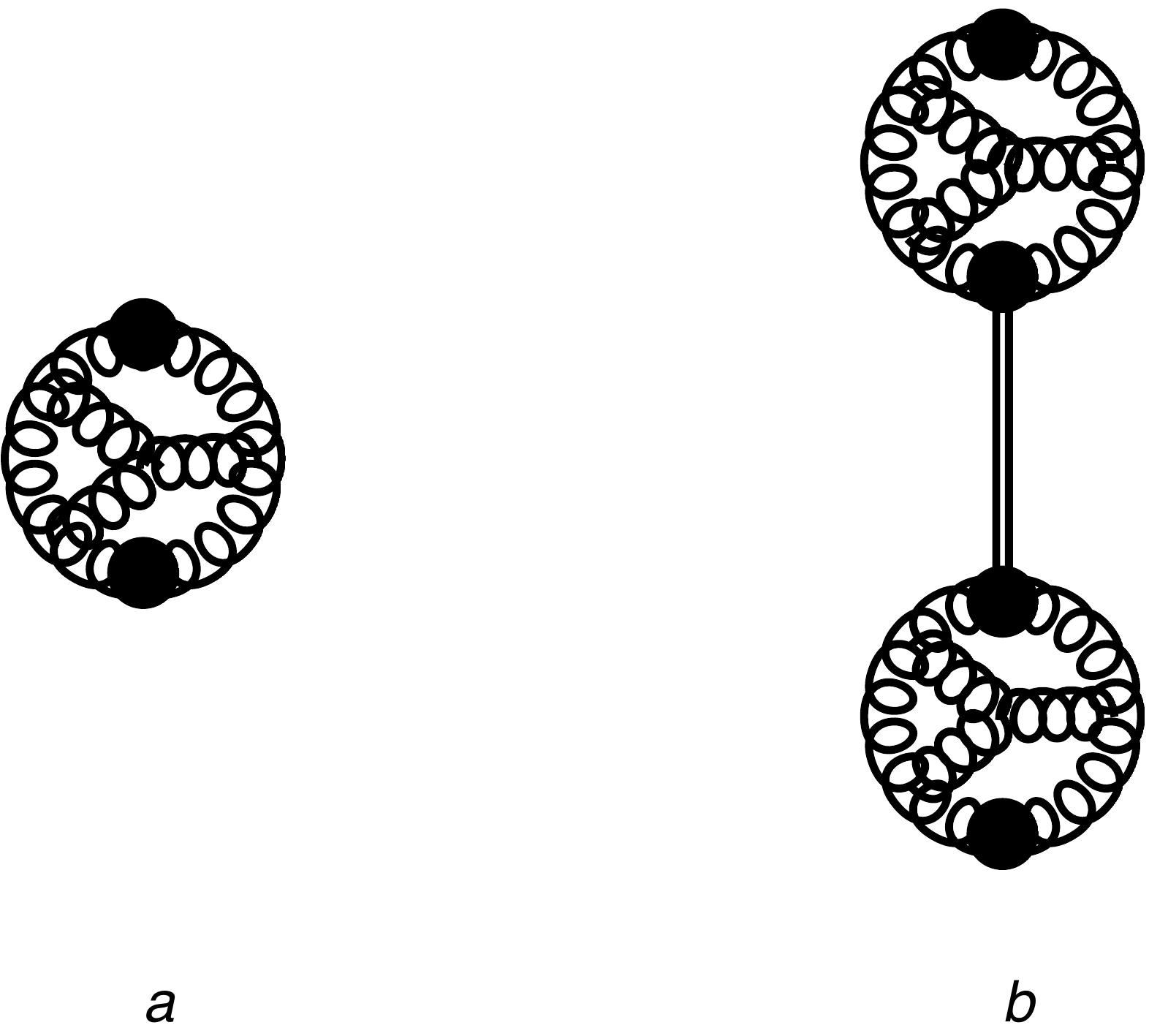}
\end{center}
\caption{a: The Yang-Mills topological susceptibility $\langle \Omega\Omega\rangle_{\rm YM}$; b: First correction to  $\langle \Omega\Omega\rangle_{\rm YM}$.
\label{figsup1}}
\end{figure}

Consider the topological susceptibility corresponding to the Fourier transform of $i\langle 0|T\Omega \Omega|0\rangle$ in Eq.\,(\ref{eq:decayconst-Omega}): 
\begin{eqnarray}
\chi(l^2) = i\int d^4x \,e^{ilx}\langle 0|T\,\Omega(x)\Omega(0)|0\rangle\,.
\end{eqnarray}
To leading order (Fig. \ref{figsup1}a), this correlator is nothing but the Fourier transform  $\chi_{\rm YM}(l^2)$ of the Yang-Mills topological susceptibility; as is well-known, it is a smooth function of momentum and doesn't have a pole~\cite{Kogut:1974kt}.

The first correction to Fig. \ref{figsup1}a, shown in Fig. \ref{figsup1}b, is given by\footnote{A classic discussion of this $1/N_c$ expansion in 
the topological susceptibility can be found in \cite{Witten:1979vv}.} 
\begin{eqnarray}
\label{eq:chi-chi-first-correction}
&&\langle 0| T\,\Omega\Omega|0\rangle|_{\rm Fig. \ref{figsup1}b} = \Big[-i \chi_{\rm YM}(l^2)\Big]\cdot \Big[-i\frac{\sqrt{2n_f}}{F_{\bar \eta}}\Big]\cdot \frac{i}{l^2}\cdot \Big[-i\frac{\sqrt{2n_f}}{F_{\bar \eta}}\Big]\cdot \Big[-i\chi_{\rm YM}(l^2)\Big]\,,
\end{eqnarray}
where we take into account the ${\bar \eta}$ vacuum propagator $\langle 0| T{\bar \eta} {\bar \eta}|0\rangle = \frac{i}{l^2}$ and the coupling between ${\bar \eta}$ and $\Omega$ as  specified by the WZW action in Eq.\,(\ref{eq:WZW-eta0}). Adding the two contributions,  we obtain,
\begin{eqnarray}
&&\langle 0| T \Omega\Omega|0\rangle|_{\rm Fig. \ref{figsup1}a + Fig. \ref{figsup1}b} = -i\chi_{\rm YM}(l^2) \Big\{1 - \frac{1}{l^2} \frac{2n_f \,\chi_{\rm YM}(l^2) }{F^2_{{\bar \eta}}} \Big\}\,.
\end{eqnarray}
Further iterating Fig.\ref{figsup1}(b) to all orders, we can express the resummed result as
\begin{eqnarray}
\label{eq:resummed-chi}
\chi(l^2) =  \chi_{\rm YM}(l^2) \frac{1}{1 + \frac{1}{l^2} \frac{2n_f\chi_{YM}(l^2)}{F^2_{\bar \eta}}}\,.
\end{eqnarray}

Note that since we are interested in the correlator in the forward limit $l\to0$, we can  rewrite it as 
\begin{eqnarray}
&&\chi(l^2) = l^2 \frac{1}{l^2 - m^2_{\eta'}} \chi_{\rm YM}(l^2) \,,
\label{topsus}
\end{eqnarray}
where we introduced
\begin{eqnarray}
&&m^2_{\eta'} \equiv - \frac{ 2\,n_f }{ F^2_{{\bar \eta}}}\chi_{\rm YM}(0) \,.
\label{metapmdef}
\end{eqnarray}
This last expression is the well-known Witten-Veneziano formula~\cite{Witten:1979vv,Veneziano:1979ec} for the mass of the $\eta'$ meson.

Taking the forward limit, we find
\begin{eqnarray}
\chi(0) = 0\,,
\label{eq:fig2nn-d}
\end{eqnarray}
which follows from topological mass generation: the WZW mixing of the massless ${\bar \eta}$ field with the topological charge density induces a massive $\eta^\prime$ field. As we saw in the previous section, combining this result with Eq.\,(\ref{eq:decayconst-Omega}) yielded  Eq.\,(\ref{eq:decayconst-Omega2}) relating the ${\bar \eta}$ decay constant $F_{\bar \eta}$ to the vacuum correlator $\langle 0|T\,\Omega \eta_0|0\rangle$ in the forward limit. We will now show that the latter is given by the slope of the topological susceptibility $\chi'(l^2)$.

\begin{figure}[htb]
\begin{center}
\includegraphics[width=80mm]{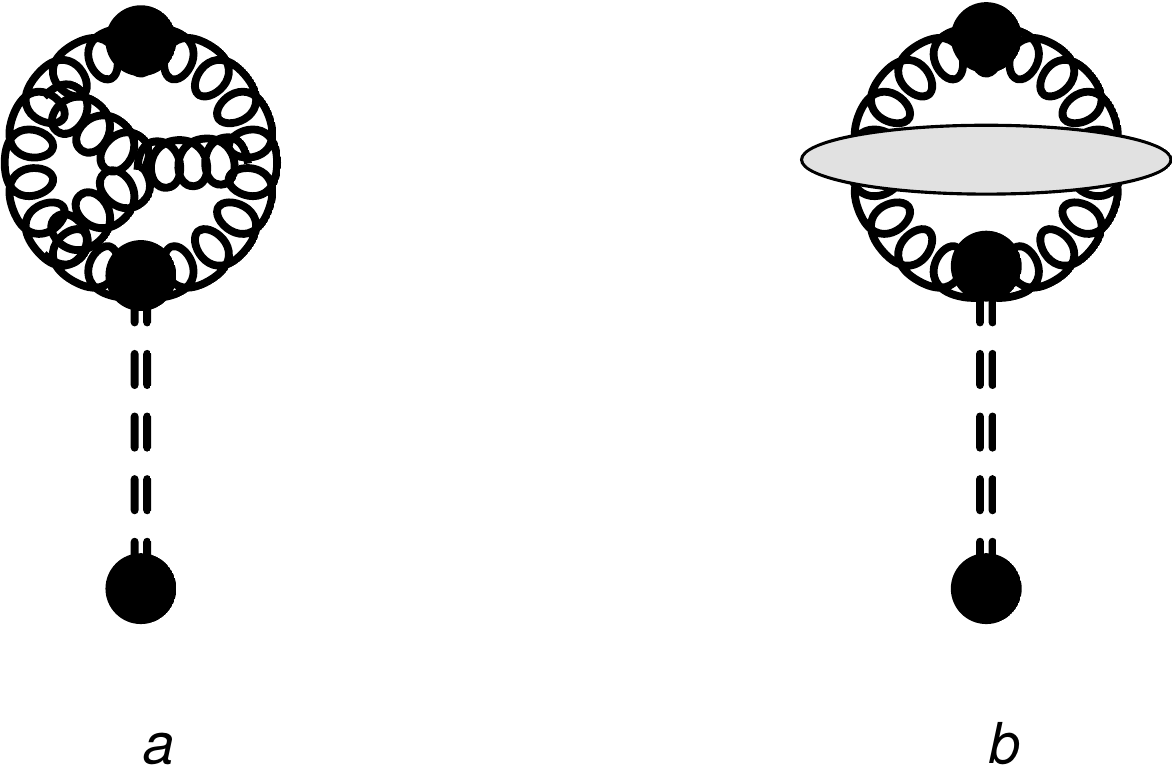}
\end{center}
\caption{a: Leading order contribution to $\langle \Omega \eta_0 \rangle$; b: The full correlator for $\langle \Omega \eta_0 \rangle$ replacing $\chi_{\rm YM}\rightarrow \chi$, as represented by the grey blob.\label{figsup2}}
\end{figure}

\subsubsection{The WZW-${\bar \eta}$ term and the correlator $\langle 0|T\Omega \eta_0 |0\rangle$}

The leading order diagram contributing to $\langle 0|T\Omega \eta_0 |0\rangle$ is shown in Fig. \ref{figsup2}a. Its computation, following our diagrammatic rules, is straightforward, and gives\footnote{Our derivation of Eq.\,(\ref{eq:WZW-eta0}) was flavor blind so 
the only difference in the coupling of $\Omega$ to $\eta_0$ relative to its coupling to ${\bar \eta}$ is to replace $n_f\rightarrow {\tilde n}_f$.},
\begin{eqnarray}
&&\langle 0|T\Omega \eta_0|0\rangle|_{\rm Fig. \ref{figsup2}a} = \Big[-i\chi_{\rm YM}(l^2)\Big] \cdot \Big[-i\frac{\sqrt{2\tilde{n}_f}}{F_{\bar \eta}}\Big] \cdot  \frac{i}{l^2}\,.
\label{mpf5}
\end{eqnarray}
To obtain the full correlator from Fig. \ref{figsup2}b, one can simply replace 
$\chi_{\rm YM}(l^2) \to \chi(l^2)$:
\begin{eqnarray}
\label{eq:chi-resum}
&&\langle 0|T\Omega \eta_0|0\rangle|_{\rm Fig. \ref{figsup2}b} = -i\frac{1}{l^2}\frac{\sqrt{2\tilde{n}_f} }{F_{\bar \eta}} \chi(l^2)\,.
\end{eqnarray}
We note that this result agrees with the parametrization of this two-point Green function in \cite{Shore:1991dv,Shore:2007yn}.

Substituting this result into the r.h.s of Eq. (\ref{eq:decayconst-Omega2}), we find
\begin{equation}
 F^2_{\bar \eta} = \lim_{l\to0} \frac{2n_f}{l^2} \chi(l^2)\,.
 \label{iv73}
\end{equation}
Expanding the topological susceptibility $\chi(l^2)$ in a Taylor series around $l^2 = 0$ as 
\begin{eqnarray}
\chi(l^2) = \chi(0) + l^2 \chi'(0) + \dots\,,
\end{eqnarray}
where $\chi'(l^2) \equiv \frac{d}{dl^2}\chi(l^2)$,
and taking into account that $\chi(0) = 0$ in the chiral limit, we can substitute the second term of the expansion into Eq. (\ref{iv73}), which yields
\begin{equation}
 F^2_{\bar \eta} = 2n_f \chi'(0)\,.
\end{equation}

As a result, using Eq.(\ref{axcursecform}), we obtain,
\begin{equation}
\langle P^\prime,S| J^\mu_5 |P,S\rangle = \sqrt{\frac{ 2}{3}}\, 2 n_f\, g_{\eta_0 NN}\,  \sqrt{ \chi'(0)}\,  S^\mu
\rightarrow \Sigma(Q^2) = \sqrt{ \frac{2}{3}}\,  \frac{2 n_f}{M_N}\, g_{\eta_0 NN} \sqrt{ \chi'(0)}\,.
\label{eq:square-root}
\end{equation}
This expression for the proton helicity in terms of square root of the slope of the QCD topological susceptibility was first obtained by Shore and Veneziano~\cite{Shore:1991dv,Shore:2007yn} from manipulations of the anomalous chiral Ward identities; we have provided here a complementary and intuitive derivation. 

  As mentioned earlier, this nontrivial result provides a simple explanation for why $\Sigma(Q^2)$ is anomalously small; the forward topological susceptibility $\chi(0)$ in the chiral limit is strictly zero, and a non-zero contribution to $\Sigma(Q^2)$ can only arise from small deviations from it, as represented by its slope at $l^2=0$, the scale for which is set by the large value of $m_{\eta^\prime}$. Narison, Shore and Veneziano~\cite{Narison:1994hv,Narison:1998aq} employed QCD sum rules to evaluate this expression obtaining results in agreement with the HERMES~\cite{Airapetian:2007mh} and COMPASS~\cite{Alekseev:2010ub} data we quoted after Eq.~(\ref{fmel4a}). A very recent update to these sum rule determinations is given in \cite{Narison:2021kny}. One can in principle compute $G_A$ in Eq.~(\ref{fmel4a}) directly on the lattice by computing the off-forward matrix element of $J_\mu^5$;  however one has to ensure that its anomalous Ward identity is satisfied~\cite{Liu:1991ni}. Alternatively, one can instead determine 
  $\Sigma$ from Eq.~(\ref{eq:square-root}) by computing $\chi^\prime(0)$ on the lattice \cite{Giusti:2001xh}; for a discussion of the current status of computations of the topological susceptibility and relevant references, we refer the reader to \cite{Bali:2021qem}. 

\subsubsection{${\bar \eta}$ effective action }
\label{sec:eta-prime-EFT}
In Eq.~(\ref{eq:WZW-eta0}), we obtained the form of the ${\bar \eta}\,\Omega$ WZW coupling from the imaginary part of the worldline effective action. Further, from Eqs. (\ref{eq:Imag-W}), (\ref{eq:worldline-Lagrangian}) and (\ref{eq:chiral-worldline}), we can deduce a kinetic term for the ${\bar \eta}$ field~\cite{Mondragon:1995va,Schubert:2001he},
\begin{eqnarray}
S_{\rm kinetic} = \int d^4 x \,\frac{1}{2}\,(\partial_\mu {\bar \eta})\, (\partial^\mu {\bar \eta}) \,.
\label{eq:action-kinetic}
\end{eqnarray}
Indeed, we employed this kinetic term in our diagrammatic analysis. 

There is an additional ``$\theta$-term" contribution from the imaginary part of the worldline effective action which has the same structure as Eq. (\ref{eq:WZW-eta0}), and is given by~\cite{Kaiser:2000gs}
\begin{eqnarray}
S_\theta = \int d^4 x\,\theta\,\Omega \,. 
\label{eq:action-theta-Omega}
\end{eqnarray}
Finally, there is also a term (see \cite{Witten:1980sp} and references therein)
representing the free energy of the $\theta$-vacuum given by 
\begin{eqnarray}
S_{\rm vac} = \frac{1}{2}\int d^4x\, \chi_{\rm YM}\, \theta^2 \,,
\label{eq:action-vac}
\end{eqnarray}
whose second derivative with respect to theta, at $\theta=0$, defines the Yang-Mills topological susceptibility.

Putting everything together, we can write the low energy ${\bar \eta}$ effective action as 
\begin{eqnarray}
\label{eq:eta0-action}
S_{{\bar \eta}} = \int d^4 x \left[\frac{1}{2}\,(\partial_\mu {\bar \eta})\,(\partial^\mu {\bar \eta}) 
+ \left(\theta -  \frac{\sqrt{2n_f}}{F_{{\bar \eta}}}{\bar \eta} \right)\,\Omega + \frac{\chi_{\rm YM}}{2}\,\theta^2\, \right]\,.
\end{eqnarray}

Since there is no kinetic term for $\theta$, it acts as a constraint and can be eliminated using the equations of motion:
\begin{eqnarray}
\frac{\delta S_{{\bar \eta}}}{\delta \theta} = 0 \rightarrow \theta  = -\frac{\Omega}{\chi_{\rm YM}} \,.
\end{eqnarray}
Plugging this back into Eq.(\ref{eq:eta0-action}), we get 
\begin{eqnarray}
S_{{\bar \eta}} = \int d^4 x \left[\frac{1}{2}\,(\partial_\mu {\bar \eta})\,(\partial^\mu {\bar \eta}) 
- \frac{\sqrt{2 n_f}}{F_{{\bar \eta}}}{\bar \eta}\,\Omega  - \frac{\Omega^2}{2\,\chi_{\rm YM}} \right]\,.
\label{eq:eta0-effective-action}
\end{eqnarray}
This form of the action is identical to that of Shore and Veneziano~\cite{Shore:2007yn} (see also \cite{Hatsuda:1989bi}) which they argue to be the simplest effective action consistent with the anomalous Ward identities of QCD. We will employ its equivalent representation in 
Eq.~(\ref{eq:eta0-action}) in our discussion in Sec.~\ref{sec:Axion}.

Defining the $\eta^\prime$ field as 
\begin{eqnarray}
\eta^\prime = \frac{F_{\eta^\prime}}{F_{{\bar \eta}}} {\bar \eta} \,,
\end{eqnarray}
where $F_{\eta^\prime}$ is the $\eta^\prime$ decay constant (see Section 3 of \cite{Shore:2007yn}) and introducing a glueball field defined as 
\begin{eqnarray}
G = \Omega + \frac{\sqrt{2 n_f}}{F_{\eta^\prime}}\,\chi_{\rm YM}\, \eta^\prime \,,
\end{eqnarray}
one can reexpress the effective action in Eq.~(\ref{eq:eta0-effective-action}) as 
\begin{eqnarray}
\label{eq:eta-prime-action}
S_{{\eta^\prime}} = \int d^4 x \left[-\frac{1}{2}\,\eta^\prime\,\left(\partial^2 + m_{\eta^\prime}^2\right)\eta^\prime - \frac{G^2}{2\chi_{\rm YM}} \right]\,,
\end{eqnarray}
where $m_{\eta^\prime}$ is the $\eta^\prime$ mass given by the Witten-Veneziano formula in Eq. (\ref{metapmdef}). Thus the mixing between ${\bar \eta}$ and $\Omega$ in Eq. (\ref{eq:eta0-action}) generates an effective action describing the physical massive $\eta^\prime$ and a non-propagating glueball field $G$, which decouples from the hadron spectrum. Note further that as 
$N_c\rightarrow \infty$, one has $\eta^\prime \rightarrow {\bar \eta}$, since $m_{\eta^\prime}\rightarrow 0$. In this ``OZI limit" of QCD, the anomaly vanishes restoring 
$U_A(1)$ and the ${\bar \eta}$ is the prodigal ninth Goldstone boson. 

Recall from Section~\ref{sec:WZW} that Eqs. (\ref{eq:WZW-eta0}), 
(\ref{eq:action-kinetic}) and (\ref{eq:action-theta-Omega}) can be understood as arising from the phase of the Dirac determinant in the QCD effective action, where the relevant low energy degrees of freedom are parametrized with scalar, pseudoscalar, vector and axial vector degrees of freedom. There is of course the path integral over the gauge field configurations to consider. In 't Hooft's~\cite{tHooft:1976snw,tHooft:1986ooh} explanation of the $U_A(1)$ problem, classical (Euclidean) instanton gauge field configurations are the dominant configurations responsible for the coupling of the topological charge density to fermion zero modes~\cite{Leutwyler:1992yt}; hence $\chi_{\rm YM}$ in Eq.~(\ref{eq:action-vac}) in this picture is saturated by the dynamics of such configurations~\cite{Schafer:1996wv}.  
However as pointed out by Veneziano~\cite{Veneziano:1979ec}, that while sufficient, instanton configurations are not required for the solution of the $U_A(1)$ problem. Indeed the discussion above did not invoke the instanton picture at all though it is consistent with it.  We will argue in the next section that while instanton-anti-instanton configurations may dominate at large $x_B$, the physics of gluon saturation suggests that other classical configurations increasingly begin to play a role on the short time scales probed by a DIS probe with decreasing  $x_B$. 

\section{Axion-like action at small $x_B$: Gluon saturation and sphaleron transitions}
\label{sec:Axion}

The triangle graph, as noted previously here, and in detail in Paper I, dominates the box diagram contributing to $g_1(x_B,Q^2)$ in both Bjorken and Regge asymptotics. The dynamics underlying the $x_B$ dependence is therefore contained in the diagrams shown in Fig~\ref{fig:2nn}, whose interplay we discussed on general grounds in the previous section. 
We will now consider these in greater detail and point to novel features that emerge at small $x_B$.

\begin{figure}[htb]
\begin{center}
\includegraphics[width=70mm]{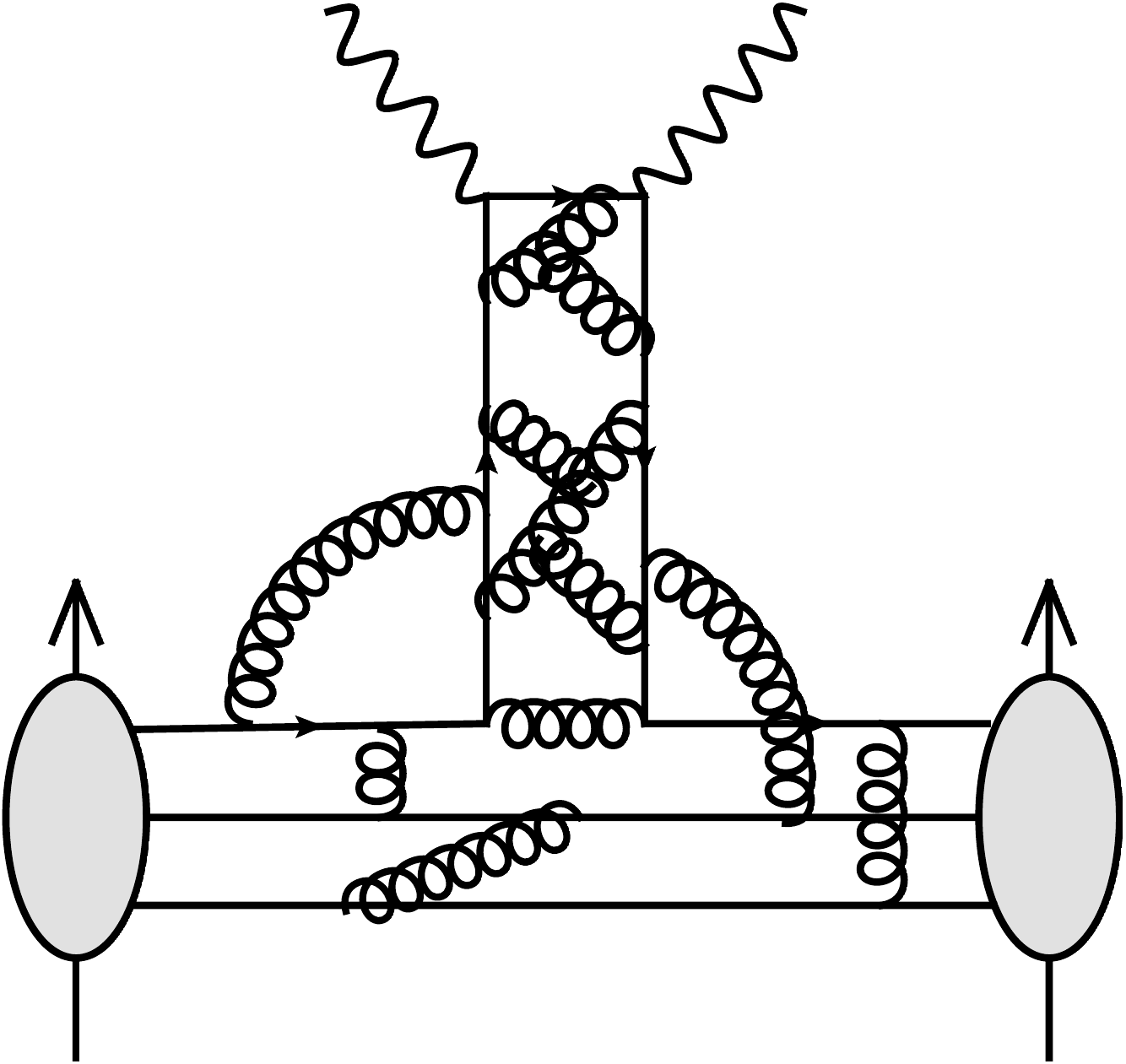}
\end{center}
\caption{Diagram illustrating isosinglet quark exchange in the t-channel. Such reggeon  contributions are allowed at both large and small $x_B$, though they are suppressed at small $x_B$. 
}
\label{fig:small-xB-Regge}
\end{figure}

A subtle point which will govern our analysis must be noted at the outset. As we observed in Eq.~(\ref{fmel4a}), it is sufficient to compute the $x_B$-dependent generalization of the coupling of the axial charge to the nucleon, as shown in Fig.~\ref{fig:2nn}(a). Alternately, one could employ the anomalous Goldberger-Treiman we derived in Eq.~(\ref{Fandetacoup}) and compute instead the $x_B$-dependent generalization of Fig.~\ref{fig:2nn}(b), specifically that of the product $F_{\eta_0}\, g_{\eta_0 NN}$.

The dynamics underlying Figs.~\ref{fig:2nn} (a) and (b) is illustrated in Fig.~\ref{fig:small-xB-Regge}. Fig.~\ref{fig:2nn} (a) corresponds to a direct axial coupling at a given $x_B$ to a valence quark in Fig.~\ref{fig:small-xB-Regge}. Though Fig.~\ref{fig:2nn} (b) is formally represented as a Feynman diagram in Fig.~\ref{fig:small-xB-Regge}, its physics (due to its sensitivity to the off-forward pole) is governed primarily by low frequency modes of the fermion determinant and is fundamentally nonperturbative. Spin diffusion along the t-channel\footnote{One may ask whether spin diffusion can occur instead due to spin precession in a background field. In Eq.\,(\ref{MLag}), these contributions to $g_1$ would correspond to terms linear in $F_{12}$. However as discussed at length in Appendix~\ref{sec:eikonal}, for operators sensitive to the anomaly, terms that are naively sub-leading (relative to this linear term) in an eikonal expansion cannot be ignored due to the off-forward pole in the t-channel. This leads to Fig.~\ref{fig:2nn} (b) being sensitive to the anomaly, as  seen clearly in Eq.\,(\ref{eq:eta0-Omega}).} can be viewed as being mediated by reggeon exchange (corresponding to the $\eta_0$) in the isosinglet sector. At large $x_B$, $g_1$ can be computed using  lattice QCD methods~\cite{Gockeler:1995wg,Liang:2018pis,Alexandrou:2019brg,Mejia-Diaz:2017hhp,Lin:2018obj,Giusti:2001xh,Bali:2021qem}. However at small $x_B$, particularly in Regge asymptotics, lattice computations are challenging due to the difficulties posed by i) computing higher moments of local operators, ii) boosting the proton to high energies on the lattice~\cite{Shanahan:2018jbm}. 

While one might consider this situation challenging for small $x_B$ computations, the  ``anomaly diagram" in Fig.~\ref{fig:2nn} (c) provides a crucial assist as we will describe shortly. Firstly, note that we showed explicitly (in the discussion culminating in Eq.~(\ref{eq:fig2nn-d})) that the other possible anomaly diagram  Fig.~\ref{fig:2nn}(d) is zero in the chiral limit. Because of the anomaly contribution in Fig.~\ref{fig:2nn}(c), we showed in  Eq.~(\ref{jmumatrom}) that the  matrix element for $J_\mu^5$ can be expressed in terms of the matrix element of the topological charge density, or equivalently, in terms of $\chi^\prime(0)$, as shown in Eq.~(\ref{eq:square-root}). The latter expression is manifestly finite in the limit $l\rightarrow 0$.

\begin{figure}[htb]
\begin{center}
\includegraphics[width=70mm]{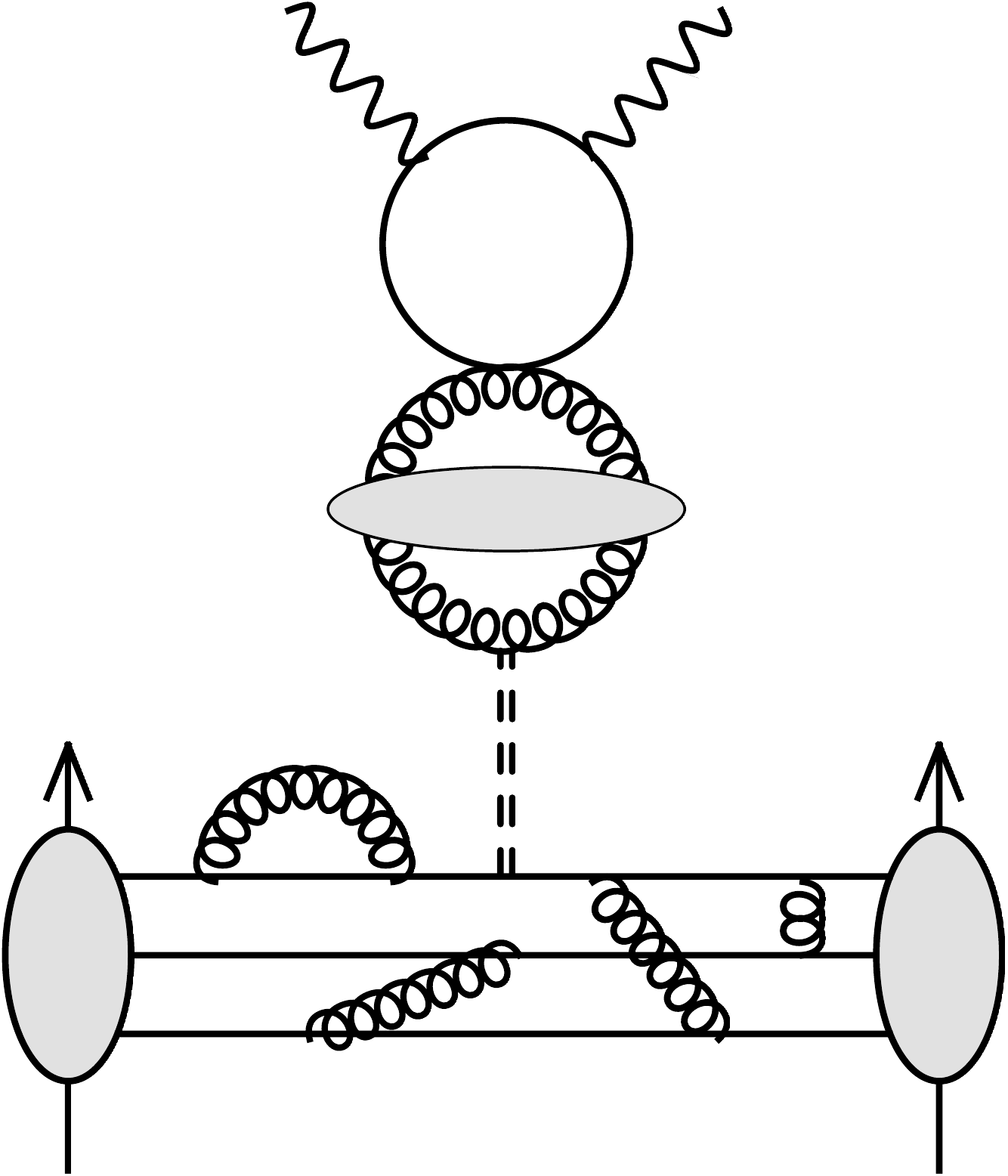}
\end{center}
\caption{Diagram illustrating the t-channel propagation of an isosinglet $\eta_0$ pseudoscalar field Bjorken and Regge asymptotics. This $\eta_0$ field has a nonperturbative distribution that couples to polarized valence partons with large momentum fractions in the proton. As it propagates along the t-channel, 
the $\eta_0$ couples to the vacuum topological charge density represented by the correlator $\langle \Omega \eta_0\rangle$ (illustrated in  Fig.~\ref{figsup2} (b)), acquiring a mass in the process before its coupling to the triangle graph via the anomaly.}
\label{fig:xB-dependence-eta0}
\end{figure}

We can further explore the structure of Eq. (\ref{jmumatrom}) by rewriting the vacuum correlator as a functional integral over $A$ and $\bar{\eta}$ fields as

\begin{eqnarray}
&&\label{eq:pol-Regge-EFT}
\langle P,S| J^\mu_5 |P,S\rangle = 2\,n_f\, i\,\int d^4y\,\int {D{\bar \eta}}\, \,{\tilde W}_{P,S}[{\bar \eta}] \, \int [DA]
\nonumber\\
&&\times \Omega(0)\,\eta_0(y)\, \exp \left(i S_{\rm YM} + i\int d^4 x \left[ \frac{1}{2}\,(\partial_\mu {\bar \eta})\,(\partial^\mu {\bar \eta})
- \frac{\sqrt{2n_f}}{F_{{\bar \eta}}}{\bar \eta}\,\Omega\right]\right) \, S^\mu\,,
\end{eqnarray}
where we generalized $g_{\eta_0 NN}$ in Eq. (\ref{jmumatrom}) by 
introducing the weight functional ${\tilde W}_{P,S}[{\bar \eta}]$; it  represents the nonperturbative distribution of ${\bar \eta}$ determined (with the  $\sqrt{\frac{2}{3}}$ normalization) from the pseudoscalar coupling of the $\eta_0$ field to the polarized proton
with the normalization 
\begin{equation}
    \int [D{\bar \eta}]\,{\tilde W}_{P,S}[{\bar \eta}]=g_{\eta_0 NN}\,.
\end{equation}

Comparing Eq.\,(\ref{eq:pol-Regge-EFT}) with the contribution of the triangle anomaly to the matrix element of the axial vector current given in Eq.\,(\ref{anomconts}), we observe that the regularization of the infrared pole is equivalent to replacing $\frac{l^\mu}{l^2}$ in the matrix element by the above functional integral.

The ${\bar \eta}\, \Omega$ term in Eq.\,(\ref{eq:pol-Regge-EFT}), after expanding out to linear order in ${\bar \eta}$, gives
\begin{equation}
    \langle {\bar \eta} {\bar \eta} \rangle\, \langle \Omega \Omega \rangle \,,
\end{equation}
where the first expectation value gives the ${\bar \eta}$-propagator $\frac{i}{l^2}$  and the second, the Yang-Mills topological susceptibility $\chi_{{\rm YM}}(l^2)$, as seen previously in Eq.\,(\ref{mpf5}). Subsequent expansion to $O({\bar \eta}^3)$, and higher odd powers in ${\bar \eta}$ then give Eq.\,(\ref{eq:chi-chi-first-correction}), illustrated by the correction to Fig. \ref{figsup1}a shown in Fig. \ref{figsup1}b, 
\begin{equation}
     \langle {\bar \eta} {\bar \eta} \rangle\, \langle \Omega \Omega \rangle  \langle {\bar \eta} {\bar \eta} \rangle\, \langle \Omega \Omega \rangle \cdots \,,
\end{equation}
with the resummation of such contributions to all orders generating the $\chi(l^2)$ in  Eq.\,(\ref{eq:resummed-chi}). 

An analogous regularization of the infrared pole in the expression for the structure function $g_1$ in Eq.\,(\ref{eq:g1-Bj-Regge}) yields
\begin{eqnarray}
&&\label{eq:g1-pol-Regge-EFT}
g_1(x_B, Q^2) =   \left(\sum_f e_f^2\right) \frac{n_f\alpha_s}{\pi M_N} i
\int d^4y\, \int^1_{x_B} \frac{dx}{x} 
~ \Big( 1  - \frac{x_B}{x} \Big) \int \frac{d\xi}{2\pi} e^{-i\xi x } \, \int {D{\bar \eta}}\, {\tilde W}_{P,S}[{\bar \eta}] \, \int [DA]
\nonumber\\
&&\times\, {\rm Tr_c} F_{\alpha\beta}(\xi n) \tilde{F}^{\alpha\beta}(0)\,\eta_0(y)\, \exp \left(i S_{\rm YM} + i\int d^4 x \left[ \frac{1}{2}\,(\partial_\mu {\bar \eta})\,(\partial^\mu {\bar \eta})
- \frac{\sqrt{2n_f}}{F_{{\bar \eta}}}{\bar \eta}\,\Omega\right]\right) \,.
\end{eqnarray}
The last two terms  in the exponential, corresponding to the dynamics of the ${\bar \eta}$, are identical to that describing the coupling of a putative axion particle~\cite{Wilczek:1977pj,Weinberg:1977ma,Kim:1986ax} to QCD matter. The underlying dynamics of the functional integral representation of Eq.\,(\ref{eq:g1-pol-Regge-EFT}) is illustrated in Fig.~\ref{fig:xB-dependence-eta0}. This expression is  consistent with 
Eq.\,(\ref{eq:eta0-effective-action}) if we assume the latter to be saturated by nonperturbative classical configurations.

We will now argue that novel dynamics  emerges in Regge asymptotics that allows one to compute the dynamics inside the blob in Fig.~\ref{fig:xB-dependence-eta0} in a weak coupling framework. This dynamics is due to the phenomenon of gluon saturation~\cite{Gribov:1984tu,Mueller:1985wy} at small $x_B$ corresponding to the close packing of gluons in the hadron. At maximal occupancies of $O(1/\alpha_s)$, the dynamics is controlled by a saturation scale $Q_S(x_B)\gg \Lambda_{\rm QCD}$, which screens color charge beyond this close packing scale. Since 
$\alpha_s(Q_S)\ll 1$ in the Regge limit,  the high occupancy of closely packed  glue within a radius $1/Q_S(x_B)$ inside the proton forms a classical lump. Further, in this limit, its dynamics can be studied systematically in weak coupling~\cite{McLerran:1993ka,McLerran:1993ni,McLerran:1994vd}. 

Gluon saturation has been studied extensively within the framework of the Color Glass Condensate (CGC) effective field theory~\cite{Iancu:2003xm,Gelis:2010nm,Kovchegov:2012mbw,Blaizot:2016qgz}. In short, large $x$ color charges in a hadron or nucleus are treated as 
static classical color charges with color charge density $\rho$ coupled to classical gauge field configurations  $A_{\rm cl}[\rho]\sim 1/g$, where $g$ is the gauge coupling. Sources and fields are separated at the scale $x_0 = \Lambda^+/P^+$; at small $x$, logarithmically enhanced gluon emissions (LLx) $\alpha_s\ln(x_0/x_0^\prime)~O(1)$ from the fields can be absorbed into a new source distribution $\rho^\prime$ at the scale $x_0^\prime = {\Lambda^\prime}^+$, and iterated, satisfying a Wilsonian renormalization group evolution equation described by a JIMWLK Hamiltonian~\cite{JalilianMarian:1997dw,JalilianMarian:1997gr,JalilianMarian:1999xt,Iancu:2000hn,Ferreiro:2001qy}. The JIMWLK Hamilitonian describes the energy evolution of the saturation scale which specifies the nonperturbative distribution $W_{Y_0}[\rho]$ of color sources at the initial rapidity scale $Y_0= \ln(1/x_0)$.

A detailed derivation in the worldline formalism of the expectation value of operators in an unpolarized proton or nucleus is given in Appendix~\ref{sec:CGC}. One obtains (see Eqs.\,(\ref{eq:CGC-EFT-unpolarized}) and (\ref{MVmodel})),
\begin{eqnarray}
\label{eq:CGC-EFT-unpolarized-Main}
\langle \mathcal{O} \rangle_{\rm unpol.} = \int \mathcal{D}\rho ~W_{Y}[\rho] ~\int \mathcal{D}A~\mathcal{O}[A]~e^{iS_{\rm CGC}[A, \rho]}\,,
\end{eqnarray}
where $Y$ is the rapidity of interest that $W_Y[\rho]$ has been evolved to,  and 
\begin{eqnarray}
S_{\rm CGC}[A, \rho] = -\frac{1}{4} \int d^4x F^{\mu\nu}_a F^a_{\mu\nu} + \frac{i}{N_c} \int d^2x_\perp ~ {\rm tr}_c \big[\rho(x_\perp)  \ln\big(U_{[\infty, -\infty]}(x_\perp)\big)\big]\,,
\label{MVmodel-Main}
\end{eqnarray}
with $U_{[\infty, -\infty]}(x_\perp)=\exp\Big[-ig \int^\infty_{-\infty} dx^+ A^-(x^+, z_\perp) \Big]$. The shock wave classical field $A_{\rm cl.}^\mu[\rho]$ corresponding to the saddle point of this effective action is well-known; we will discuss it further in Sec.~\ref{sec:top-shock-wave}.

\begin{figure}[htb]
\begin{center}
\includegraphics[width=150mm]{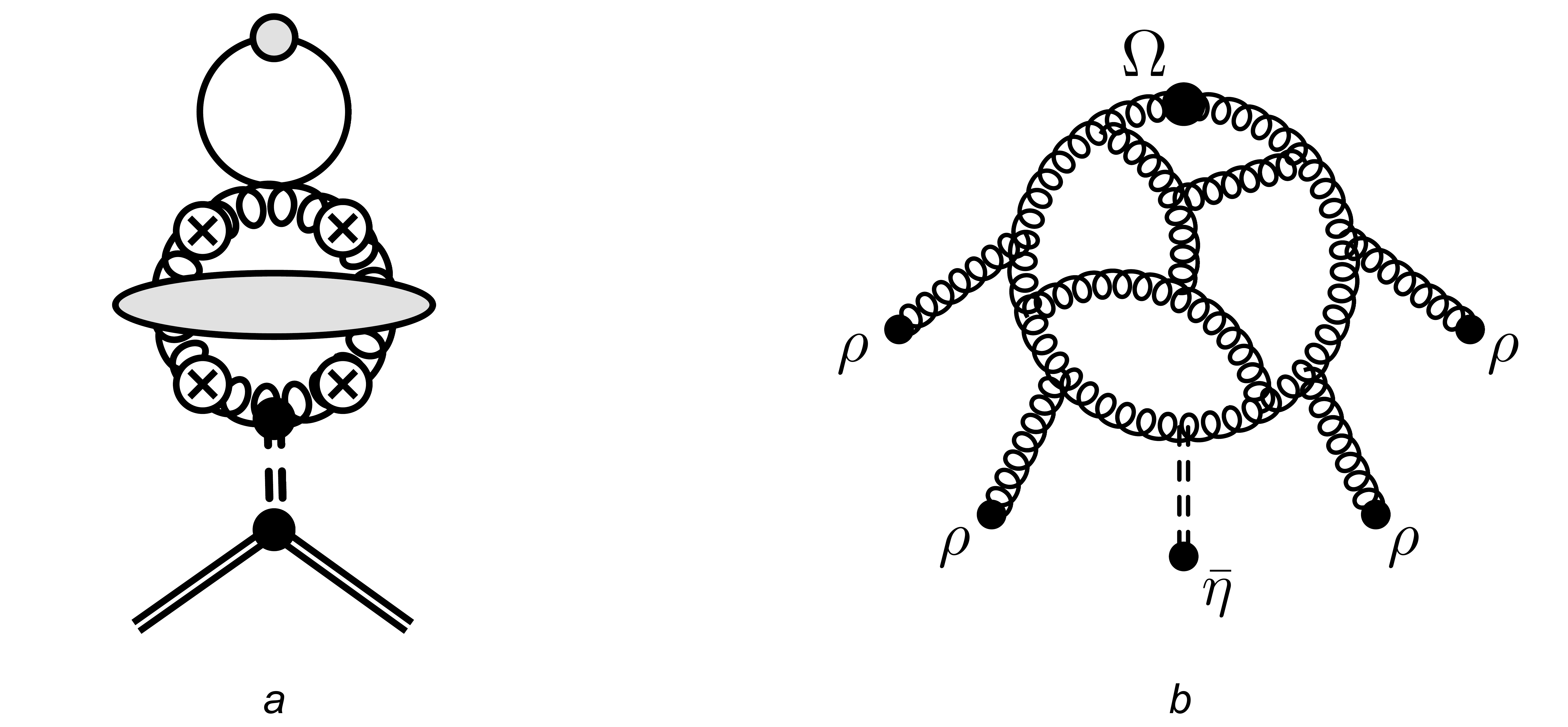}
\end{center}
\caption{Left: Diagram similar to Fig.~\ref{fig:2nn} (c) but now including the coupling to color  sources with each $\rho$ insertion denoted by the ``operator" symbol. Right: The correlator $\langle \Omega {\bar \eta}\rangle$ in the diagram on the left in the background field generated by static classical color sources $\rho$ at small $x_B$.
}
\label{fig:rho-coupling}
\end{figure}

Our interest here is in deriving the spin-dependent effective action in the Regge limit of $x_B\rightarrow 0$. From our general discussion in Appendix~\ref{sec:CGC}, in addition to the evolution of the initial density matrix of the polarized proton in coordinate/momentum phase space, and in color, we must consider its evolution in spin and flavor. As we have discussed at length, the evolution of the density matrix in the flavor isosinglet  sector is governed by the imaginary part of the worldline effective action, specifically the WZW term in Eq.\,(\ref{eq:WZW-eta0}) and the corresponding kinetic term for the ${\bar \eta}$ field in Eq.\,(\ref{eq:action-kinetic}). Combining these with the CGC effective action (\ref{MVmodel-Main}) we obtain
\begin{eqnarray}
\label{eq:CGC-EFT-polarized-Main}
\langle \mathcal{O} \rangle_{\rm pol.}^{\rm Regge} = \int \mathcal{D}\rho \,W_{Y}[\rho] \int {D{\bar \eta}}\, {\tilde W}_{P,S}[{\bar \eta}] \int \mathcal{D}A~\mathcal{O}[A]~e^{iS_{\rm pCGC}[A, \rho,{\bar\eta}]}\,,
\end{eqnarray}
where the spin-polarized CGC effective action is 
\begin{eqnarray}
\label{eq:CGC-EFT-unpolarized-Main-action}
&&S_{\rm pCGC}[A, \rho,{\bar\eta}] = S_{\rm CGC}[A, \rho] + \int d^4 x \left[ \frac{1}{2}\,(\partial_\mu {\bar \eta})\,(\partial^\mu {\bar \eta})
- \frac{\sqrt{2n_f}}{F_{{\bar \eta}}}{\bar \eta}\,\Omega\right]\,.
\end{eqnarray}
In particular,  Eq.\,(\ref{eq:g1-pol-Regge-EFT}) for $g_1$ in the Regge limit is\footnote{As discussed in Paper I, a consistent treatment in the worldline formalism would set $x_B=0$ in the argument of Eq.~(\ref{eq:g1-pol-Regge-EFT}).}
\begin{eqnarray}
&&\label{eq:g1-pol-Regge-CGC-EFT}
g_1^{\rm Regge}(x_B, Q^2) =   \left(\sum_f e_f^2\right) \frac{n_f\alpha_s}{\pi M_N}
i \int d^4 y\,\int^1_{x_B} \frac{dx}{x} 
~ \int \frac{d\xi}{2\pi} e^{-i\xi x } \,\int \mathcal{D}\rho ~W_{Y}[\rho] \, \int {D{\bar \eta}}\, {\tilde W}_{P,S}[{\bar \eta}] \, \int [DA]
\nonumber\\
&&\times\, {\rm Tr_c} F_{\alpha\beta}(\xi n) \tilde{F}^{\alpha\beta}(0)\,\eta_0(y)\, \exp \left(i S_{\rm CGC} + i\int d^4 x \left[ \frac{1}{2}\,(\partial_\mu {\bar \eta})\,(\partial^\mu {\bar \eta})
- \frac{\sqrt{2n_f}}{F_{{\bar \eta}}}{\bar \eta}\,\Omega\right]\right) \,.
\end{eqnarray}
This functional integral, describing spin diffusion of $S^\mu$ at small $x_B$, is illustrated in Fig.~\ref{fig:rho-coupling}. 

The path integral over $\rho$ causes $g_1^{\rm Regge}$  to differ qualitatively from the corresponding expression in the Bjorken limit. In Fig.~\ref{fig:rho-coupling} (left), the coupling of the color sources to $\chi(l^2)$ is illustrated with the ``operator" symbol here representing the $\rho$ insertions; the figure on the right represents the correlator 
$\langle \Omega {\bar \eta}\rangle$ inside Fig.~\ref{fig:rho-coupling} (left). 

In the absence of the coupling to the ${\bar \eta}$ field, the topological charge density 
$\Omega\propto {\bf E}\cdot {\bf B}$ in the CGC is zero~\cite{Kharzeev:2001ev}; we note though that a finite $\Omega$ is generated in a nuclear collision~\cite{Kharzeev:2001ev,Lappi:2006fp,Kharzeev:2007jp,Fillion-Gourdeau:2008abi,
Mace:2016svc,Jokela:2020ibz}. The interplay of the axion-like coupling with the CGC was also  considered recently for the case of a physical axion interacting with the CGC, which bears strong similarity with our problem of the ${\bar \eta}$ interacting with the CGC gauge fields~\cite{Jokela:2020ibz}. It is also analogous to the problem of an axion or axion-like field propagating in a hot non-Abelian plasma, which is 
relevant in a number of cosmological contexts~\cite{McLerran:1990de,Berghaus:2020ekh}.
The axion dynamics considered in \cite{McLerran:1990de} and \cite{Jokela:2020ibz}, respectively, are especially relevant because, as we will now discuss, they bookend two model  approaches to computing the effective action in Eq.\,(\ref{eq:g1-pol-Regge-CGC-EFT}) that apply in different kinematic regimes of interest.

\subsection{Spin diffusion via over-the-barrier topological transitions}
\label{sec:sphaleron}

\begin{figure}[htb]
\begin{center}
\includegraphics[width=140mm]{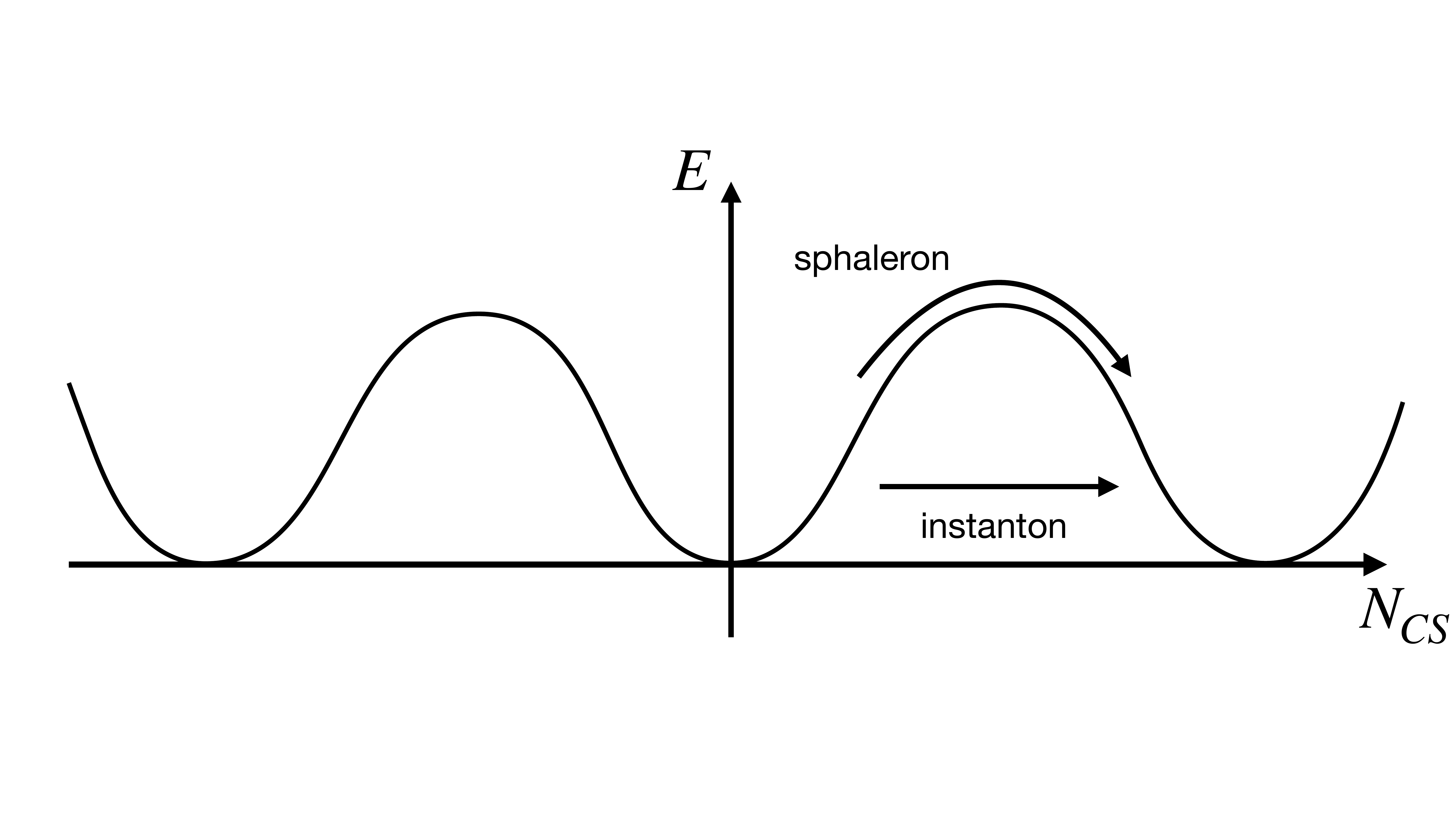}
\end{center}
\caption{The vacuum energy of $\theta$-vacua as function of the Chern-Simons number $N_{\rm CS}$ labeled by positive and negative integers. The arrows denote instanton/anti-instanton tunneling configurations and over-the-barrier sphaleron configurations; both of these effects result in a change of the Chern-Simons number. 
}
\label{fig:vacuum-energy}
\end{figure}
    In the first approach, the saddle point in Eq.\,(\ref{eq:pol-Regge-EFT}) of the path integral in over gauge fields, after an analytic continuation to Euclidean space\footnote{The color sources $\rho$ are static on the relevant time scales.}, is given by instanton classical fields $A_\mu^{\rm inst.}$ satisfying $F_{\mu\nu}=\pm {\tilde F}_{\mu\nu}$. As we noted previously, without the term in Eq.\,(\ref{MVmodel-Main}) coupling the gauge fields to $\rho$, the effective action in Eq.\,(\ref{eq:pol-Regge-EFT}) should be identical to the effective action in Sec.~\ref{sec:eta-prime-EFT} with the instanton fields saturating the topological charge density~\cite{Schafer:1996wv}. Recall that this action reproduces the results of Sec.~\ref{sec:Shore-Veneziano}. 
    
In this approach, the $J\cdot A$ coupling of the color sources to the gauge fields, along with the average over the color density matrix $W_Y[\rho]$ can be considered analogous to a thermal average\footnote{The evolution equation for $W_Y[\rho]$  satisfies the Kossakowski-Lindblad form of the density matrix for open quantum systems~\cite{Armesto:2019mna}.}, with the saturation scale $Q_S$ playing an analogous role to the temperature.   
    
The computation then follows along the lines\footnote{One must understand ${\bf E}\equiv {\bf E(\rho)}$  and $ {\bf B}\equiv  {\bf B(\rho)}$ in the Hamiltonian employed in that  derivation.} of \cite{McLerran:1990de}. From the equations of motion in Eq.\,(\ref{eq:g1-pol-Regge-CGC-EFT}), one has
    \begin{equation}
        \frac{\sqrt{2n_f}}{F_{\eta^\prime}} \Omega = -\partial_\mu \partial^\mu  {\eta^\prime}\,,
    \label{eq:MMS1}
    \end{equation}
    The explicit derivation in \cite{McLerran:1990de}, performed in the real-time Schwinger-Keldysh formalism, gives after thermal averaging (here corresponding to the averaging of sources),
    \begin{equation}
        \frac{\partial^2 {\eta^\prime}}{\partial t^2}= -\gamma \frac{\partial { \eta^\prime}}{\partial t} - m_{\eta^\prime}^2 {\eta^\prime}\,.
        \label{eq:MMS2}
    \end{equation}
    A subtle point, discussed at length in \cite{McLerran:1990de}, is that the coupling of the gauge fields to the color charges does not alter topological mass generation whereby  ${\bar \eta}\rightarrow \eta^\prime$. Both ${\bar \eta}$ and $\eta^\prime$ couple identically to $\Omega$, with the only difference being the strength of the coupling given by the difference in their respective decay constants.

In particular, note that the only difference in the equations of motion relative to 
    that derived from Eq.\,(\ref{eq:eta-prime-action}) is the term with the friction coefficient $\gamma$. This term reflects the drag on $\eta^\prime$ propagation due to the coupling of the color sources to the gauge field. In our picture, this is fundamentally what causes the quenching of the coefficient ($g_1$) of the spin four-vector $S^\mu$ reflecting the efficiency of spin diffusion\footnote{In high energy DIS, it is more convenient to represent Eq.\,(\ref{eq:MMS1}) in lightcone coordinates. Clearly, this choice of coordinates should not alter our discussion of the physics of spin diffusion.}. 

The underlying dynamics is illustrated in Fig.~\ref{fig:vacuum-energy}. In `t Hooft's picture~\cite{tHooft:1976snw,tHooft:1986ooh}, tunneling instanton-anti-instanton  configurations generate  the nontrivial Yang-Mills topological susceptibility which, we have seen, are responsible for the large $\eta^\prime$ mass. The effect of the coupling to large 
$x$ sources, and the averaging over $W_Y[\rho]$ is to introduce the saturation momentum $Q_S > \Lambda_{\rm QCD}$, which can lead to over-the-barrier sphaleron transitions as shown in Fig.~\ref{fig:vacuum-energy}. For the finite temperature case, the friction coefficient $\gamma$ is proportional to the sphaleron transition rate~\cite{McLerran:1990de}: $\gamma = 2n_f\Gamma_{\rm sphaleron}/F_{\bar \eta}^2 T$, where\footnote{The Chern-Simons current $K^\mu = \frac{g^2}{8\pi^2}\varepsilon^{\mu\nu\rho\sigma} {\rm Tr}\left(A_\nu \partial_\rho A_\sigma-\frac{2ig}{3} A_\nu A_\rho A_\sigma\right)$, which satisfies $\partial_\mu K^\mu = \Omega$. } 
\begin{equation}
    \Gamma_{\rm sphaleron} = \lim_{\delta t\rightarrow \infty}\frac{1}{V \delta t} \langle \left(N_{\rm CS}(t+\delta t) - N_{\rm CS}(t)\right)^2\rangle \,,
\end{equation}
with $N_{\rm CS}(t) = \int d^3 x \,K^0$.
Here $V$ denotes the three dimensional volume of the system. At finite temperature, 
$\Gamma_{\rm sphaleron} = \kappa \alpha_s^5\, T^4$, where $\kappa$ is a nonperturbative constant~\cite{Moore:2010jd}. In the CGC, from parametric arguments alone\footnote{Interestingly, in numerical simulations of the hot and dense Glasma~\cite{Kharzeev:2001ev,Lappi:2006fp,Mace:2016svc} produced in a nuclear collision, one finds that the sphaleron transition rate 
scales with the string tension of a spatial Wilson loop in the Glasma~\cite{Mace:2016svc}. }, one can deduce 
that $\Gamma_{\rm sphaleron}\propto Q_S^4$ and $\gamma \propto 2n_f\frac{Q_S^3}{F_{\bar \eta}^2}$. Parametrically, for $t\sim 1/Q_S$, the interaction time of the probe with the shock wave, the first term on the r.h.s will dominate over the second for when $\gamma^2 > m_{\eta^\prime}^2$, or 
equivalently, when $\frac{2n_f}{F^2_{\bar \eta}} Q^6_S > \chi_{\rm YM} (0)$. When the friction term dominates, 
${\bar \eta}\propto m_{\eta^\prime}\,\exp(-\gamma t)$ $\color{Red}$ . From Eq.\,(\ref{eq:MMS1}), we then have 
\begin{equation}
\langle \Omega \eta_0\rangle \propto F_{\bar \eta} \, \chi_{\rm YM}(0)\, \left(\frac{Q_S^2}{F_{\bar \eta}^2}\right)^3 \exp\left(- 4n_f\,C\,\frac{Q_S^2}{F_{\bar \eta}^2}\right)\,, 
\end{equation}
for $t\sim \frac{1}{Q_S}$, with $\langle\cdots\rangle$ denoting the average over the 
path integrals in Eq.\,(\ref{eq:g1-pol-Regge-CGC-EFT}). Here $C$ is a nonperturbative constant and we have employed Eq.\,(\ref{metapmdef}), the  Witten-Veneziano formula.
Substituting this expression in Eq\,(\ref{eq:g1-pol-Regge-CGC-EFT}), 
we obtain\footnote{In writing this expression we have assumed that the four-volume corresponding to the $\eta^\prime$ field is sensitive only to scales $1/Q_S$ over which a 
sphaleron transition takes place inducing the drag, but is homogeneous over longer spacetime scales, as suggested by Eq.\,(\ref{eq:MMS2}). Since the sphaleron transition rate is defined per unit four volume, the two 
factors effectively cancel. We have also assumed that the phase does not contribute at small $x_B$.} 
\begin{eqnarray}
 g_1^{\rm Regge}(x_B, Q^2) \propto  \frac{Q_S^2 m_{\eta^\prime}^2}{ F_{\bar\eta}^3 M_N}\,\exp\left(- 4\,n_f C\,\frac{Q_S^2}{F_{\bar\eta}^2}\right)\,,
\end{eqnarray}
 We have not specified the prefactors of the expression in this model computation (along the lines of \cite{McLerran:1990de}) of the effective action in Eq.~(\ref{eq:g1-pol-Regge-CGC-EFT}) because, unless $C$ is much smaller than $O(1)$, they do not affect the takeaway message that $g_1$ is exponentially quenched with increasing $Q_S$, already for $Q_S$ of a few hundred MeV.

A detailed derivation of the arguments outlined above and predictions for polarized DIS measurements at the EIC~\cite{Accardi:2012qut,Aschenauer:2017jsk} will be the subject of Paper III~\cite{PaperIII}. The kinematic regime where they are valid is where the coupling of color sources to the gauge fields can be treated as a perturbation to the instanton-anti-instanton configurations populating the QCD vacuum. More specifically,  one requires small $x_B$ values where the color sources can be approximated as classical color charge configurations but one still has $Q_S < m_{\eta^\prime}$. We will now turn our attention to the strict Regge regime where $x_B\rightarrow 0$, giving $Q_S\gg m_{\eta^\prime}$.

\subsection{Spin diffusion through topological shock wave configurations}
\label{sec:top-shock-wave}
\begin{figure}[htb]
\begin{center}
\includegraphics[width=180mm]{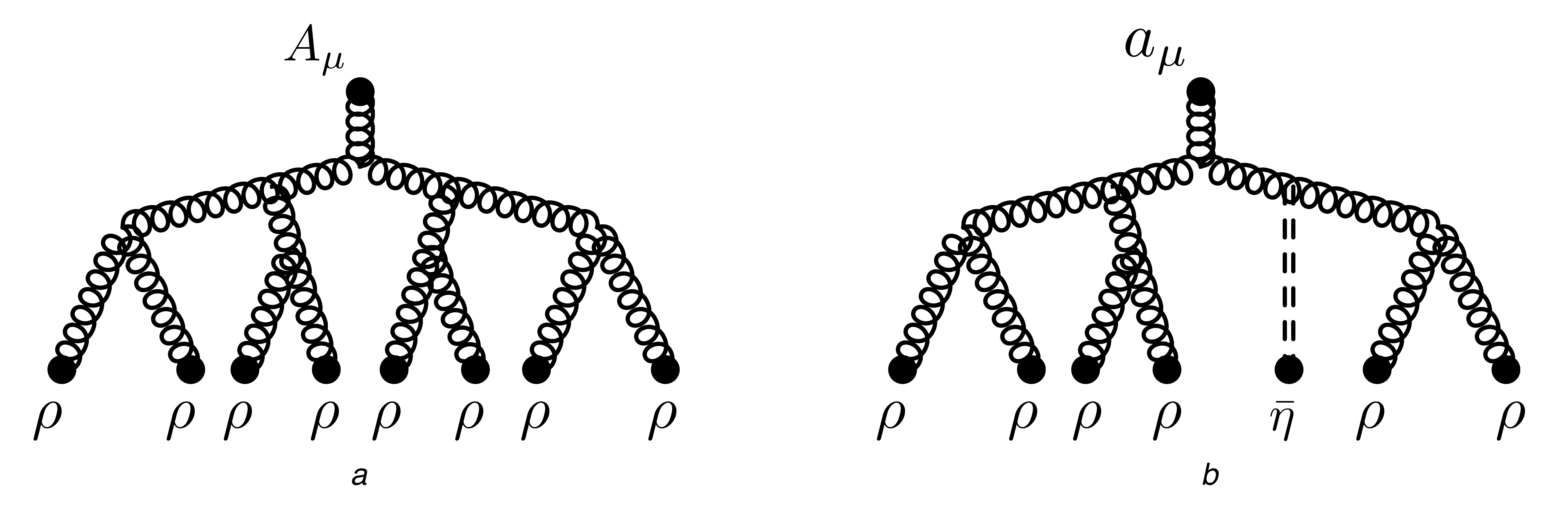}
\end{center}
\caption{
Left: The shock wave gauge field configuration $A_\mu$ of the MV model~\cite{McLerran:1993ka,McLerran:1993ni} generated by static classical color sources $\rho$. Right: The modified shock wave configuration 
$A_\mu  + a_\mu$, where $a_\mu$ is generated by the axion-like current resulting from the WZW ${\bar \eta}\,\Omega$ coupling.
}\label{fig:omega-shock-wave}
\end{figure}

In Regge asymptotics, the coupling term in Eq.\,(\ref{MVmodel-Main}) is as large as the Yang-Mills action, with both being $O(1/g^2)$. It can no longer be thought of as a perturbation to the instanton-anti-instanton configurations. The saddle point solution that minimizes the action in  Eq.\,(\ref{MVmodel-Main}) corresponds to the CGC shock wave solutions~\cite{McLerran:1993ka,McLerran:1993ni,JalilianMarian:1996xn,Kovchegov:1996ty}.  These configurations by themselves do not carry any topological charge. However the presence of the ${\bar \eta}\,\Omega$ term in Eq.\,(\ref{eq:pol-Regge-EFT}) changes this result qualitatively. 

Similarly to the discussion in \cite{Jokela:2020ibz}, the equations of motion are 
\begin{eqnarray}
 D_\mu F^{\mu\nu} = J^\nu + j_{\rm ax}^\nu \,\,\,;\,\,\, D_\mu {\tilde F}^{\mu\nu} = 0\,,
\end{eqnarray}
with 
\begin{eqnarray}
 j_{\rm ax}^\nu = -\partial_\mu {\bar \eta}\,{\tilde F}^{\mu\nu}\,\,\,{\rm with}\,\,\,
 D_\mu j_{\rm ax}^\mu = 0\,.
\end{eqnarray}
In contrast to Sec.~\ref{sec:sphaleron}, we will not consider the backreaction on the ${\bar \eta}$ fields from $\Omega$, since the 
shock wave fields are much shorter-lived on the relevant time scales in Regge asymptotics.

In the absence of the axion current $j_{\rm ax}^\mu$, the equations of motion can be solved exactly. For a nucleus moving with large $P^+\rightarrow \infty$, for the corresponding lightcone current given by $J^\mu = \delta^{\mu+} \rho(x^-,x_\perp)$, one obtains (in Lorenz gauge $\partial_\mu A_{\rm cl}^\mu=0$)
\begin{equation}
\label{eq:cov-gauge}
    A_{\rm cl}^+ = -\frac{\rho}{\nabla_\perp^2}\,\,\,;\,\,\,\, A_{\rm cl}^-=0\,\,\,\,;\,\,\,\, A_{\rm cl}^i=0 \,.
\end{equation}
These gauge configurations are static in $x^+$ since $D_\mu J^\mu=0$, where $D_\mu$ represents the covariant derivative. They are also pure gauge configurations satisfying $F^{ij}=0$, with $i,j=1,2$. 

Turning on the current $j_{\rm ax}^\mu$ induces a gauge field $a_{\rm cl}^\mu$; solving the equations of motion for $A_{\rm cl}^\mu +a_{\rm cl}^\mu$, is equivalent to solving the small fluctuation equations,
\begin{equation}
    \left( D^2 g^{\nu\mu} -D^\nu D^\mu - 2ig F_{\rm cl}^{\nu\mu}\right) a_{{\rm cl},\mu} = j_{{\rm ax}, {\rm cl}}^\nu \,,
\end{equation}
where the covariant derivatives are understood to be those of the classical fields, and 
$j_{{\rm ax}, {\rm cl}}^\nu= -\partial_\mu {\bar \eta}\, {\tilde F}_{\rm cl}^{\mu\nu}$.
Here ${\tilde F}_{\rm cl}^{\mu\nu}$ is dual field strength in $A^-=0$ gauge, whose only nonzero component is
\begin{equation}
\label{eq:axial-dual-Bfield}
{\tilde F}_{\rm cl}^{+i,a}= \epsilon^{ij}\,U^{ab}\, \partial_j A_{\rm cl}^{+,b}\,,
\end{equation}
expressed in terms of the field strength in Lorenz gauge, with $A_{\rm cl}^{+,a}$ given 
in Eq.\,(\ref{eq:cov-gauge}) and $U^{ab}$ is an adjoint matrix which will be defined below shortly. 

Since this equation is linear in $a_{{\rm cl},\mu}$, it is straightforward to solve, with 
\begin{equation}
    a_{{\rm cl},\mu}(x) = \int d^4 y\, G_{\mu\nu}(x-y)\, j_{{\rm ax}, {\rm cl}}^\nu(y)\,,
\end{equation}
with 
\begin{equation}
    \left( D^2 g^{\mu\rho} -D^\mu D^\rho - 2ig F_{\rm cl}^{\mu\rho}\right)G_{\rho\nu}(x-y) = i\delta^\mu_\nu\delta^{(4)}(x-y) \,,
\end{equation}
where $G_{\mu\nu}(x-y)$ is the shock wave gluon propagator in the axial gauge. Its Fourier transform\footnote
{We define $$G^0(x-y) = \int \frac{d^4 p}{(2\pi)^4} e^{ip\cdot (x-y)} G^0 (p) \,\,\, {\rm and}\,\,\,
G(x,y) = \int \frac{d^4p}{(2\pi)^4} \frac{d^4 q}{(2\pi)^4} e^{i p\cdot x +i q\cdot y} 
    {\tilde G}(p,q)\,.$$
} can be written as~\cite{McLerran:1994vd,Ayala:1995kg,Balitsky:2001mr}
\begin{equation}
    {\tilde G}_{\mu\nu;ab}(p,q) = {\tilde G}_{\mu\rho;ac}^0(p) {\cal T}_g^{\rho\sigma;cd}(p,q) {\tilde G}_{\sigma\nu;db}^0(q)\,,
\end{equation}
with the free propagator in $a^-=0$ gauge ($n^\mu = \delta^{\mu+}$, satisfying $n\cdot a = a^-$), 
\begin{equation}
    {\tilde G}_{\mu\nu;ab}^0(p) = \frac{i}{p^2+i\epsilon}\left( -g_{\mu\nu} + \frac{p_\mu n_\nu + p_\nu n_\mu}{p\cdot n}\right)\delta_{ab} \,,
\end{equation}
and the effective vertex 
\begin{equation}
    {\cal T}_g^{\mu\nu;ab}(p,q) = -2\pi \delta(p^--q^-) (2p^-)\, {\rm sign}(p^-) \,g^{\mu\nu}\int d^2 {\bf z}_\perp e^{-i(p-q)_\perp\cdot {\bf z}_\perp} \left(U^{ab}\right)^{{\rm sign}(p^-)} ({\bf z}_\perp) \,.
\end{equation}
The dependence of the shock wave propagator on the color sources $\rho$ is contained in the adjoint Wilson line
\begin{equation}
    U(x_\perp) = {\cal P}_- \exp\left(- i g \int dz^- \frac{1}{\nabla_\perp^2} \rho^a (z^-,x_\perp) T^a\right) \,,
\end{equation}
where ${\cal P}_-$ denotes path ordering in $x^-$ variable and $T^a$ are matrices representing the adjoint generators of the $SU(3)$ color algebra. 

Following \cite{Jokela:2020ibz}, we can write 
\begin{equation}
\label{eq:Om-final-1}
    \Omega = \frac{\alpha_s}{2\pi}\,\left(\partial_+ a_i^a \right)\,{\tilde F}_{\rm cl}^{+i,a}\,,
\end{equation}
where the parenthesis denotes the electric field induced by the ${\bar \eta}$ field and the other term is the CGC background magnetic field given by Eq.\,(\ref{eq:axial-dual-Bfield}). This result for $\Omega$, illustrated in Fig.~\ref{fig:omega-shock-wave}, is a nontrivial functional of $\rho$, and linear 
in ${\bar \eta}$. 

Since the only nonzero component the classical dual field strength tensor is ${\tilde F}^{+i} = -\epsilon^{ij}\,F_{-j}$, and since our choice of gauge gives $a^-=0$,
we only obtain the contribution from $a_i$ given by
\begin{equation}
\label{eq:Om-final-2}
    \partial_+ a_i = -\int d^4 y\, (\partial_+ G_{ij})\,  (\partial_+{\bar \eta})\,
    {\tilde F}_{\rm cl}^{+j}(y) \,.
\end{equation}
Substituting Eq.\,(\ref{eq:Om-final-2}) into Eq.\,(\ref{eq:Om-final-1}), we now have all the ingredients to compute $\langle \Omega\rangle$. 
Thus similarly to Sec.~\ref{sec:sphaleron}, we can write 
\begin{eqnarray}
\label{eq:Omega-CGC}
 \langle \Omega \eta_0\rangle \propto \langle (\partial_+{\bar \eta}){\bar \eta}\rangle_{{\bar \eta}}\,\langle  {\tilde F}_{\rm cl}^{+i}(x)\, {\tilde F}_{\rm cl}^{+j}(y)\rangle_{\rho}\,,
\end{eqnarray}
where the second term on the r.h.s is the correlator of transverse magnetic fields. The detailed computation of the r.h.s is quite subtle and will be a subject addressed further in Paper III~\cite{PaperIII}. However on general grounds, one expects the correlator of 
magnetic fields in the CGC to obey
\begin{eqnarray}
\label{eq:eta-prop-CGC}
 \langle  {\tilde F}_{\rm cl}^{+i}(x)\, {\tilde F}_{\rm cl}^{+j}(y)\rangle_{\rho} \propto 
 \exp\left(- Q_S^2 (x-y)^2\right) \approx \exp\left(- {\bar C}\,\frac{Q_S^2}{F_{\bar \eta}^2}\right)\,,
\end{eqnarray}
where ${\bar C}$ is a constant factor that must be computed and the typical scale for the propagation of the ${\bar \eta}$ field can reasonably be set to be its decay constant. Hence as in the case of the model computation in Sec.~\ref{sec:sphaleron}, from 
Eqs.\,(\ref{eq:Omega-CGC}) and (\ref{eq:eta-prop-CGC}), one expects $g_1(x_B,Q_S^2)$ in the Regge limit to be similarly exponentially suppressed with increasing $Q_S$. As noted, a more detailed computation of the path integral in Eq.\,(\ref{eq:g1-pol-Regge-CGC-EFT}), and phenomenological consequences thereof, beyond these simple model estimates, will be discussed separately in Paper III~\cite{PaperIII}.

\section{Summary and outlook}
In our previous paper~\cite{Tarasov:2020cwl}, we computed the contribution of the box diagram to the polarized structure function $g_1(x_B,Q^2)$ employing the worldline representation of the fermion determinant in QCD. We demonstrated that the isosinglet triangle anomaly dominates the structure of the box diagram in both the Bjorken and Regge asymptotics of QCD. Specifically, $S^\mu\,g_1$ is proportional to forward limit $l^\mu\rightarrow 0$ of the off-forward matrix element of the nonlocal operator $\frac{l^\mu}{l^2}\,\Omega$ , where $\Omega$ is the topological charge density in a polarized proton with the spin four-vector $S^\mu$. That this result holds when $x_B\rightarrow 0$ is remarkable and strongly suggestive of  the  fundamental role of the topology of the QCD vacuum in the proton's spin. 

In this paper, we significantly developed the framework introduced in \cite{Tarasov:2020cwl}. A major focus was to demonstrate how the off-forward pole of the anomaly cancels in the forward limit and the consequences thereof. In order to do so, we reaffirmed that the anomaly arises from the imaginary part of the worldline effective action, which corresponds to the phase of the fermion determinant. We then discussed a generalization of the worldline effective action that takes into account the coupling of  fermion modes to low energy scalar, pseudoscalar, vector and axial vector degrees of freedom. Limiting ourselves to isosinglet contributions to the imaginary part of the worldline effective action, we showed explicitly (in Appendix \ref{sec:WZW-appendix}) the existence of a  Wess-Zumino-Witten 
  term that couples the topological charge density $\Omega$ to a massless isosinglet pseudoscalar field ${\bar \eta}$. While this particular WZW contribution is well-known in the chiral perturbation theory literature, our derivation of this term in the worldline framework is new. 

We then demonstrated the fundamental role played by the WZW term in the cancellation of the anomaly pole in the off-forward matrix element of the isosinglet axial vector current 
$J_\mu^5$ in the polarized proton; in the forward limit, this matrix element determines the proton's helicity $\Sigma(Q^2)$. We first identified the axial vector and pseudoscalar contributions to this 
matrix element and derived the anomalous Goldberger-Treiman 
relation  that connects the two. Specifically, as first suggested by Veneziano, the axial vector charge $G_A$ representing the direct coupling of $J_\mu^5$ to the polarized proton can be equated to the product of the isosinglet coupling to the proton times its decay constant. 

Due to the WZW term, this pseudoscalar exchange can also be mediated through the anomaly, specifically the QCD topological susceptibility $\chi(l^2)$. The leading contribution to this quantity is the Yang-Mills topological susceptibility $\chi_{\rm YM}(l^2)$, which is of the order of typical nonperturbative QCD scales. However, as shown by Witten and Veneziano, higher order $1/N_c$ contributions to $\chi_{\rm YM}(l^2)$ from ${\bar \eta}$ exchange mediated via the WZW term results in the topological generation of the $\eta^\prime$ mass, resolving the $U_A(1)$ problem. The resulting topological screening of the ${\bar \eta}$ pole then ensures that, in the chiral limit, $\chi(l^2)\rightarrow 0$ when $l^2\rightarrow 0$. We showed explicitly how this topological screening results in the cancellation of the pole of the anomaly and recovered the striking result of \cite{Shore:1990zu,Shore:1991dv} that $\Sigma(Q^2)\propto \sqrt{\chi'(0)}$, where $\chi'(0)$ is the slope of the QCD topological susceptibility in the forward limit. In the topological screening picture, the fact that $\chi(0) =0$ in the chiral limit provides a natural explanation of the ``spin puzzle" of why the measured isosinglet axial charge is much smaller than its octet counterpart~\cite{Veneziano:1989ei,Shore:2007yn,Kuhn:2008sy}. In other words, the underlying physics that resolves the $U_A(1)$ problem also resolves the proton's spin puzzle. 

In \cite{Tarasov:2020cwl}, we showed that $g_1$ in the Regge limit is represented by the same matrix element that contributes in Bjorken asymptotics. 
However the computation of the matrix element in the two limits is quite different because the former is strongly influenced by the physics of gluon saturation which introduces a large emergent scale $Q_S(x)$. Using the insights provided by the worldline effective action outlined in Appendices \ref{sec:WZW-appendix}-\ref{sec:eikonal}, we constructed an axion-like effective action for $g_1$ that captures the physics of gluon saturation and is consistent with anomalous chiral Ward identities. The underlying dynamics of this action is controlled by the saturation scale and the Yang-Mills topological susceptibility $\chi_{\rm YM}$, or equivalently, the $\eta^\prime$ mass. In the absence of the coupling to a large number of color sources (represented by $Q_S$), our formulation is compatible with a picture of spin diffusion mediated by instanton-anti-instanton configurations, as suggested by 't Hooft's explanation of the $U_A(1)$ problem.

However when classical sources begin to play a role at small $x_B$, the additional presence of a dynamical momentum scale $Q_S$
can induce over-the-barrier sphaleron-like topological transitions (as opposed to the instanton-anti-instanton tunneling transitions) between different $\theta$-vacua labeled by integer-valued Chern-Simons numbers. This is very similar to the temperature induced sphaleron transitions that have been studied previously. In particular, one can map our axion-like effective action in this case to earlier work \cite{McLerran:1990de} describing the propagation of an axion in a hot QCD plasma. Similarly to that study, the presence of classical color sources does not prevent topological mass generation of the $\eta^\prime$; it however experiences a drag force that strongly impacts spin diffusion. 

This picture of spin diffusion is plausible for $x_B$ values in the kinematic window corresponding to $\Lambda_{\rm QCD} < Q_S < m_{\eta^\prime}$. However at very small $x_B$ when $Q_S \gg m_{\eta^\prime}$, our effective action suggests that it is more likely that static shock wave CGC configurations dominate over instanton/sphaleron-like configurations. In this case, the WZW ${\bar \eta}\,\Omega$ coupling generates a current that induces a nontrivial topological charge density in the gluon shock wave. The problem here, with essential modifications, is similar to recent work~\cite{Jokela:2020ibz} on the interaction of a putative axion with QCD matter in the presence of classical color sources. The influence of such a perturbation on the shock wave diminishes rapidly with increasing $Q_S$ leading to a rapid quenching of spin diffusion in the Regge limit. 

This first study can be quantified further to provide concrete predictions for $g_1$ at small $x_B$. It is equally important is to understand how renormalization works in our framework analogously to previous work in perturbative QCD~\cite{Kodaira:1979pa,Vogelsang:1990ug,Zijlstra:1993sh,Vogelsang:1995vh,deFlorian:2019egz}. While the anomaly equation holds both for bare and renormalized quantities~\cite{Espriu:1982bw}, the correspondence at small $x_B$ between the smearing of the topological charge density and that of the gluon shock wave needs to be better understood. Work in this direction is in progress. Not least, for quantitative precision, we will need to extend our computation beyond the chiral limit and take into account the influence of light quark masses. Following the pioneering work in \cite{Leutwyler:1996sa,HerreraSiklody:1996pm,Kaiser:2000gs}, there has been considerable work in the chiral perturbation theory of the $U(3)$ nonet both on the phenomenology~\cite{Gan:2020aco} of $\eta^\prime$ mixings and decays and on high order precision computations in this framework~\cite{Vonk:2019kwv}. With regard to the latter, 
finite temperature computations are especially relevant~\cite{GomezNicola:2019myi}. 

It will also be important to identify other signatures of the topological screening picture, given the exciting possibility that $g_1$ at small $x_B$
could be sensitive to sphaleron-like transitions. One possibility discussed previously in the literature is to measure semi-inclusive hadron production in DIS, off polarized proton and deuteron targets, in the target fragmentation region~\cite{Shore:1997tq,deFlorian:1997th}. Specifically, it was argued that first $x_B$ moments of so-called ``fracture" functions~\cite{Grazzini:1997ih} (of the momentum fraction of the nucleon carried by the hadron) satisfies the following. i) It is sensitive to the ratio of the isosinglet and isotriplet axial charges; this ratio, a quantitative measure of OZI violation, is proportional to $\chi'(0)$.  ii) It is independent of the target. Our work suggests that such an OZI suppression may be strongly sensitive to $x_B$ and in this kinematic regime, may also have a target dependence due to the differing color charge densities probed. For a discussion of effects of the anomaly in the context of quark fragmentation, see \cite{Kang:2010qx}. 
One can also pursue in parallel similar signatures in polarized proton-proton collisions\footnote{We thank Werner Vogelsang for a discussion on this point.}. These and other such possible phenomenological consequences with be pursued separately.

We now turn to our outlook on the possible implications of this work for QCD spin studies. The presence of the off-forward infrared pole of the anomaly in $\Sigma$ (and, as we have argued, in $g_1$) suggests i) that $\Sigma$ can be expressed as an intrinsically nonlocal operator, and ii) $\Sigma$ has a nontrivial dependence on infrared physics, in particular that governing the physics of the $\eta^\prime$. Indeed, our final result for $\Sigma$ is proportional to the square root of the slope of the topological susceptibility in the forward limit.  Our work (and preceding work in \cite{Veneziano:1989ei,Shore:1990zu}) therefore brings into question the applicability\footnote{A corollary to this question would be to better understand in this topological screening framework the role of entanglement~\cite{Beane:2019loz} in the proton's spin.} of QCD factorization for quantities such as $g_1$ that are sensitive to the anomaly~\cite{Collins:1992xw}. Likewise, similarly to our concerns regarding the applicability of collinear kinematics, we discuss at length in Appendix~\ref{sec:eikonal} why the eikonal expansion, frequently employed in studies of spin at small $x_B$, is not applicable to observables sensitive to the anomaly. This is  because the latter couples to zero modes of the Dirac operator~\cite{Leutwyler:1992yt}, the physics of which is missed when higher order terms in this expansion are omitted. Not least in importance  are the possible implications of this work for spin sum rules and their  interpretation~\cite{Aidala:2012mv,Ji:2020ena}. 

Looking beyond QCD spin, the possibility of laboratory measurements of sphaleron transitions is of great interest in phenomena spanning a wide range in energy scales. As is well-known, sphaleron transitions are conjectured to play a key role in electroweak baryogenesis~\cite{Kuzmin:1985mm,Cohen:1993nk,Riotto:1999yt}. Sphaleron-like topological transitions can also produce a 
Chiral Magnetic Effect (CME)~\cite{Kharzeev:2007jp,Fukushima:2008xe} in heavy-ion collisions. In this case, topological transitions in drive charge separation in an external magnetic field, signatures of which can be extracted in the heavy-ion experiments. While the CME has been ruled out~\cite{STAR:2021mii} at the highest energies studied at the 
Relativistic Heavy Ion Collider (RHIC), prospects for its measurement exist at lower energies~\cite{An:2021wof}. Complicating the extraction of a definitive signal of this effect is the large experimental and theoretical background~\cite{Koch:2016pzl} in the complex environment of the RHIC collisions. Our study suggests that polarized DIS measurements at small $x_B$ may provide an alternate route to empirically establish the existence of such topological transitions in nature.

\section*{Acknowledgements}
We thank  Constantia Alexandrou, Elke Aschenauer, Silas Beane, Daniel Boer, Gia Dvali, Renee Fatemi, Stefano Forte, Yoshitaka Hatta, Bob Jaffe, Xiangdong Ji, Yuri Kovchegov, Elliot Leader, Kei-Fei Liu, Rob Pisarski, Werner Vogelsang and Feng Yuan for  conversations that have influenced this work. 

A.T.’s work is supported by the U.S. Department of Energy, Office of Science, Office of Nuclear Physics under Award Number DE-SC0004286. R.V.'s work is supported by the U.S. Department of Energy, Office of Science, Office of Nuclear Physics, under Contracts No. de-sc0012704.
A.T and R.V.'s work on this topic is also supported by the U.S. Department of Energy, Office of Science, Office of Nuclear Physics,  within the framework of the TMD Theory Topical Collaboration.

\appendix

\section{Detailed derivation of the WZW action for the ${\bar \eta}$}
\label{sec:WZW-appendix}

We begin with the expression in Eq.(\ref{eq:Imag-W}):
\begin{eqnarray}
W_\mathcal{I} = - \frac{i}{32}\int^1_{-1}d\alpha \int^\infty_0 dT \mathcal{N} \int_P \mathcal{D}x \mathcal{D}\psi~ {\rm tr} ~\chi \bar{\omega}(0) \exp\Big[-\int^T_0 d\tau \mathcal{L}_{(\alpha)}(\tau)\Big]
\label{WIinit}
\end{eqnarray}

To isolate the WZW term of interest, 
we will first  expand $W_{\mathcal{I}}$ up to  order $\Pi A^2$ in a presence of the scalar field\footnote{This scalar field can be understood, via a Legendre transform, as the functional derivative of the Wess-Zumino effective action~\cite{Zumino,Wess:1971yu} with respect to the chiral condensate. We will take this field to be constant for the rest of the discussion.}  $\Phi$, assuming that all fields commute with each other. We start by expanding $W_{\mathcal{I}}$ to the linear power in $\Pi$ and taking the trace of two component matrices:\footnote{We will set the einbein $\mathcal{E}=2$ at the end of the derivation.},
\begin{eqnarray}
&&W_{\mathcal I}[\Pi] = \frac{\mathcal{E}\Phi}{4} \int^1_{-1}d\alpha \int^\infty_0 dT \mathcal{N} \int_P \mathcal{D}x \mathcal{D}\psi~ \exp\Big[-\int^T_0 d\tau \Big\{ \frac{\dot{x}^2}{2\mathcal{E}} + \frac{1}{2}\psi\dot{\psi} - i \dot{x}^\mu A_\mu + \frac{i\mathcal{E}}{2}\psi^\mu \psi^\nu F_{\mu\nu}  \Big\} \Big] \exp\Big[- T \frac{\mathcal{E}\alpha^2 \Phi^2}{2} \Big]
\nonumber\\
&&\times {\rm tr}_c \Big(    \psi_{5}(\tau_0) \Pi(x_0) - \psi^\mu(\tau_0) \dot{x}_{\mu}(\tau_0) \int^T_0 d\tau_1 \psi^\nu(\tau_1) \psi_{5}(\tau_1) \partial_\nu \Pi(x_1) \Big)
\label{WIPi}
\end{eqnarray}
where we use the shorthand notation $x_i\equiv x(\tau_i)$. The first term in this equation corresponds to expansion of the insertion $\bar{\omega}(0)$ in Eq. (\ref{WIinit}) to the leading power in $\Pi$ while the second term is obtained by expansion of the exponential factor.

Now expanding the exponential factor to the order $A^2$ we obtain,
\begin{eqnarray}
&&W_{\mathcal I}[\Pi A^2] = -\frac{\mathcal{E}\Phi}{4} \int^1_{-1}d\alpha \int^\infty_0 dT \mathcal{N} \int_P \mathcal{D}x \mathcal{D}\psi~\exp\Big[-\int^T_0 d\tau \Big\{ \frac{\dot{x}^2}{2\mathcal{E}} + \frac{1}{2}\psi\dot{\psi}  \Big\} \Big] \exp\Big[- T \frac{\mathcal{E}\alpha^2 \Phi^2}{2} \Big]
\nonumber\\
&&\times {\rm tr}_c \Big( \psi_{5}(\tau_0) \Pi(x_0) - \psi^\mu(\tau_0) \dot{x}_{\mu}(\tau_0) \int^T_0 d\tau_1 \psi^\nu(\tau_1) \psi_{5}(\tau_1) \partial_\nu \Pi(x_1)\Big) V_2 V_3
\label{func1}
\end{eqnarray}
where the interaction with the background gluon field is defined by the worldline vertex,
\begin{eqnarray}
&&V_i \equiv \int^T_0 d\tau_i \Big( \dot{x}^\rho(\tau_i) + \mathcal{E}\psi^\rho(\tau_i) \psi^\alpha(\tau_i) \partial_\alpha \Big) A_\rho(x_i)
\end{eqnarray}

To calculate the functional integrals in Eq. (\ref{func1}), it is convenient to rewrite the background fields in terms of their Fourier transforms,
which gives\footnote{In the second term of the equation, we used rotational invariance  to replace the integration over $\tau_1$ by $\tau_0$, and subsequently substituted $\tau_0\leftrightarrow \tau_1$.},
\begin{eqnarray}
&&W_{\mathcal I}[\Pi A^2] = -\frac{\mathcal{E}\Phi}{4} \int^1_{-1}d\alpha \int^\infty_0 dT \mathcal{N} \int_P \mathcal{D}x \mathcal{D}\psi \exp\Big[-\int^T_0 d\tau \Big\{ \frac{\dot{x}^2}{2\mathcal{E}} + \frac{1}{2}\psi\dot{\psi}  \Big\} \Big] \exp\Big[- T \frac{\mathcal{E}\alpha^2 \Phi^2}{2} \Big]
\nonumber\\
&&\times \prod_{k=0,2,3}\int \frac{d^4 p_k}{(2\pi)^4}\, {\rm tr}_c \psi_{5}(\tau_0) \Big(
 1 
 + i p_{0\nu} \psi^\nu(\tau_0)\int^T_0 d\tau_1 \psi^\mu(\tau_1) \dot{x}_{\mu}(\tau_1)  \Big) \,\Pi(p_0)\,e^{ip_0 x_0}\, \tilde{V}_2 \,e^{ip_2 x_2} \,\tilde{V}_3 \,e^{ip_3 x_3}\,,
 \label{func2}
\end{eqnarray}
where
\begin{eqnarray}
\tilde{V}_i \equiv \int^T_0 d\tau_i \Big[ \dot{x}^\rho(\tau_i) + i \mathcal{E}\psi^\rho(\tau_i) \psi^\alpha(\tau_i) p_{i\alpha} \Big] A_\rho(p_i)
\end{eqnarray}

With this form for Eq. (\ref{func2}), we can begin to perform the functional integration over the coordinate and Grassmann worldlines. Considering first the Grassmanian integrals,
since they satisfy periodic boundary conditions,  we first separate their zero modes as
\begin{eqnarray}
&&\psi(\tau) = \psi + \xi(\tau)\,,
\end{eqnarray}
and substitute
\begin{eqnarray}
&&\int_P \mathcal{D} \psi \exp\Big[-\int^T_0 d\tau \frac{1}{2}\psi\dot{\psi} \Big] \to \int d^5\psi \int_P \mathcal{D} \xi \exp\Big[-\int^T_0 d\tau \frac{1}{2}\xi\dot{\xi} \Big]
\end{eqnarray}

The integration over the zero modes in Eq.(\ref{func2}) can be easily done using the Grassmann identity, 
\begin{eqnarray}
\int d^5\psi \,\psi^\mu \psi^\nu \psi^\rho \psi^\sigma \psi^5 = \epsilon^{\mu\nu\rho\sigma}\,,
\label{zmodeint}
\end{eqnarray}
where the convention we use for the Levi-Civita tensor in Euclidean space is $\epsilon_{1234} = 1$.

The remaining integral over $\xi$ can be straightforwardly calculated (see Refs. \cite{Strassler:1992zr,DHoker:1995aat,Schubert:2001he} for details) using
\begin{eqnarray}
&&\int_P \mathcal{D} \xi \exp\Big[-\int^T_0 d\tau \frac{1}{2}\xi\dot{\xi} \Big] \xi^{\mu}(\tau_1)\xi^{\nu}(\tau_2) = -\frac{1}{2}g^{\mu\nu}\dot{G}_B(\tau_1, \tau_2)\,,
\label{otmodeint}
\end{eqnarray}
where the derivative of the bosonic worldline propagator is
\begin{eqnarray}
&&\dot{G}_B(\tau_1, \tau_2) = {\rm sign}(\tau_1 - \tau_2) - \frac{2(\tau_1 - \tau_2)}{T}\,. 
\end{eqnarray}
Note that before applying Eqs. (\ref{zmodeint}) and (\ref{otmodeint}) , one should properly arrange the variables in Eq.(\ref{func2}), taking into account the anti-commuting property of Grassmann variables.

After a rather laborious calculation, we obtain,
\begin{eqnarray}
&&W_{\mathcal I}[\Pi A^2] = -\frac{\mathcal{E}^2 \Phi}{4} \int^1_{-1}d\alpha \int^\infty_0 dT \mathcal{N} \int_P \mathcal{D}x \exp\Big[-\int^T_0 d\tau \frac{\dot{x}^2}{2\mathcal{E}} \Big] \exp\Big[- T \frac{\mathcal{E}\alpha^2 \Phi^2}{2} \Big] \prod_{k=0,2,3}\int \frac{d^4 p_k}{(2\pi)^4} \int^T_0 d\tau_2 \int^T_0 d\tau_3
\nonumber\\
&&\times \,{\rm tr}_c\,  \Big( \mathcal{E} p_{2\alpha} p_{3\beta} \epsilon^{\rho\alpha\sigma\beta}
- p_{0\nu} \int^T_0 d\tau_1 \dot{x}_{1\mu}
 \mathcal{S}^{\mu\nu;\sigma\rho}_{\tau_1,\tau_2,\tau_3}(p_2,p_3) \Big)\,\Pi(p_0)\,e^{ip_0 x_0}\, A_\rho(p_2)\,e^{ip_2 x_2} \,A_\sigma(p_3)\, e^{ip_3 x_3}\,,
 \label{func3}
\end{eqnarray}
where
\begin{eqnarray}
&&\mathcal{S}^{\mu\nu;\sigma\rho}_{\tau_1,\tau_2,\tau_3}(p_2,p_3)\equiv p_{2\alpha} \dot{x}^\sigma_3 \epsilon^{\mu \rho \alpha \nu }  + p_{3\beta} \dot{x}^\rho_2 \epsilon^{\mu \sigma \beta \nu } + i \frac{\mathcal{E}}{2} p_{2\alpha} p_{3\beta}  \epsilon^{\mu \rho \alpha \sigma } g^{\beta\nu}\dot{G}_B(\tau_3, \tau_0) 
 \nonumber\\
 &&- i \frac{\mathcal{E}}{2} p_{2\alpha} p_{3\beta} \epsilon^{\mu \rho \alpha \beta} g^{\sigma\nu} \dot{G}_B(\tau_3, \tau_0) + i \frac{\mathcal{E}}{2} p_{2\alpha} p_{3\beta} \epsilon^{\mu \rho \sigma \beta } g^{\alpha\nu} \dot{G}_B(\tau_2, \tau_0)
 - i \frac{\mathcal{E}}{2} p_{2\alpha} p_{3\beta} \epsilon^{\mu \rho \sigma \nu } g^{\alpha\beta} \dot{G}_B(\tau_2, \tau_3)
 \nonumber\\
 &&+ i \frac{\mathcal{E}}{2} p_{2\alpha} p_{3\beta} \epsilon^{\mu \rho \beta \nu } g^{\alpha\sigma} \dot{G}_B(\tau_2, \tau_3)
- i \frac{\mathcal{E}}{2} p_{2\alpha} p_{3\beta} \epsilon^{\mu \alpha \sigma \beta } g^{\rho\nu} \dot{G}_B(\tau_2, \tau_0) + i \frac{\mathcal{E}}{2} p_{2\alpha} p_{3\beta} \epsilon^{\mu \alpha \sigma \nu } g^{\rho\beta} \dot{G}_B(\tau_2, \tau_3) 
\nonumber\\
&&- i \frac{\mathcal{E}}{2} p_{2\alpha} p_{3\beta} \epsilon^{\mu \alpha \beta \nu } g^{\rho\sigma} \dot{G}_B(\tau_2, \tau_3)
 + i \frac{\mathcal{E}}{2} p_{2\alpha} p_{3\beta} \epsilon^{\rho \alpha \sigma \beta } g^{\mu\nu} \dot{G}_B(\tau_1, \tau_0) - i \frac{\mathcal{E}}{2} p_{2\alpha} p_{3\beta} \epsilon^{\rho \alpha \sigma \nu } g^{\mu\beta} \dot{G}_B(\tau_1, \tau_3)
 \nonumber\\
 &&+ i \frac{\mathcal{E}}{2}p_{2\alpha} p_{3\beta} \epsilon^{\rho \alpha \beta \nu } g^{\mu\sigma} \dot{G}_B(\tau_1, \tau_3)
 - i \frac{\mathcal{E}}{2} p_{2\alpha} p_{3\beta} \epsilon^{\rho \sigma \beta \nu } g^{\mu\alpha} \dot{G}_B(\tau_1, \tau_2)  
+ i \frac{\mathcal{E}}{2} p_{2\alpha} p_{3\beta} \epsilon^{\alpha \sigma \beta \nu } g^{\mu\rho} \dot{G}_B(\tau_1, \tau_2)\,.
\end{eqnarray}

The integration over the coordinate worldlines can be performed in a similar way. We begin again by separating out the zero modes as, 
\begin{eqnarray}
x(\tau_i) = \bar{y} + y(\tau_i)\,,
\label{zeromodebos}
\end{eqnarray}
and replacing
\begin{eqnarray}
&&\int_P \mathcal{D}x \to \int d^4\bar{y}\int_P \mathcal{D}y \,.
\end{eqnarray}

Performing the change of variables (Eq.(\ref{zeromodebos})) in Eq. (\ref{func3}), the integral over the zero modes $\bar{y}$ yields the overall 4-momentum conserving delta-function:
\begin{eqnarray}
&&\int d^4\bar{y} \,e^{ip_0 \bar{y}}e^{ip_2 \bar{y}}e^{ip_3 \bar{y}} = (2\pi)^4\delta^4(p_0 + p_2 + p_3)\,.
\end{eqnarray}
The remaining integral over $y(\tau)$ can be calculated using the identity 
\begin{eqnarray}
&&\mathcal{N} \int_P \mathcal{D}y \exp\Big[-\int^T_0 d\tau \frac{\dot{y}^2}{2\mathcal{E}} \Big]\, \mathcal{Y} \,e^{ip_0 y_0 }e^{ip_2 y_2 }e^{ip_3 y_3} = \frac{1}{(2\pi\mathcal{E}T)^{2}}\, \langle \mathcal{Y}\, e^{ip_0 y_0 }e^{ip_2 y_2 }e^{ip_3 y_3}\rangle\,,
\end{eqnarray}
where $\mathcal{Y}$ is an arbitrary product of factors $\dot{y}^\alpha_i$ (such as those appearing in Eq. (\ref{func3})) and the notation $\langle\mathcal{Y} e^{ip_0 y_0 }e^{ip_2 y_2 }e^{ip_3 y_3}\rangle$ denotes all possible Wick contractions between the trajectories $y$ in the product $\mathcal{Y} e^{ip_0 y_0 }e^{ip_2 y_2 }e^{ip_3 y_3}$. The resulting Wick contractions can be calculated using 
\begin{eqnarray}
\langle y^\mu(\tau_1) y^\nu(\tau_2)\rangle = - g^{\mu\nu} G_B(\tau_1, \tau_2)\,\, \,{\rm with}\,\,\,
G_B(\tau_1, \tau_2) = \frac{\mathcal{E}}{2}|\tau_1 - \tau_2| - \mathcal{E}\frac{(\tau_1-\tau_2)^2}{2T}\,,
\end{eqnarray}
and
\begin{eqnarray}
\langle y^\mu(\tau_1) e^{iky(\tau_2)}\rangle = i\langle y^\mu(\tau_1) y^\nu(\tau_2)\rangle\, k_\nu e^{iky(\tau_2)}\,.
\end{eqnarray}
After the computations using these Wick contractions, the expectation value of the remaining exponential factors should be performed, and give, 
\begin{eqnarray}
&&\langle e^{ip_0y_0} e^{ip_2y_2} e^{ip_3y_3}\rangle = \exp\Big[ p_0 \cdot p_2 G_B(\tau_0, \tau_2) + p_0\cdot p_3 G_B(\tau_0, \tau_3) + p_2\cdot p_3 G_B(\tau_2, \tau_3) \Big]\,.
\nonumber
\end{eqnarray}

As a result of these manipulations, and assuming that the background gluon fields are on the mass-shell, we obtain the formula:
\begin{eqnarray}
&&W_{\mathcal I}[\Pi A^2] = - \frac{\mathcal{E}^2 \Phi}{4} \int^1_{-1}d\alpha \int^\infty_0 dT \frac{1}{(2\pi\mathcal{E}T)^{2}}  \exp\Big[- T \frac{\mathcal{E}\alpha^2 \Phi^2}{2} \Big] \prod_{k=0,2,3}\int \frac{d^4 p_k}{(2\pi)^4} \int^T_0 d\tau_2 \int^T_0 d\tau_3
\nonumber\\
&&\times  {\rm tr}_c  \big( \mathcal{E} 
+ \int^T_0 d\tau_1 \mathcal{T}_{\tau_1,\tau_2,\tau_3}(p_2,p_3) \big) \epsilon^{\rho\alpha\sigma\beta} p_{2\alpha} A_\rho(p_2) p_{3\beta} A_\sigma(p_3)\,\Pi(p_0)
 \nonumber\\
&&  \times \exp\Big( p_2\cdot p_3 [ - G_B(\tau_0, \tau_2) -  G_B(\tau_0, \tau_3) + G_B(\tau_2, \tau_3) ] \Big) (2\pi)^4\delta^4(p_0 + p_2 + p_3)\,,
\end{eqnarray}
where 
\begin{eqnarray}
&&\mathcal{T}_{\tau_1,\tau_2,\tau_3}(p_2,p_3) \equiv - \frac{\partial^2}{\partial\tau_1 \partial\tau_3} G_B(\tau_1, \tau_3)
 - \frac{\partial^2}{\partial\tau_1\partial\tau_2} G_B(\tau_1, \tau_2) - \frac{\mathcal{E} p_2\cdot p_3 }{2} \Big\{ \big( \dot{G}_B(\tau_0, \tau_1) +  \dot{G}_B(\tau_1, \tau_3)  + \dot{G}_B(\tau_3, \tau_0) \big)^2
\nonumber\\
&&+ \big(  \dot{G}_B(\tau_0, \tau_1) + \dot{G}_B(\tau_1, \tau_2) + \dot{G}_B(\tau_2, \tau_0) \big)^2 -  \big( \dot{G}_B(\tau_2, \tau_1) +  \dot{G}_B(\tau_1, \tau_3) +  \dot{G}_B(\tau_3, \tau_2) \big)^2 - \dot{G}^2_B(\tau_3, \tau_0)
 -  \dot{G}^2_B(\tau_2, \tau_0)
\nonumber\\
 &&  +  \dot{G}^2_B(\tau_2, \tau_3)  
- \dot{G}_B(\tau_0, \tau_1) \dot{G}_B(\tau_3, \tau_0)
  -  \dot{G}_B(\tau_1, \tau_3) \dot{G}_B(\tau_3, \tau_0) - \dot{G}_B(\tau_0, \tau_1) \dot{G}_B(\tau_2, \tau_0) - \dot{G}_B(\tau_2, \tau_0) \dot{G}_B(\tau_1, \tau_2)
 \nonumber\\
 && +  \dot{G}_B(\tau_3, \tau_2) \dot{G}_B(\tau_2, \tau_1) +  \dot{G}_B(\tau_1, \tau_3) \dot{G}_B(\tau_3, \tau_2) \Big\}\,.
\end{eqnarray}

We integrate the terms with second derivatives over proper time variable by parts. Taking into account,
\begin{eqnarray}
&&\int^T_0 d\tau_1 \dot{G}_B(\tau_1, \tau_0) = \int^T_0 d\tau_1 \dot{G}_B(\tau_1, 0) = \int^T_0 d\tau_1 \Big[ {\rm sign} (\tau_1) - \frac{2\tau_1}{T} \Big] = 0\,,
\end{eqnarray}
reparametrizing the proper time variable $\tau = u T$, and using the following relations between the worldline propagators:
\begin{eqnarray}
 \dot{G}_B(u_1, u_2) + \dot{G}_B(u_2, u_4) + \dot{G}_B(u_4, u_1) &=& -G_F(u_1, u_2)G_F(u_2, u_4)G_F(u_4, u_1)\,,\nonumber\\
1 - \dot{G}^2_B(u_i, u_j) &=& 4 \,G_B(u_i, u_j)\,,
\label{b2dot}
\end{eqnarray}
we obtain\footnote{For convenience, we have followed convention to fix the einbein $\mathcal{E} = 2$}
\begin{eqnarray}
&&W_{\mathcal I}[\Pi A^2] = - \frac{2\Phi}{(4\pi )^{2}} \int^1_{-1}d\alpha \int^\infty_0 dT \prod_{k=0,2,3}\int \frac{d^4 p_k}{(2\pi)^4}  \int^1_0 du_2 \int^1_0 du_3 {\rm tr}_c \big(  1 - 2 T B^2 \big)
\nonumber\\
&&\times \epsilon^{\rho\alpha\sigma\beta} p_{2\alpha} A_\rho(p_2) p_{3\beta} A_\sigma(p_3) \Pi(p_0)
 \exp\big( - T (\Phi^2 \alpha^2  + B^2 ) \big) (2\pi)^4\delta^4(p_0 + p_2 + p_3)
\end{eqnarray}
where $B^2 \equiv p_2\cdot p_3 ( G_B(u_0, u_2) +  G_B(u_0, u_3) - G_B(u_2, u_3) )$. Note that two terms in this equation corresponds to two terms in Eq. (\ref{WIPi}).

Integration over period of the worldline $T$ is trivial, and gives, 
\begin{eqnarray}
&&W_{\mathcal I}[\Pi A^2] = - \frac{2\Phi}{(4\pi )^{2}} tr_c \int^1_{-1}d\alpha \prod_{k=0,2,3}\int \frac{d^4 p_k}{(2\pi)^4} \epsilon^{\rho\alpha\sigma\beta} p_{2\alpha} A_\rho(p_2) p_{3\beta} A_\sigma(p_3) \Pi(p_0)  
\nonumber\\
&&\times \int^1_0 du_2 \int^1_0 du_3 \Big(    \frac{1}{\alpha^2 \Phi^2 + B^2 } -  \frac{2 B^2}{(\alpha^2 \Phi^2 + B^2 )^2}  \Big) (2\pi)^4\delta^4(p_0 + p_2 + p_3)\,.
\end{eqnarray}

The integration over $\alpha$ can be done using the equations
\begin{eqnarray}
&&\int^1_0 d\alpha \frac{1}{\alpha^2 \Phi^2 + B^2 } 
 = \frac{1}{\Phi^2}\frac{\Phi}{B}\arctan(\frac{\Phi}{B})\,,
\end{eqnarray}
and
\begin{eqnarray}
&&\int^1_0 d\alpha \frac{ \alpha^2 \Phi^2}{(\alpha^2 \Phi^2 + B^2 )^2} = \frac{1}{2\Phi^2}\Big(-\frac{1}{1 + B^2/\Phi^2} + \frac{\Phi}{B}\arctan(\frac{\Phi}{B})\Big)\,.
\end{eqnarray}

We see that while both terms in Eq. (\ref{WIPi}) contain a singularity corresponding to $B\to 0$ their sum is finite:
\begin{eqnarray}
&&W_{\mathcal I}[\Pi A^2] = \frac{1}{4\pi^{2}}\frac{1}{\Phi} {\rm tr}_c \prod_{k=0,2,3}\int \frac{d^4 p_k}{(2\pi)^4} \epsilon^{\rho\alpha\sigma\beta} p_{2\alpha} A_\rho(p_2) p_{3\beta} A_\sigma(p_3) \Pi(p_0)
\nonumber\\
&&\times\int^1_0 du_2 \int^1_0 du_3 \frac{1}{1 + B^2/\Phi^2} (2\pi)^4\delta^4(p_0 + p_2 + p_3)\,.
\end{eqnarray}

Keeping only the leading term of expansion in powers of $\Phi$, which is dominant in the limit $\Phi\to\infty$, we get
\begin{eqnarray}
&&W_{\mathcal I}[\Pi A^2] = \frac{1}{4\pi^{2}} \frac{1}{\Phi} {\rm tr}_c \prod_{k=0,2,3}\int \frac{d^4 p_k}{(2\pi)^4} \epsilon^{\rho\alpha\sigma\beta} p_{2\alpha} A_\rho(p_2) p_{3\beta} A_\sigma(p_3) \Pi(p_0) (2\pi)^4\delta^4(p_0 + p_2 + p_3)\,.
\end{eqnarray}

Substituting the Fourier transformation of the background field we can rewrite this equation as
\begin{eqnarray}
&&W_{\mathcal I}[\Pi A^2] = -\frac{1}{8\pi^{2}} \frac{1}{\Phi} \,{\rm tr}_c \int d^4x \,\Pi(x)\, F_{\mu\nu}(x) \tilde{F}^{\mu\nu}(x) .
\label{resappA}
\end{eqnarray}
where, as previously for the triangle, we replaced the derivatives of the gluon fields by  field strength tensors.

\section{Worldline derivation of the CGC effective action}
\label{sec:CGC}

The expectation value of an arbitrary operator $\mathcal{O}$
in the proton can be represented as the trace 
\begin{eqnarray}
\langle \mathcal{O} \rangle = {\rm Tr}\{\mathcal{O} \mathcal{R}\}\,,
\label{expvalue}
\end{eqnarray}
of the operator convoluted with the density matrix  $\mathcal{R}$ of the target hadron, with the Schwinger-Keldysh path integral~\cite{Schwinger:1960qe,Keldysh:1964ud} defined as 
\begin{eqnarray}
{\cal Z} = {\rm Tr}\left({\cal R}\right) = {\rm Tr}\left({\cal U}_{0,-\infty}\,{\cal R}_{\rm init.}\, {\cal U}_{-\infty,0}\right) = \int dA_1 dA_2
\int d \Psi_1d \Psi_2 \langle A_1, \Psi_1 | {\cal R}_{\rm{init}} | A_2, \Psi_2\rangle\int\limits_{A_1}^{A_2} \mathcal{D}A \int\limits_{\Psi_1}^{\Psi_2} \mathcal{D} \Psi \mathcal{D} \bar{\Psi} \,e^{i S_\mathcal{C}}\,.
\label{eq:SK}
\end{eqnarray}
The density matrix is obtained from the  real-time evolution from $t= -\infty$ of the initial density matrix $\mathcal{R}_{\rm init.}$ on the Schwinger-Keldysh double time contour ${\mathcal C}$ with ${\cal U}_{t,t^\prime} = \exp(-i H(t-t^\prime))$, where $H$ is the QCD Hamiltonian. This can be expressed as the path integral on the r.h.s, where $S_{\mathcal{C}}$ is the QCD action on the double time contour, and $A_1,\Psi_1$ ($A_2$,$\Psi_2$), are respectively the gauge and fermion fields defined on the upper (lower) part of the contour. 

Note that 
$\langle A_1,\Psi_1 | {\cal R}_{\rm init.} | A_2, \Psi_2\rangle$ denotes matrix elements of the initial density matrix of non-interacting quarks and gluons at $t\rightarrow -\infty$,  with ${\cal R}_{\rm init.} = {\cal R}_{\rm YM} \otimes {\cal R}_{\rm valence}$, where ${\cal R}_{\rm  YM} = | 0 \rangle \langle 0|$ is the Yang-Mills vacuum and ${\cal R}_{\rm valence}$ is the three valence quark state with the proton's quantum numbers. We can rewrite Eq.~(\ref{eq:SK}) as 
\begin{align}
Z=  \int dA_1dA_2 \langle A_1 | \hat{\rho}_\text{YM} | A_2\rangle  \int\limits_{A_1}^{A_2}  \mathcal{D}A  \, Z_f[A] \, \exp{ \{ i S_\mathcal{C}^{\text{YM}} \}}\,,
\end{align}
where 
\begin{align}\label{eq:quarkpartQCD}
Z_f[A]\equiv \int  & d \Psi_1d \Psi_2 \langle \Psi_1 | {\cal R}_{\rm valence}|  \Psi_2\rangle
\int\limits_{\Psi_1}^{\Psi_2} \mathcal{D} \Psi \mathcal{D} \bar{\Psi} \,\exp {  \big\{ i S_{\mathcal{C}}^q \big\} } \,,
\end{align}
with the Dirac action $ S_{\mathcal{C}}^q\equiv \int d^4z_{\mathcal{C}}   \bar{\Psi}(i  \slashed{D}[A]-m) \Psi $ and the Yang-Mills action 
$S^{\text{YM}}_{\mathcal{C}} \equiv -\frac{1}{2}\int d^4z_\mathcal{C} \, \text{tr}_c F^{\mu\nu}F_{\mu\nu}$.

We will focus here on the fermionic path integral $Z_f [A]$ in the gauge field background. It was shown explicitly in Appendix A of \cite{Mueller:2019qqj} how this expression can be mapped to on to an initial value problem describing the evolution of the density matrix, expressed in terms of the worldline bosonic and Grassmannian variables, on the the corresponding Schwinger-Keldysh contour. We will here, and in the following sub-sections, not repeat the derivation in \cite{Mueller:2019qqj} (see also \cite{Mueller:2019gjj}) but only employ salient features relevant to our discussion.  

We consider first the density matrix of a single valence quark before generalizing to that for the hadron. 
At asymptotic negative infinity, its initial density matrix  can be expressed as the direct product of the density matrices corresponding to the quark worldline's color, its (bosonic) coordinate/momentum and its (fermion) spin degrees of freedom:
\begin{eqnarray}
\mathcal{R}_{\rm q\,init.} = \mathcal{R}_{\rm q\, init.}^c\otimes \mathcal{R}_{\rm q\,init.}^b \otimes \mathcal{R}_{\rm q\, init.}^s\,.
\label{dens-quark}
\end{eqnarray}

At high energies, it is appropriate to consider the motion of the quark along the light-cone direction, which has the initial coordinate space representation, 
\begin{eqnarray}
\mathcal{R}_{\rm q~ init.}^b = \int d^4z_{\rm max} \int d^4z_{\rm min}~ e^{iP^+(z^-_{\rm max} - z^-_{\rm min})} e^{-iP_\perp (z_{\perp\rm max} - z_{\perp\rm min})}~ |z_{\rm min})(z_{\rm max}| \,.
\end{eqnarray}
Here $P^{+},P_\perp$ are lightcone momenta of the quark  and we have neglected $P^-\sim 1/P^+$ assuming $P^+/M_p \gg 1$ ($M_p$ being the proton mass), for problems of interest. The coordinates $z_{\rm max}$ and $z_{\rm min}$, as we will soon see, are the boundaries of the quark's worldline. Further, the spin density matrix can be expressed as~\cite{Berezin:1976eg,Mueller:2019gjj,Tarasov:2019rfp}  
\begin{equation}
    {\cal R}_{\rm init.}^s = \frac{1}{ 4 } \Big(1+\lambda \,\psi_5\Big) \Big(1 + 2\psi^+\psi^-\Big)\,.
\end{equation}

In terms of this initial density matrix, the worldline path integral representation of the density matrix $\mathcal{R}_q[A]$ can therefore be expressed as~\cite{Mueller:2019qqj},
\begin{eqnarray}
&&\mathcal{R}_q[A] = {\rm Tr}_c\,\mathcal{R}^c_{\rm q~ init.}\,\otimes\, \int d^4 z_{\rm max} ~ \int d^4 z_{\rm min}~ e^{iP^+(z^-_{\rm max} - z^-_{\rm min})}~ e^{-iP_\perp (z_{\perp\rm max} - z_{\perp\rm min})}\, 
\nonumber\\
&&\times \int^\infty_0 dT \int^{z_{\rm max}}_{z_{\rm min}} \mathcal{D}z \int\mathcal{D}\psi\,\, 
\frac{1}{ 4 } \Big(1+\lambda \,\psi_5\Big) \Big(1 + 2\psi^+\psi^-\Big)
\exp\Big[-\int^T_0 d\tau \Big(\frac{1}{4}\dot{z}^2 + \frac{1}{2}\psi\dot{\psi} + igA\dot{z} - ig\psi F\psi\Big)\Big]\,,
\label{rhoinit}
\end{eqnarray}
where the color trace is over that of the initial quark density matrix convoluted with the color matrix of the exponent in the fundamental representation. 

\begin{figure}[htb]
\begin{center}
\includegraphics[width=150mm]{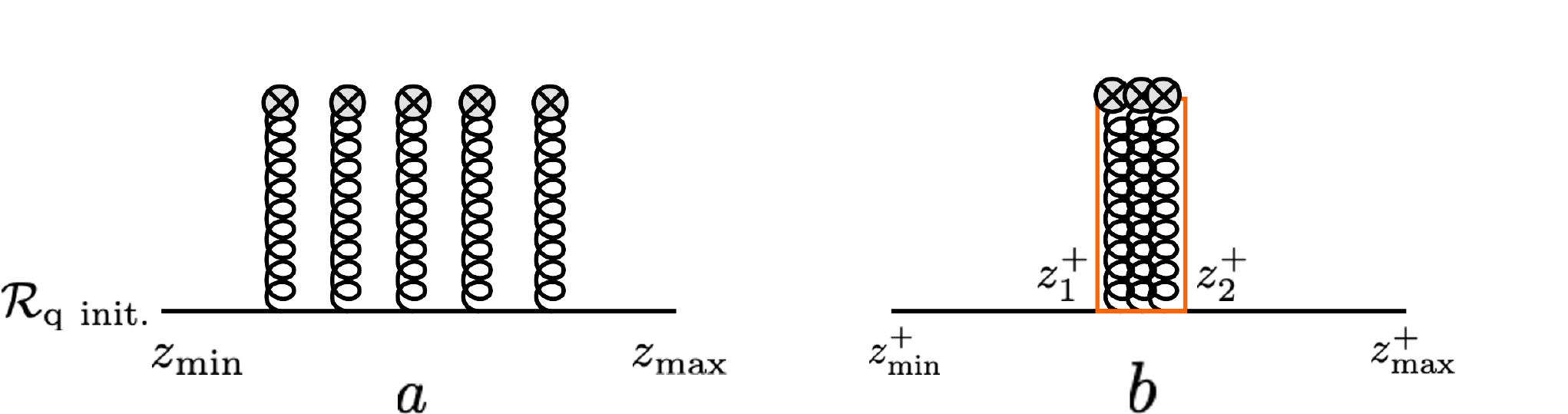}
\end{center}
\caption{a) Evolution of the initial density matrix $\rho_{\rm init}$ in the worldline representation. b) The shock-wave approximation for the small-x emission.
\label{fig:1}}
\end{figure}

Note that $A_\mu$ here is the small $x$ gauge field (and $F^{\mu\nu}$ the corresponding field strength) that dresses the quark world-line in the course of its real-time evolution. This is illustrated in Fig. \ref{fig:1}a. For a hadron with $P^+\rightarrow \infty$, the interaction of this gauge field with the projectile is nearly instantaneous as  illustrated in Fig. \ref{fig:1}b, with  $z^-_1$ and $z^-_2$ are initial and final points dilineating the width $\sim 1/P^+$. We will now discuss below this shockwave structure of gauge fields in the Regge limit.

As a warmup to the polarized case of interest, we will first  use the spin averaged worldline quark density matrix to   motivate the construction of the small $x$ effective action for unpolarized hadrons\footnote{This EFT was also derived previously from Wong's equations for quarks~\cite{JalilianMarian:2000ad},  obtained as the saddle point approximation to the one-loop worldline QCD  effective action--see \cite{Mueller:2019gjj}
and references therein.}.
The interaction of small $x$ gauge fields with the  quark worldline in the Regge limit is characterized by a strict ordering of the longitudinal momentum components on the lightcone. In general, if the large $x$ source is characterized by the momentum scale $P = (P^+, P^-, P_\perp)$ and the small $x$ emission by the scale $k = (k^+, k^-, k_\perp)$, the ordering between the longitudinal components is
\begin{eqnarray}
P^+ \gg k^+,\ \ \  P^- \ll k^-\,.
\label{orderingsx}
\end{eqnarray}
\begin{figure}[htb]
\begin{center}
\includegraphics[width=150mm]{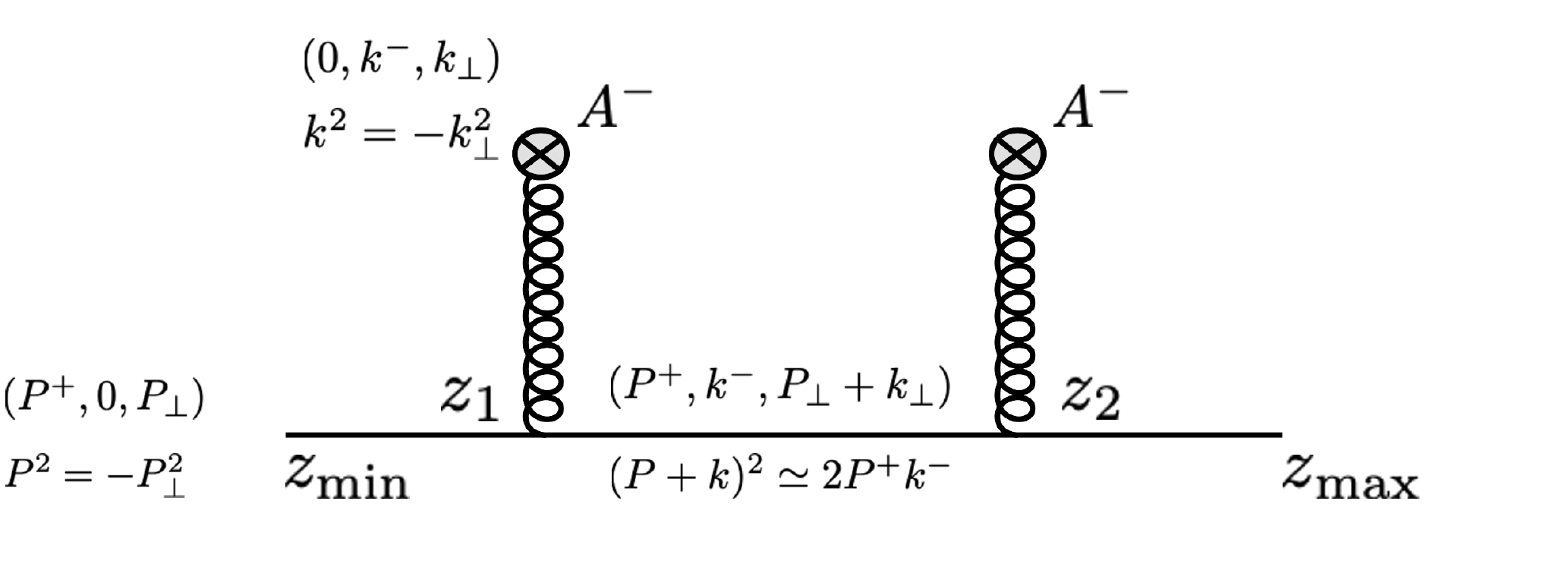}
\end{center}
\caption{The small-x emission of the target in the second order of the perturbative expansion.\label{fig:2}}
\end{figure}
The Feynman $x$ for the emitted gluon field is $x = k^+/P^+ \to 0$; hence, $k^2 \simeq - k^2_\perp$.
Since $P^+$ and $k^-$ are large, the virtuality of the quark in the course of emission is $(P+k)^2 \simeq 2P^+ k^-$.

For a target propagating very close to the lightcone, the  $A^-$ component dominates in the emission of the accompanying small $x$ gauge fields when we choose the axial $A^+ = 0$ gauge. The interaction term $A_\mu \dot{z}^\mu$ in the worldline action in Eq. (\ref{rhoinit}), can therefore be simplified to
\begin{eqnarray}
A_\mu \dot{z}^\mu \simeq A^- \dot{z}^+\,,
\label{eikonalapprox1}
\end{eqnarray}
since in the eikonal approximation $\dot{z}_\perp \sim 0$ and $A_\perp$ is also suppressed relative to $A^-$. Likewise, the spin precession term  $\psi^\nu \psi^\nu F_{\mu\nu}$ in the worldline action simplifies to 
\begin{eqnarray}
\psi^\mu \psi^\nu F_{\mu\nu} \simeq 2\psi^+ \psi_i F^{-i}\,.
\label{eikonalapprox2}
\end{eqnarray}

Consequently, we can rewrite the worldline evolution of the initial density matrix in Eq.~(\ref{rhoinit}) as
 \begin{eqnarray}
&&\mathcal{R}_q[A] = {\rm tr_c} ~\mathcal{R}^c_{\rm q~init.}  \int d^4 z_{\rm max} ~ \int d^4 z_{\rm min}~ e^{iP^+(z^-_{\rm max} - z^-_{\rm min})}~ e^{-iP_\perp (z_{\perp\rm max} - z_{\perp\rm min})}\int^\infty_0 dT \int^{z_{\rm max}}_{z_{\rm min}} \mathcal{D}z
\nonumber\\
&&\times  \int\mathcal{D}\psi~  \frac{1}{ 4 } \Big(1+\lambda \,\psi_5\Big) \Big(1 + 2\psi^+\psi^-\Big) ~
\exp\Big[-\int^T_0 d\tau \Big(\frac{1}{4}\dot{z}^2 + \frac{1}{2}\psi\dot{\psi} + ig\dot{z}^+ A^- - 2ig\psi^+ \psi_i F^{-i}\Big)\Big]
\label{rhoinit2}
\end{eqnarray}
One can further simplify Eq. (\ref{rhoinit2}) by taking into account the fact that computing the functional integral over Grassmann variables will yield zero for the interaction term $\psi^+ \psi_i F^{-i}$. In the leading eikonal approximation, for spin averaged observables, we obtain
\begin{eqnarray}
&&\mathcal{R}_q[A] = {\rm tr_c} ~\mathcal{R}^c_{\rm q~init.}  \int d^4 z_{\rm max} ~ \int d^4 z_{\rm min}~ e^{iP^+(z^-_{\rm max} - z^-_{\rm min})} e^{-iP_\perp (z_{\perp\rm max} - z_{\perp\rm min})}
\nonumber\\
&&\times \int^\infty_0 dT \int^{z_{\rm max}}_{z_{\rm min}} \mathcal{D}z \int\mathcal{D}\psi~ \
\exp\Big[-\int^T_0 d\tau \Big(\frac{1}{4}\dot{z}^2 + \frac{1}{2}\psi\dot{\psi} + igA^- \dot{z}^+ \Big)\Big]\,.
\label{rhoinitlo}
\end{eqnarray}
Thus in this approximation, one can deduce from the exponential factor in Eq. (\ref{rhoinitlo}) that the interaction of the quark with the background field, without the spin-dependent terms, is equivalent to that of a scalar particle. Spin effects will come in starting at sub-eikonal order from the transverse field strengths $F^{ij}$. However as we will soon discuss in Appendix~\ref{sec:eikonal}, for spin observables sensitive to the anomaly, such an eikonal approximation breaks down.

Returning to the unpolarized case, in Fig. \ref{fig:2},  since $k^+ \ll P^+$, and 
\begin{eqnarray}
(z^-_{\rm max} - z^-_{\rm min}) \sim (z^-_2 - z^-_1) \sim \frac{1}{P^+} \to 0\,,
\label{kinapp}
\end{eqnarray}
 we can neglect the dependence of the background gauge field on $z^-$:  $A_\mu \equiv A_\mu(z^+, z_\perp)$. Further, since 
\begin{eqnarray}
(z^+_{\rm max} - z^+_{\rm min}) \sim \frac{1}{P^-} \to \infty\,,
\label{kinapp2}
\end{eqnarray}
and taking into account the momentum ordering in Eq.~(\ref{orderingsx}), 
\begin{eqnarray}
(z^+_2 - z^+_1) \sim \frac{1}{k^-} \ll (z^+_{\rm max} - z^+_{\rm min}) \,,
\label{kinapp3}
\end{eqnarray}
where $z^+_1$ and $z^+_2$ are initial and final points of the emission of the gauge field, as illustrated in Fig. \ref{fig:1}b,  we can simplify the phase factor in Eq. (\ref{rhoinitlo}) to 
\begin{eqnarray}
&&\exp\Big[-ig \int^T_0 d\tau A^-\dot{z}^+ \Big] = \exp\Big[-ig \int^{z^+_2}_{z^+_1} dz^+ A^-(z^+, z_\perp) \Big] \simeq \exp\Big[-ig \int^\infty_{-\infty} dx^+ A^-(x^+, z_\perp) \Big] \,.
\label{approxsw}
\end{eqnarray}
This is the lightlike ``Wilson line" representing the color rotation of the quark in the background field.

Substituting Eq. (\ref{approxsw}) into Eq. (\ref{rhoinitlo}), splitting the functional integral over coordinates into two segments corresponding to before and after the instantaneous emission of the gauge field at  point $z$, and after some algebra\footnote{We employed here the identity $\int_0^T d\tau\, {\dot x}^+ \,{\rm sign}({\dot x}^+)\delta(x^+)=1$.}, we obtain, 
\begin{eqnarray}
&&\mathcal{R}_q[A] = {\rm tr_c} ~ \mathcal{R}^c_{\rm q~init.} ~ \int d^4 z_{\rm max} ~ \int d^4 z_{\rm min}~ e^{iP^+(z^-_{\rm max} - z^-_{\rm min})}~ e^{-iP_\perp (z_{\perp\rm max} - z_{\perp\rm min})} \int dz^- \int d^2 z_{\perp}
\label{rhoinitlo2}\\
&&\times ~ \int^\infty_0 dT_2 \int^{z_{\rm max}}_{z} \mathcal{D}z  ~ \dot{z}^+
 \exp\Big[-\int^{T_2}_0 d\tau \frac{1}{4}\dot{z}^2 \Big] \exp\Big[-ig \int^\infty_{-\infty} dx^+ A^-(x^+, z_\perp) \Big] \int^\infty_0 dT_1 \int^{z}_{z_{\rm min}} \mathcal{D}z ~ \exp\Big[-\int^{T_1}_0 d\tau \frac{1}{4}\dot{z}^2 \Big]
\nonumber
\end{eqnarray}
The structure of Eq. (\ref{rhoinitlo2}) is clear: the worldline propagates without the emission from $z_{\rm min}$ to $z$,  emits the small $x$ background field at a point $z$  and subsequently freely propagates from point $z$ to $z_{\rm max}$.

The functional integrals in Eq. (\ref{rhoinitlo2}) can be performed straightforwardly using the techniques discussed at length in \cite{Tarasov:2019rfp}. 
We obtain\footnote{We used here the Schwinger notation, namely, $|x)$ is an eigenvector for the coordinate operator $\hat{x}^\mu$: $\hat{x}^\mu|x) = x^\mu|x)$. The corresponding canonical conjugate momentum operator is $\hat{p}^\mu$: $\hat{p}^\mu |p) = p^\mu |p)$, and the commutator $[\hat{p}^\mu, \hat{x}^\nu] = -i g^{\mu\nu}$, giving 
 \begin{eqnarray}
&&(x|y) = \delta^4(x-y);\ \ \ \ (p|q) = (2\pi)^4\delta(p-q);\ \ \ \ \int d^4x |x)(x| = 1\ \ \ \ \int\frac{d^4p}{(2\pi)^4} |p)(p| = 1;\ \ \ \ (x|p) = e^{-ipx}
\label{Shwnot}
 \end{eqnarray}
With Eq. (\ref{Shwnot}), the last relation in Eq. (\ref{int1}) can be obtained easily. Note that for brevity we drop the hats in the notation of operators.
We further defined in Eq.~(\ref{rhoinitlo3})
\begin{eqnarray}
(x_{2\perp}|f(p_\perp)|x_{1\perp}) \equiv \int \frac{d^2p_\perp}{(2\pi)^2} e^{ip_\perp (x_{2\perp} - x_{1\perp})}f(p_\perp)\,,
\label{fourier1}
\end{eqnarray}
where $f(p_\perp)$ is an arbitrary function of the operator $p_\perp$.
},
\begin{eqnarray}
\int^\infty_0 dT \int^{z_1}_{z_2} \mathcal{D}z ~ \exp\Big[-\int^T_0 d\tau \frac{1}{4}\dot{z}^2 \Big]  = \int \frac{d^4p}{(2\pi)^4} e^{-ip(z_1 - z_2)} \frac{1}{p^2} = (z_1| \frac{1}{p^2} |z_2)\,,
\label{int1}
\end{eqnarray}
Similarly, 
 \begin{eqnarray}
 \int^\infty_0 dT \int^{z_1}_{z_2} \mathcal{D}z  ~ \dot{z}^+_2
 \exp\Big[-\int^{T}_0 d\tau \frac{1}{4}\dot{z}^2 \Big] = -i (z_1|\frac{2p^+}{p^2}|z_2)
 \label{int2}
\end{eqnarray}

Substituting Eqs. (\ref{int1}) and (\ref{int2}) into Eq. (\ref{rhoinitlo2}), integrating over longitudinal coordinates, and analytically continuing into the Minkowski space, we get
\begin{eqnarray}
&&\mathcal{R}_q[A] = {\rm tr_c} \int d^2 z_{\perp \rm max} \int d^2 z_{\perp \rm min} \int d^2z_\perp ~ e^{-iP_\perp (z_{\perp\rm max} - z_{\perp\rm min})}
\nonumber\\
&&\times~ \mathcal{R}^c_{\rm q~init.} ~  (z_{\perp\rm max} | \frac{1}{ p^2_{\perp} } |z_\perp) \exp\Big[-ig \int^\infty_{-\infty} dx^+ A^-(x^+, z_\perp) \Big] (z_\perp |\frac{1}{ p^2_{\perp} } | z_{\perp \rm min} )~ 2P^+ \int dx^- \,,
\label{rhoinitlo3}
\end{eqnarray}
where the integral $\int dx^-$ represents momentum conservation of $P^+$. 

From Eq. (\ref{rhoinitlo3}), we see that in the high energy limit, the only effect of the background field is to color rotate the initial density matrix by an infinite lightlike Wilson line.  
We can rewrite it more compactly as 
\begin{eqnarray}
&&\mathcal{R}_q[A] = \int d^2z_\perp~ {\rm tr_c}\Big[ \rho_q(z_\perp)~ \exp\Big[-ig \int^\infty_{-\infty} dx^+ A^-(x^+, z_\perp) \Big] \Big]\,,
\label{rhoinitlo4}
\end{eqnarray}
where the quark color matrix $\rho_q(z_\perp) = \rho^a_q(z_\perp) t^a$ absorbs all the terms in  Eq. (\ref{rhoinitlo3}) with the exception of the Wilson line. 

The above is the leading eikonal expression for the quark density matrix. There is of course a non-perturbative distribution of the distribution of ``valence partons" in the hadron carrying large fractions of its momentum. For a large nucleus and/or for large parton occupancies in the proton, they can be represented by a classical distribution of color charges, and the trace replaced by a path integral over a weighted distribution of their color charge densities~\cite{McLerran:1993ka,McLerran:1993ni,Jeon:2004rk}. Hence in the Regge limit, Eq.~(\ref{expvalue}) can be expressed as, 
\begin{eqnarray}
\label{eq:CGC-EFT-unpolarized}
\langle \mathcal{O} \rangle = \int \mathcal{D}\rho ~W[\rho] ~\int \mathcal{D}A~\mathcal{O}[A]~e^{iS[A, \rho]}\,.
\end{eqnarray}
Here the first path integral representing the initial density matrix denotes a static statistical distribution of classical color charges with the non-perturbative weight $W[\rho]$, whose kinematics is separated in lightcone momenta from that of the dynamical fields, represented by 
the second path integral over the small $x$ gauge fields with the effective action,
\begin{eqnarray}
S[A, \rho] = -\frac{1}{4} \int d^4x F^{\mu\nu}_a F^a_{\mu\nu} + \frac{i}{N_c} \int d^2x_\perp ~ {\rm tr}_c \big[\rho(x_\perp)  \ln\big(U_{[\infty, -\infty]}(x_\perp)\big)\big]\,.
\label{MVmodel}
\end{eqnarray}
The second term in the effective action represents the eikonal coupling of large $x$ and small $x$ modes, with the ``exponential of  $\ln\big(U_{[\infty, -\infty]}(x_\perp)$" corresponding to the phase in Eq.~(\ref{rhoinitlo4}); this particular representation of the interaction term was first discussed at length in \cite{JalilianMarian:2000ad} and its interpretation further discussed in \cite{Caron-Huot:2013fea}. 

For a large nucleus, $W[\rho]$ is given by a Gaussian (MV) distribution  of classical color charges~\cite{McLerran:1993ka,McLerran:1993ni,Jeon:2004rk}, whose dimensionful variance introduces the saturation scale $Q_S$; for a large nucleus; in the Regge limit, this scale is much larger than intrinsically non-perturbative scales in the nucleus, justifying the application of the systematic weak coupling techniques of the CGC EFT~\cite{Gelis:2010nm,Kovchegov:2012mbw}.
Our discussion  in this Appendix therefore corresponds to a  rederivation of the small $x$ CGC EFT for an unpolarized hadron in the worldline formalism.

\section{Breakdown of the eikonal expansion for $g_1(x_B,Q^2)$}
\label{sec:eikonal}

We will discuss here why the high energy eikonal expansion of operators 
(in powers of $1/P^+$), a powerful tool in high energy scattering which enormously simplified the discussion in Appendix~\ref{sec:CGC}, breaks down completely for quantities that are sensitive to the chiral anomaly. A clear case in point is that of $g_1(x_B,Q^2)$. Simply put, this breakdown\footnote{While our focus here is on eikonal approximations, our discussion also applies to other kinematic approximations such as collinear kinematics often employed in perturbative QCD.} occurs  because the anomaly couples to zero modes of the Dirac operator that must be treated in exact kinematics. 

The importance of the careful treatment of kinematics to uncover the anomaly is of course well-known\footnote{See also the nice pedagogical reviews \cite{Bilal:2008qx,Vasquez-Mozo,AlvarezGaume:2005qb,Adler:2004qt}.} from the seminal work of Adler, Bell, Jackiw and Bardeen~\cite{Adler:1969gk,Adler:1969er,Bell:1969ts,Bardeen:1974ry} but to the best of our knowledge has not been adddressed in this high energy context. To illustrate its importance, we will revisit our derivation in Paper I of the antisymmetric part of the box diagram and point out where the sub-eikonal terms discussed in the literature appear and why they are insufficient to reproduce the anomaly.

The antisymmetric component of the DIS polarization tensor can, to one loop accuracy, be expressed as~\cite{Tarasov:2020cwl} 
\begin{eqnarray}
{\tilde{\Gamma}}^{\mu\nu}_A[k_1, k_3] &=& \frac{e^2 e_f^2 }{2}\int^\infty_0\frac{dT}{T} ~{\rm Tr_c} \int \mathcal{D}x\int \mathcal{D}\psi 
\Big[ V^\mu_1(k_1)V^\nu_3(k_3) - (\mu\leftrightarrow\nu)\Big] \label{photamp}\\
&\times& \exp\Big\{-\int^T_0 d\tau \Big(\frac{1}{4} \dot{x}^2 + \frac{1}{2}\psi_\mu\dot{\psi}^\mu + ig\dot{x}^\mu A_\mu - ig \psi^\mu \psi^\nu F_{\mu\nu}\Big)\Big\}\,,
\nonumber
\end{eqnarray}
where the vertex 
\begin{eqnarray}
V^\mu_i(k_i) \equiv \int^T_0 d\tau_i  (\dot{x}^\mu_i + 2i \psi^\mu_i k_j\cdot\psi_j ) e^{ik_i\cdot x_i}\,,
\end{eqnarray}
corresponds to the interaction of a  worldline  with the external electromagnetic current, and $x_i\equiv x(\tau_i)$, $\psi_i\equiv \psi(\tau_i)$.

\begin{figure}[htb]
 \begin{center}
 \includegraphics[width=90mm]{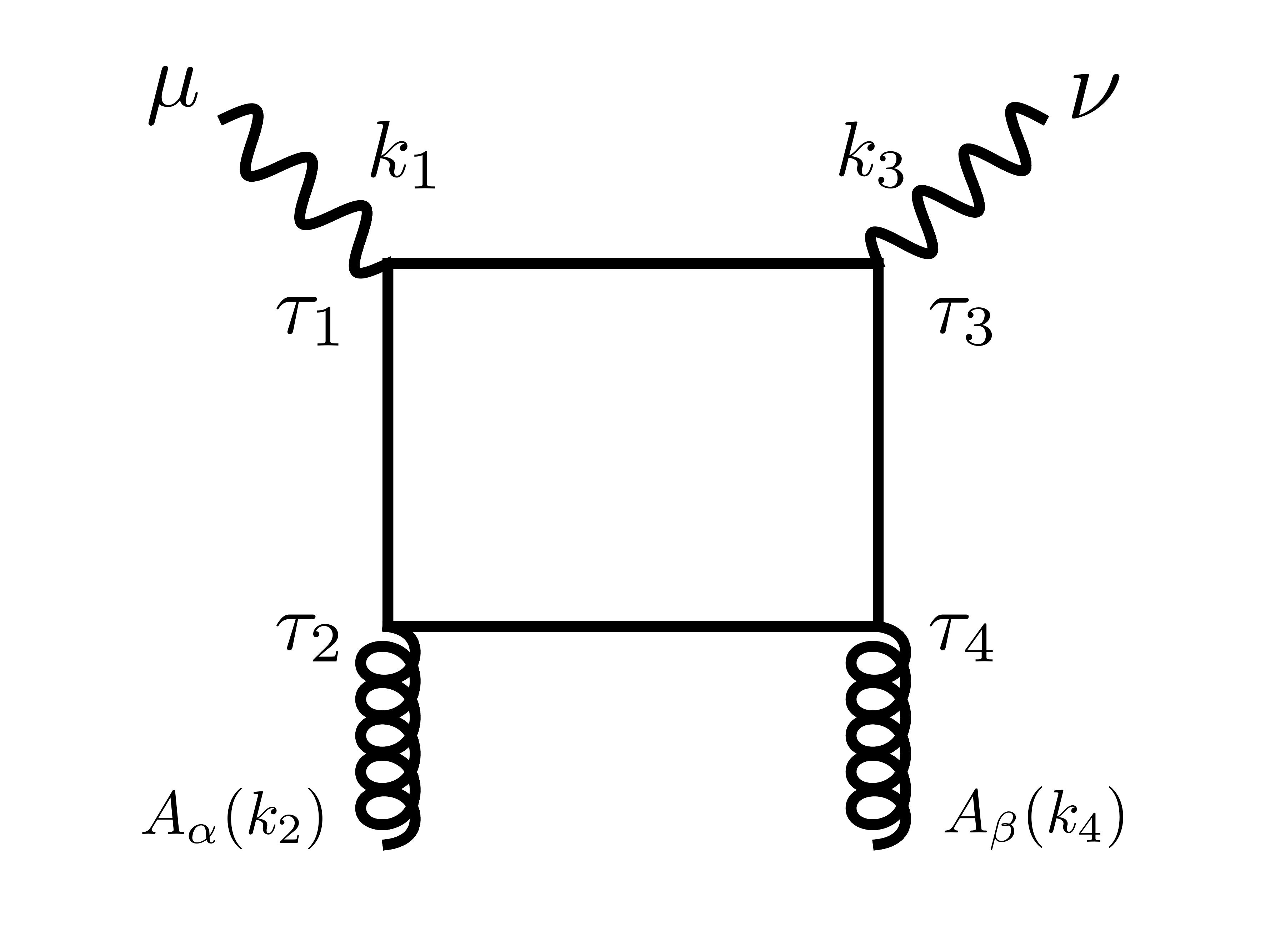}
 \end{center}
 \caption{\label{fig:boxgraph}The box diagram $\Gamma^{\mu\nu\alpha\beta}_A[k_1, k_3, k_2, k_4]$ for polarized DIS.}
 \end{figure}

To recover the general expression for the box diagram in Fig.~\ref{fig:boxgraph}, one expands the third (Wilson line) and fourth (spin precession) terms of the exponential in the above expression to second order in the gauge fields: 
\begin{eqnarray}
&&{\tilde{\Gamma}}^{\mu\nu}_A[k_1, k_3] = (-ig)^2 \frac{e^2 e_f^2 }{2}\int^\infty_0\frac{dT}{T} ~{\rm Tr_c} \int \mathcal{D}x\int \mathcal{D}\psi~\exp\Big\{-\int^T_0 d\tau \Big(\frac{1}{4} \dot{x}^2 + \frac{1}{2}\psi \cdot\dot{\psi} \Big)\Big\}
\\
&&\times~\Big[ V^\mu_1(k_1)V^\nu_3(k_3) \int^T_0 d\tau_2 \Big( \dot{x}^\alpha_2 A_\alpha(x_2) +  2 \psi^\alpha_2 \psi^\lambda_2 \partial_\lambda A_{\alpha}(x_2)\Big) \int^T_0 d\tau_4 \Big( \dot{x}^\beta_4 A_\beta(x_4) + 2 \psi^\beta_4 \psi^\eta_4 \partial_\eta A_{\beta}(x_4) \Big) - (\mu\leftrightarrow\nu)\Big]\,.
\nonumber
\label{eq:box-quadratic}
\end{eqnarray}
Further expressing the gauge fields in terms of their Fourier transforms,  
\begin{eqnarray}
A_\alpha(x_2) = \int \frac{d^4k_2}{(2\pi)^4} e^{ik_2\cdot x_2} {\tilde A}_\alpha(k_2);\ \ \ \ \ A_\beta(x_4) = \int \frac{d^4k_4}{(2\pi)^4} e^{ik_4\cdot x_4} {\tilde A}_\beta(k_4)\,,
\label{fourbackground}
\end{eqnarray}
one obtains,
\begin{eqnarray}
&&\Gamma^{\mu\nu}_A[k_1, k_3] = \int \frac{d^4k_2}{(2\pi)^4} \int \frac{d^4k_4}{(2\pi)^4}~
\Gamma^{\mu\nu\alpha\beta}_A[k_1, k_3, k_2, k_4]~ {\rm Tr_c}({\tilde A}_\alpha(k_2) {\tilde A}_\beta(k_4))\,,
\end{eqnarray}
where the expression for the box diagram is 
\begin{eqnarray}
&&\Gamma^{\mu\nu\alpha\beta}_A[k_1, k_3, k_2, k_4] \equiv -\frac{g^2 e^2 e_f^2 }{2}\int^\infty_0\frac{dT}{T} ~ \int \mathcal{D}x\int \mathcal{D}\psi~\exp\Big\{-\int^T_0 d\tau \Big(\frac{1}{4} \dot{x}^2 + \frac{1}{2}\psi\cdot\dot{\psi} \Big)\Big\}
\nonumber\\
&&\times \Big[V^\mu_1(k_1)V^\nu_3(k_3)V^\alpha_2(k_2) V^\beta_4(k_4) - (\mu\leftrightarrow\nu)\Big]\,.
\label{eq:box1}
\end{eqnarray}

In this derivation, we have made no approximation with regard to keeping leading or subleading components of the gauge field or field strength tensor but expanded all terms to quadratic order in the gauge fields\footnote{We have not specified a choice of gauge either; in computations of the anomaly, one often employs the Fock-Schwinger gauge $x_\mu A^\mu=0$~\cite{AlvarezGaume:1983ig}. With this gauge choice, writing $x= {\bar x} + x^\prime$, where ${\bar x},x^\prime$ denote zero and non-zero modes respectively, one can write $A_\mu(x)\approx \frac{1}{2} {x^\prime}^\nu\, F_{\mu\nu}({\bar x}) $~\cite{Mueller:2017arw}.}. For our discussion, this expansion to quadratic order is fully sufficient. Further expansion to higher orders will not change our results and conclusions - we will address this point at the end of our discussion. 

One can further rewrite Eq.~(\ref{eq:box1}) as 
\begin{eqnarray}
&&\Gamma^{\mu\nu\alpha\beta}_A[k_1, k_3, k_2, k_4] = -\frac{g^2e^2 e_f^2 }{2}\int^\infty_0\frac{dT}{T} ~ \int \mathcal{D}x\int \mathcal{D}\psi~\exp\Big\{-\int^T_0 d\tau \Big(\frac{1}{4} \dot{x}^2 + \frac{1}{2}\psi\cdot \dot{\psi} \Big)\Big\}
\nonumber\\
&&\times \prod^4_{k=1}\int^T_0 d\tau_k~ \Big[\sum^9_{n=1}\mathcal{C}^{\mu\nu\alpha\beta}_{n;(\tau_1,\tau_2,\tau_3,\tau_4)}[k_1, k_3, k_2, k_4] - (\mu\leftrightarrow\nu)\Big]e^{i\sum^4_{i=1}k_i x_i} \,.
\end{eqnarray}
where the coefficients $\mathcal{C}^{\mu\nu\alpha\beta}_{n;(\tau_1,\tau_2,\tau_3,\tau_4)}[k_1, k_3, k_2, k_4]$ are functions of  coordinate ($x_i \equiv x(\tau_i)$) and Grassmann variables ($\psi_i\equiv \psi(\tau_i)$), which depend on the proper time coordinates $\tau_i$ of the interaction of the worldlines with the external virtual photon and gluon fields. 

These coefficients were worked out in full generality in Appendix A of \cite{Tarasov:2020cwl}. We showed that the Bjorken limit ($Q^2\rightarrow \infty$ and 
$x_B={\rm fixed}$) and the Regge limit ($x_B\rightarrow 0$, $Q^2= {\rm fixed}$) corresponded to taking $\tau_1\rightarrow \tau_3$ and $\tau_2\rightarrow \tau_4$ respectively in the sum of all these contributions. In Appendix C of \cite{Tarasov:2020cwl}, we showed explicitly for the Bjorken limit that this gave,
\begin{eqnarray}
\Gamma^{\mu\nu\alpha\beta}_A[k_1, k_3, k_2, k_4]\Big|_{Q^2\to \infty}
&=& - 2\frac{g^2e^2 e_f^2 }{\pi^2}\epsilon^{\mu\nu\eta}_{\ \ \ \ \kappa} (k_{1\eta} - k_{3\eta}) (2\pi)^4\delta^{(4)}(\sum^4_{i=1} k_i)
 ~\frac{( k^\kappa_{2} + k^\kappa_{4} ) \epsilon^{\alpha\beta\sigma\lambda}  k_{2\sigma} k_{4\lambda} }{(k_{2} + k_4)^2  }  
 \nonumber\\
&\times&  \frac{1}{2 k_1 \cdot k_2}\Big[ \big( 1  + \frac{k^2_{1}}{2 k_1 \cdot k_2} \big)\ln\Big[ \frac{2 k_1 \cdot k_2 + k^2_{1}}{k^2_{1}}\Big]  - 1 \Big]\,.
\label{Gammaaftb}
\end{eqnarray}
We further showed that, up to overall kinematic factors, this result for the box diagram agreed with the Adler-Bell-Jackiw (ABJ) result~\cite{Adler:1969gk,Bell:1969ts} for the triangle graph of the chiral anomaly. In particular, $\frac{( k^\kappa_{2} + k^\kappa_{4} )}{(k_{2} + k_4)^2  }\equiv \frac{l^\kappa}{l^2}$, the infrared pole of the anomaly. 

We  emphasize that to recover the anomaly in this derivation it was essential that we not make any approximations to the kinematics of the internal loop variables besides those imposed by the asymptotic kinematics of the external probe ($Q^2$, $x_B$). Performing kinematic approximations of the internal variables (collinear/eikonal) that do not respect the anomalous functional chiral Ward identity will miss the fundamental physics of the anomaly with its accompanying infrared pole. 

Though this observation is general, we can illustrate it further in the context of the small $x$ eikonal expansion we discussed previously in Appendix~\ref{sec:CGC}. In the strict eikonal limit, only the $A^+$ component of the current contributes\footnote{To avoid confusion, note that the large $A^-$ fields emitted by the proton with $P^+\rightarrow \infty$ correspond to a large $A^+$ field for the currents in the box.}, and the interaction of the worldline current with the external gauge field can be expressed as~\cite{Tarasov:2019rfp},
\begin{eqnarray}
j_W^{A^+}(x_i) = \big(\dot{x}^-_i + 2  \psi^-_i \psi^m_i  \partial_m \big) \delta(x^-_i) {\tilde V}(x_{\perp,i})\,,
\label{eq:W-current}
\end{eqnarray}
with  ${\tilde V}(x_{\perp,i})\equiv U_{[\infty, -\infty]}(x_{\perp,i})$, where 
\begin{eqnarray}
U_{[x,y]}(x_{\perp,i}) = \exp\Big\{-ig \int^x_y dx_i^- A^+(x_i^-, x_{\perp,i})\Big\}\,.
\end{eqnarray}

For unpolarized DIS, the product of the currents in Eq.~(\ref{eq:W-current}) at $\tau_2$ and $\tau_4$ gives the well-known ``dipole model" expression 
\begin{eqnarray}
{\rm Tr_c} \left({\tilde V}(x_\perp)\, {\tilde V}^\dag (y_\perp)\right)\,,
\end{eqnarray}
in the polarization tensor. For polarized DIS, the leading sub-eikonal contribution corresponds to $A_\perp\sim \frac{1}{P^+}\neq 0$, and finite transverse field strengths $F_{12}$. The corresponding worldline current can be written as 
\begin{eqnarray}
j_W^{A_\perp}(x_i) = \dot{x}^-_i \delta(x^-_i){\tilde V}(x_{\perp,i}) + 2 \,i g\, \psi^1_i \psi^2_i \,{\tilde V}_{\rm pol.}(x_{\perp,i})\,,
\label{efcurwithph}
\end{eqnarray}
where 
\begin{equation}
{\tilde V}_{\rm pol.}(x_{\perp,i}) = U_{[\infty, x]}(x_\perp)F_{12}(x_i)U_{[x, -\infty]}(x_\perp) \,.
\end{equation}
This expression for the ``polarized Wilson line" (which we see is very straightforward to obtain in the worldline formalism) was introduced in the context of polarized DIS at small $x$ previously in \cite{Kovchegov:2015pbl} and interpreted as above in \cite{Kovchegov:2018znm}. (See also 
\cite{Chirilli:2021lif} for a recent discussion of such Wilson line operators.)

It is now clear that one will obtain contributions of the type 
\begin{eqnarray}
 j_W^{A^+}(x_2) j_W^{A_\perp}(x_4) \rightarrow {\rm Tr_c} \left({\tilde V}(x_\perp)\, {\tilde V}_{\rm pol.}^\dag (y_\perp)\right)\,,
\label{eq:sub-eikonal}
\end{eqnarray}
which give the leading sub-eikonal ``polarized dipole" contributions to the antisymmetric spin-dependent part of the polarization tensor. The small $x$ QCD evolution of such operators has been discussed at length in the literature \cite{Kirschner:1983di,Kirschner:1994vc,Bartels:1995iu,Bartels:1996wc,Kovchegov:2015pbl,Kovchegov:2016weo,Kovchegov:2017jxc,Kovchegov:2018znm,Kovchegov:2020hgb,Cougoulic:2019aja,Boussarie:2019icw,Chirilli:2021lif}. 

However for operators sensitive to the anomaly (such as $g_1$ and its moments), keeping only the leading contributions in Eq.~(\ref{eq:sub-eikonal}) is problematic because it misses, already at quadratic order in the fields, terms like
\begin{equation}
   (\partial_\lambda A_1 (x_2))\,(\partial_\eta A_2 (x_4))\,,
\end{equation}
that contribute to Eq.~(\ref{eq:box1}) and, as we argued, are essential to reproduce the anomaly. Formally, these terms are of order $1/P^+$ relative to the leading sub-eikonal contribution in Eq.~(\ref{eq:sub-eikonal}) so one might imagine such an omission to be appropriate when $P^+\rightarrow \infty$. However this omission misses the anomaly entirely\footnote{One may argue that the anomaly is not sensitive to small $x_B$; this argument is not tenable since it must be recovered for any $x_B$ in the limit $Q^2\rightarrow \infty$.} which, as we have argued, is sensitive to both small $x_B$ and large $x_B$. Since the isosinglet axial vector current $J_{\mu}^5$ (which satisfies the anomaly equation) gives the dominant contribution to $g_1$ in both Bjorken and Regge asymptotics, a fundamental piece of physics is missed by restricting oneself to sub-eikonal contributions alone. 

What gives? One way to understand the breakdown of the eikonal power counting is to note that the contribution of the sub-sub-eikonal terms to the infrared divergence  $l^\mu/l^2\rightarrow 0$ can compensate for their relative suppression with $P^+$ for any finite $P^+$. A deeper reason is that the contribution of the anomaly results from zero modes that correspond to a global phase of the Dirac determinant in the QCD path integral that, as we discussed at length in Section~\ref{sec:WZW}, are best represented as collective modes; for a comprehensive discussion, we refer the reader to \cite{Leutwyler:1992yt}. 

It is important to appreciate that these observations don't just apply to the Regge limit but to the Bjorken limit as well. In the latter case, the temptation would be to argue that the anomaly contribution is a ``twist-four" contribution which would be suppressed relative to the leading twist anomaly free expression; this argument is equally fallacious because the twist expansion, like the eikonal expansion, does not apply to the zero modes that contribute to the anomaly. 

Finally, we return to the topic of higher order terms in Eq.~(\ref{eq:box1}). Employing so-called Wess-Zumino consistency conditions, it can be shown\footnote{For a formal derivation, see Section 9.3 in
\cite{Bilal:2008qx}. For a pedagogical demonstration using Feynman diagrams, we refer the reader to the lecture notes of Vasquez-Mozo~\cite{Vasquez-Mozo}.} that the order $O(A^3)$ terms combine with the quadratic terms to give the $F {\tilde F}$ structure of Eq.\,(\ref{eq:g1-Bj-Regge}), where 
$F_{\mu\nu}$ and ${\tilde F}_{\mu\nu}$ correspond respectively to the nonlinear QCD field strength tensor and its dual~\cite{Bardeen:1974ry}. Further commentary and references to the extension to the non-Abelian case of the Abelian Adler-Bardeen theorem~\cite{Adler:1969er}  can be found in Section 2.4 of Adler's historical review of the topic~\cite{Adler:2004qt}.

\bibliography{wlines}

\end{document}